\documentclass[aps,preprint,floatfix,nofootinbib]{revtex4}
\usepackage{epsfig,graphicx,color}

\begin{document}
\def\be{\begin{eqnarray}}
\def\en{\end{eqnarray}}
\def\la{\langle}
\def\ra{\rangle}
\def\non{\nonumber}
\def\ov{\overline}
\newcommand{\jhep}[1]{{\it JHEP}\ #1:}

\title{Flavor SU(3) symmetry and QCD factorization \\
in $B \to PP$ and $PV$ decays}

\author{Hai-Yang Cheng\footnote{Email: phcheng@phys.sinica.edu.tw}
and Sechul Oh\footnote{Email: scoh@phys.sinica.edu.tw}}

\affiliation{
Institute of Physics, Academia Sinica, Taipei 115, Taiwan }

\date{\today}% It is always \today, today,

\begin{abstract}
Using flavor SU(3) symmetry, we perform a model-independent analysis of charmless
$\bar B_{u,d} (\bar B_s) \to PP, ~PV$ decays.
All the relevant topological diagrams, including the presumably subleading diagrams, such as the
QCD- and EW-penguin exchange diagrams and flavor-singlet weak annihilation ones, are introduced.
Indeed, the QCD-penguin exchange diagram turns out to be important in understanding the data for
penguin-dominated decay modes.
In this work we make efforts to bridge the (model-independent but less quantitative) topological
diagram or flavor SU(3) approach and the (quantitative but somewhat model-dependent) QCD factorization
(QCDF) approach in these decays, by explicitly showing how to translate each flavor SU(3) amplitude
into the corresponding terms in the QCDF framework.
After estimating each flavor SU(3) amplitude numerically using QCDF, we discuss various
physical consequences, including SU(3) breaking effects and some useful SU(3) relations among decay
amplitudes of $\bar B_s \to PV$ and $\bar B_d \to PV$.
\end{abstract}

\maketitle %
%%%%%%%%%%%%%%%%%%%%%%%%%%%%%%%%%%%%%%%%%%%%%%%%%%%%%%%%%%%%%%%%%%%%%%%%%%%%%%%%%%%%%%%%%%
%%%%%%%%%%%%%%%%%%%%%%%%%%%%%%%%%%%%%%%%%%%%%%%%%%%%%%%%%%%%%%%%%%%%%%%%%%%%%%%%%%%%%%%%%%
\section{Introduction}

A large number of hadronic $B_{u,d}$ decay events have been collected  at the $B$ factories which enable
us to  make accurate measurements of branching fractions (BFs) and direct {\it CP} asymmetries for many
modes.  With the advent of the LHCb experiment, a tremendous amount of new experimental data on $B$
decays is expected to be obtained. In particular, various decay processes of heavier $B_s$ and $B_c$
mesons as well as very rare $B$ decay modes are expected to be observed.

In earlier\textbf{} works on hadronic decays of $B$ mesons, the factorization hypothesis, based on the
color transparency argument, was usually assumed to estimate the hadronic matrix elements which are
inevitably involved in theoretical calculations of the decay amplitudes for these processes. Under the
factorization assumption, the matrix element of a four-quark operator is expressed as a product of a
decay constant and a form factor. Naive factorization is simple but fails to describe color-suppressed
modes. This is ascribed to the fact that color-suppressed decays receive sizable nonfactorizable
contributions that have been neglected in naive factorization. Another issue is that the decay amplitude
under naive factorization is not truly physical because the renormalization scale and scheme dependence
of the Wilson coefficients $c_i(\mu)$ are not compensated by that of the matrix element
$\langle M_1M_2|O_i|B\rangle(\mu)$.
In the improved ``generalized factorization" approach~\cite{genFA1,genFA2}, nonfactorizable effects are
absorbed into the parameter $N_c^{\rm eff}$, the effective number of colors. This parameter can be
empirically determined from experiment.

With the advent of heavy quark effective theory, nonleptonic $B$ decays can be analyzed systematically
within the QCD framework.
There are three popular approaches available in this regard: QCD factorization (QCDF)~\cite{BBNS99},
perturbative QCD (pQCD)~\cite{pQCD} and soft-collinear effective theory (SCET)~\cite{SCET}.
In QCDF and SCET, power corrections of order $\Lambda_{\rm QCD}/m_b$ are often plagued by the end-point
divergence that in turn breaks the factorization theorem.  As a consequence, the estimate of power
corrections is generally model dependent and can only be studied in a phenomenological way.
In the pQCD approach, the endpoint singularity is cured by including the parton's transverse momentum.

%This naive factorization approach ignores the non-factorizable contributions from the soft
%interactions in the
%initial and final states. In order to compensate the non-factorizable contributions, the naive
%factorization scheme has been generalized by introducing the effective number of color $N_c$ as a
%phenomenological parameter. In this generalized factorization, the renormalization scheme and scale
%dependence in the hadronic matrix elements has been resolved.
%Theoretically, the QCD factorization (QCDF) approach has provided a novel method to study hadronic $B$
%decays. In this approach, the naive factorization contributions become the leading term and as
%sub-leading contributions, radiative corrections from hard gluon exchange can be systematically
%calculated by using the perturbative QCD method in the heavy quark limit, where suppressed power
%corrections of $O(\Lambda_{\rm QCD}/ m_b)$ are neglected. Since the %nonfactorizable contributions in
%the naive factorization, such as the contributions from hard scattering with the spectator quark in
%the $B$ meson and the contributions from weak annihilation, can be perturbatively computed, the
%phenomenological parameter $N_c$ used in the generalized factorization scheme is no longer needed to
%compensate the non-factorizable contributions.

Because a reliable evaluation of hadronic matrix elements is very difficult in general, an alternative
approach which is essentially model independent is based on the diagrammatic
approach~\cite{Chau,CC86,CC87}. In this approach, the topological diagrams are classified according to
the topologies of weak interactions with all strong interaction effects included.
Based on flavor SU(3) symmetry, this model-independent analysis enables us to extract the topological
amplitudes and sense the relative importance of different underlying decay mechanisms.  When enough
measurements are made with sufficient accuracy, we can extract the diagrammatic amplitudes from
experiment and compare to theoretical estimates, especially checking whether there are any significant
final-state interactions or whether the weak annihilation diagrams can be ignored as often asserted in
the literature. The diagrammatic approach was applied to hadronic $B$ decays first in~\cite{Chau91}.
Various topological amplitudes have been extracted from the data
in~\cite{Chiang,Chiang06,ChiangPV,Chiang:2003pm} after making some reasonable approximations.

Based on SU(3) flavor symmetry,  the decay amplitudes also can be decomposed into linear combinations
of the SU(3)$_{\rm F}$ amplitudes which are SU(3) reduced matrix
elements~\cite{Zeppenfeld:1980ex,Gronau:1994rj,Oh:1998wa,Deshpande:2000jp}.
This approach is equivalent to the diagrammatic approach when SU(3) flavor symmetry is imposed to the
latter.

%There exists a different approach which is purely based on flavor SU(3) symmetry
%(SU(3)$_{\rm F}$)~\cite{Gronau:1994rj,Oh:1998wa}.
%In this method, the decay amplitudes of two body $B$ decays are decomposed into linear combinations of
%the SU(3)$_{\rm F}$ amplitudes which are reduced matrix elements defined in Ref.~\cite{Gronau:1994rj}.
%This approach, though more general, lacks detailed predictions that one can make using the QCDF, pQCD
%or SCET approach.

In this work we make efforts to bridge these two different approaches, using QCDF and flavor SU(3)
symmetry, in $\bar B_{u,d} (\bar B_s) \to PP, ~PV$ decays.
For this aim, we first introduce all the relevant topological diagrams, including the presumably
subleading diagrams, such as the QCD- and EW-penguin exchange ones and flavor-singlet weak annihilation
ones,
%\footnote{These diagrams were ignored in Ref.~\cite{Gronau:1994rj}}
some of which turn out to be important especially in penguin-dominant decay processes.
Then all these decay modes are analyzed by using the {\it intuitive} topological diagrams and expressed
in terms of the SU(3)$_{\rm F}$ amplitudes.
Each SU(3)$_{\rm F}$ amplitude is \emph{translated into} the corresponding terms in the framework of
QCDF. Applying these relations, one can easily find the rather {\it sophisticated} results of the
relevant decay amplitudes calculated in the QCDF framework.
The magnitude and the strong phase of each SU(3)$_{\rm F}$ amplitude are numerically estimated in QCDF.
We further discuss some examples of the applications, including the effects of SU(3)$_{\rm F}$ breaking
and useful SU(3)$_{\rm F}$ relations among decay amplitudes.

This paper is organized as follows. In Sec.~II, we introduce topological quark diagrams relevant to
$\bar B_{u,d} (\bar B_s) \to PP, ~PV$ decays and the framework of QCDF.  The explicit SU(3)$_{\rm F}$
decomposition of the decay amplitudes and the relations between the SU(3)$_{\rm F}$ amplitudes and
the QCDF terms are presented.
In Sec.~III, we make a numerical estimation of the SU(3)$_{\rm F}$ amplitudes and discuss its
consequences and some applications.  Our conclusions are given in Sec.~IV.

%%%%%%%%%%%%%%%%%%%%%%%%%%%%%%%%%%%%%%%%%%%%%%%%%%%%%%%%%%%%%%%%%%%%%%%%%%%%%%%%%%%%%%%%%%
%%%%%%%%%%%%%%%%%%%%%%%%%%%%%%%%%%%%%%%%%%%%%%%%%%%%%%%%%%%%%%%%%%%%%%%%%%%%%%%%%%%%%%%%%%
\section{Flavor SU(3) analysis and QCD factorization}

It has been established sometime ago that a least model-dependent analysis of heavy meson decays can
be carried out in the so-called quark-diagram approach.~\footnote{It is also referred to as the flavor-flow
diagram or topological-diagram approach in the literature.}
In this diagrammatic scenario, all two-body nonleptonic weak decays of heavy mesons can be expressed in
terms of six distinct quark diagrams~\cite{Chau,CC86,CC87}:~\footnote{Historically, the quark-graph
amplitudes $T,\,C,\,E,\,A,\,P$ named in~\cite{Gronau:1994rj} were originally denoted by
$A,\,B,\,C,\,D,\,E$, respectively, in~\cite{CC86,CC87}. }
$T$, the color-allowed external $W$-emission tree diagram; $C$, the color-suppressed internal
$W$-emission diagram; $E$, the $W$-exchange diagram; $A$, the $W$-annihilation diagram; $P$, the
horizontal $W$-loop diagram; and $V$, the vertical $W$-loop diagram.  (The one-gluon exchange
approximation of the $P$ graph is the so-called ``penguin diagram''.)  For the analysis of charmless
$B$ decays, one adds the variants of the penguin diagram such as the electroweak penguin and the
penguin annihilation and singlet penguins, as will be discussed below.
It should be stressed that these diagrams are classified according to the topologies of weak
interactions with all strong interaction effects encoded, and hence they are {\it not} Feynman graphs.
All quark graphs used in this approach are topological and meant to have all the strong interactions
included, {\it i.e.}, gluon lines are included implicitly in all possible ways.  Therefore, analyses
of topological graphs can provide information on final-state interactions (FSIs).

\begin{figure}[t]
\vspace*{1ex}
\includegraphics[width=4in]{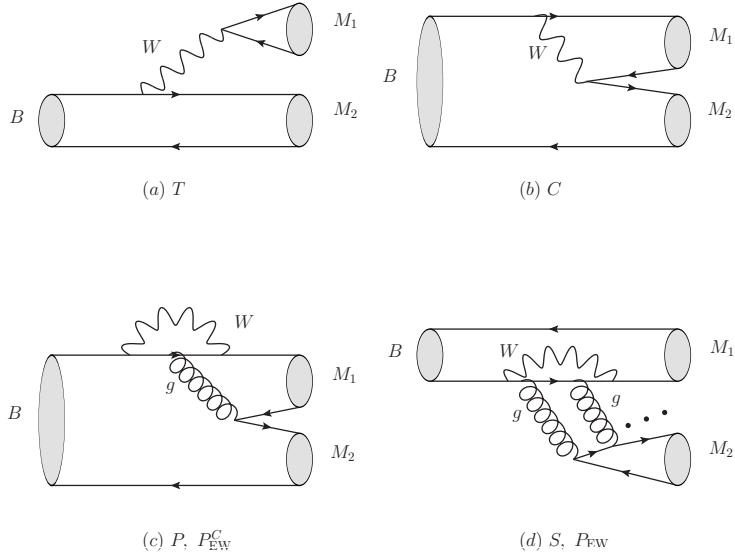}
\vspace*{-1ex}
\caption{Topology of possible diagrams:
(a) Color-allowed tree $[T]$, (b) Color-suppressed tree $[C]$,
(c) QCD-penguin $[P]$, (d) Singlet QCD-penguin $[S]$ diagrams with 2 (3) gluon lines for
$M_2$ being a pseudoscalar meson $P$ (a vector meson $V$).
The color-suppressed EW-penguin $[P_{\rm EW}^C]$ and color-favored EW-penguin
$[P_{\rm EW}]$ diagrams are obtained by replacing the gluon line from (c) and all the gluon lines
from (d), respectively, by a single $Z$-boson or photon line. }
\label{Fig1}
\end{figure}

\begin{figure}[th]
\vspace*{1ex}
\includegraphics[width=4in]{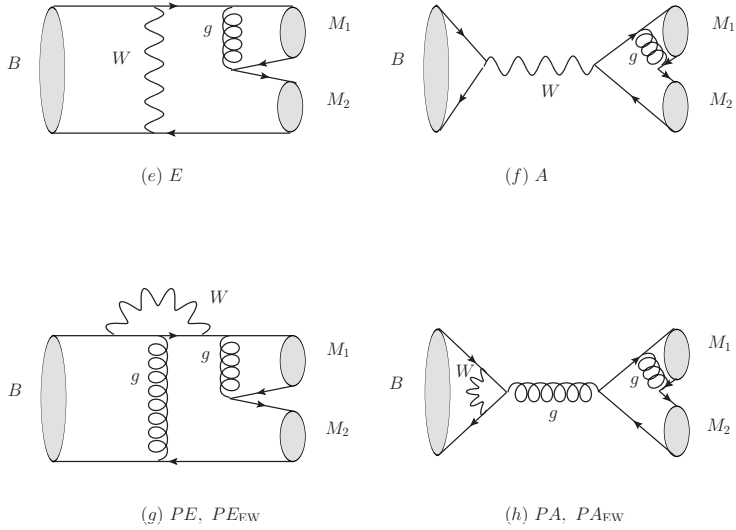}
\vspace*{-1ex}
\caption{(e) $W$-exchange $[E]$, (f) $W$-annihilation $[A]$, (g) QCD-penguin exchange $[PE]$,
(h) QCD-penguin annihilation $[PA]$ diagrams.
The EW-penguin exchange $[PE_{\rm EW}]$ and EW-penguin annihilation $[PA_{\rm EW}]$
diagrams are obtained from (g) and (h), respectively, by replacing the left gluon line
by a single $Z$-boson or photon line.
The gluon line of (e) and (f) and the right gluon line of (g) and (h) can be attached
to the fermion lines in all possible ways.}
\label{Fig2}
\end{figure}

\begin{figure}[th]
\vspace*{1ex}
\includegraphics[width=4in]{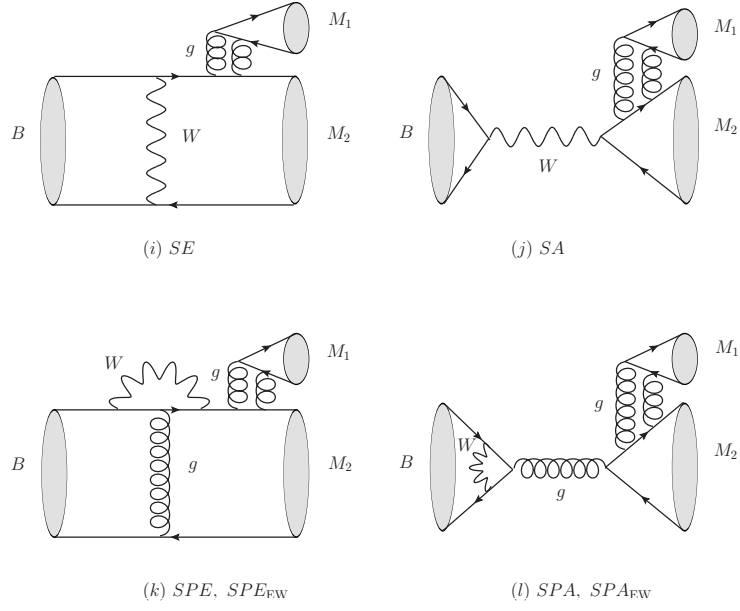}
\vspace*{-1ex}
\caption{(i) Flavor-singlet $W$-exchange $[SE]$, (j) Flavor-singlet $W$-annihilation $[SA]$,
(k) Flavor-singlet QCD-penguin exchange $[SPE]$,
(l) Flavor-singlet QCD-penguin annihilation $[SPA]$ diagrams.
The Flavor-singlet EW-penguin exchange $[SPE_{\rm EW}]$ and flavor-singlet EW-penguin annihilation
$[SPA_{\rm EW}]$ diagrams are obtained from (k) and (l), respectively, by replacing
the leftest gluon line by a single $Z$-boson or photon line.
The double gluon lines of (i), (j), (k) and (l) are shown for the case of $M_1 = P$.
They are replaced by three gluon lines when $M_1 = V$.  Each of the gluon lines of (i), (j),
(k) and (l) can be separately attached to the fermion lines in all possible ways.}
\label{Fig3}
\end{figure}

In SU(3)$_{\rm F}$ decomposition of the decay amplitudes for $\bar B_{u,d} (\bar B_s) \to M_1 M_2$
(with $M_1 M_2 = P_1 P_2, ~PV, ~VP$) modes~\cite{Gronau:1994rj}, we represent the decay amplitudes in
terms of topological quark diagram contributions.
The topological amplitudes which will be referred to as SU(3)$_{\rm F}$  amplitudes hereafter,
corresponding to different topological quark diagrams, as shown in
Figs.~\ref{Fig1}-\ref{Fig3}, can be classified into three distinct groups as follows:  \\
\noindent
(i) {\it Tree and penguin amplitudes} \\
\mbox{}~~~
$T$ : color-favored tree amplitude (equivalently, external $W$-emission),  \\
\mbox{}~~~
$C$ : color-suppressed tree amplitude (equivalently, internal $W$-emission),  \\
\mbox{}~~~
$P$ : QCD-penguin amplitude,  \\
\mbox{}~~~
$S$ : singlet QCD-penguin amplitude involving SU(3)$_{\rm F}$-{\it singlet} mesons
 (e.g., $\eta^{(\prime)}, ~\omega, ~\phi$), \\
\mbox{}~~~
$P_{\rm EW}$ : color-favored EW-penguin amplitude, \\
\mbox{}~~~
$P_{\rm EW}^C$ : color-suppressed EW-penguin amplitude, \\
(ii) {\it Weak annihilation amplitudes} \\
\mbox{}~~~
$E$ : $W$-exchange amplitude, \\
\mbox{}~~~
$A$ : $W$-annihilation amplitude, \\
\mbox{}~~~
($E$ and $A$ are often jointly called ``weak annihilation''.) \\
\mbox{}~~~
$PE$ : QCD-penguin exchange amplitude, \\
\mbox{}~~~
$PA$ : QCD-penguin annihilation amplitude, \\
\mbox{}~~~
$PE_{\rm EW}$ : EW-penguin exchange amplitude, \\
\mbox{}~~~
$PA_{\rm EW}$ : EW-penguin annihilation amplitude, \\
\mbox{}~~~
($PE$ and $PA$ are also jointly called ``weak penguin annihilation''.) \\
(iii) {\it Flavor-singlet weak annihilation amplitudes}: all involving SU(3)$_{\rm F}$-{\it singlet}
mesons~\footnote{The singlet amplitudes $SE$ and $SA$ were first discussed in~\cite{Li,hairpin} as the
disconnected hairpin amplitudes and denoted by $E_h$ and $A_h$, respectively, in \cite{hairpin}.}
\\
\mbox{}~~~
$SE$ : singlet $W$-exchange amplitude, \\
\mbox{}~~~
$SA$ : singlet $W$-annihilation amplitude,  \\
\mbox{}~~~
$SPE$ : singlet QCD-penguin exchange amplitude,  \\
\mbox{}~~~
$SPA$ : singlet QCD-penguin annihilation amplitude,  \\
\mbox{}~~~
$SPE_{\rm EW}$ : singlet EW-penguin exchange amplitude,  \\
\mbox{}~~~
$SPA_{\rm EW}$ : singlet EW-penguin annihilation amplitude.  \\

%%%%%%%%%%%%%%%%%%%%%%%%%%%%%%%%%%%
%%%%%%%%%%%%%%%%%%%%%%%%%%%%%%%%%%%

%%%%%%%%%%%%%%%%%%%%%%%%%%%%%%%%%%%
%%%%%%%%%%%%%%%%%%%%%%%%%%%%%%%%%%%

%%%%%%%%%%%%%%%%%%%%%%%%%%%%%%%%%%%
%%%%%%%%%%%%%%%%%%%%%%%%%%%%%%%%%%%

Within the framework of QCD factorization \cite{Beneke:1999br}, the effective Hamiltonian matrix
elements for $\bar B \to M_1 M_2$ $(M_1 M_2 = P_1 P_2, PV)$ are written in the form
\begin{equation}
 \langle M_1 M_2 | {\cal H}_{\rm eff} |\bar B \rangle
 = \frac{G_F}{\sqrt2} \sum_{p=u,c}  \lambda_p^r ~
  \langle M_1 M_2| {\cal T_A}^p + {\cal T_B}^p |\bar B \rangle ~,
\label{fac}
\end{equation}
where the Cabibbo-Kobayashi-Maskawa (CKM) factor $\lambda_p^r \equiv V_{pb} V_{pr}^*$ with $r = s, d$.
${\cal T_A}^p$ describes contributions from naive factorization, vertex corrections, penguin contractions
and spectator scattering expressed in terms of the flavor operators $a_i^p$, while ${\cal T_B}^p$ contains
annihilation topology amplitudes characterized by  the annihilation operators $b_j^p$.
The flavor operators $a_i^p$ are basically the Wilson coefficients in conjunction with
short-distance nonfactorizable corrections such as vertex corrections and hard spectator interactions.
In general, they have the expressions \cite{Beneke:1999br, Beneke:2003zv}
\begin{eqnarray}
 a_i^p (M_1M_2) =
 \left( c_i + {c_{i \pm 1} \over N_c} \right) N_i (M_2)
 + {c_{i \pm 1} \over N_c}~ {C_F \alpha_s \over 4 \pi}
  \Big[ V_i (M_2) +{4 \pi^2 \over N_c} H_i (M_1 M_2) \Big] +P_i^p (M_2) ~,
\label{eq:ai}
\end{eqnarray}
where $i=1,\cdots,10$, the upper (lower) signs apply when $i$ is odd (even), $c_i$ are the Wilson
coefficients, $C_F=(N_c^2-1)/(2N_c)$ with $N_c=3$, $M_2$ is the emitted meson and $M_1$ shares the same
spectator quark with the $B$ meson.
The quantities $N_i (M_2) = 0$ or 1 for $i = 6,8$ and $M_2=V$ or else, respectively.
The quantities $V_i (M_2)$ account for vertex corrections, $H_i (M_1M_2)$ for hard spectator
interactions with a hard gluon exchange between the emitted meson and the spectator quark of the $B$
meson and $P_i(M_2)$ for penguin contractions.

The weak annihilation contributions to the decay $\bar B \to M_1 M_2$ $(M_1 M_2 = P_1 P_2, ~PV, ~VP)$
can be described in terms of the building blocks $b_j^p$ and $b_{j, {\rm EW}}^p$~:
\begin{eqnarray}
 \frac{G_F}{\sqrt2} \sum_{p=u,c} \lambda_p^r
  ~\langle M_{1}M_2| {\cal T_B}^p |\bar B \rangle
 &=& i \frac{G_F}{\sqrt{2}} \sum_{p=u,c} \lambda_p^r ~
  f_B f_{M_1} f_{M_2} \sum_j (d_j b_j^p +d'_j b_{j,{\rm EW}}^p ).
\label{weak_ann}
\end{eqnarray}
The building blocks have the expressions~\cite{Beneke:1999br}
\begin{eqnarray}
 b_1 &=& {C_F \over N_c^2} c_1 A_1^i, \qquad \quad
  b_3^p = {C_F \over N_c^2} \left[ c_3 A_1^i +c_5 (A_3^i +A_3^f) +N_c c_6 A_3^f \right], \nonumber \\
 b_2 &=& {C_F \over N_c^2 }c_2 A_1^i, \qquad \quad
  b_4^p = {C_F \over N_c^2} \left[ c_4 A_1^i +c_6 A_2^f \right], \nonumber \\
 b_{\rm 3,EW}^p &=& {C_F \over N_c^2} \left[ c_9 A_1^i +c_7 (A_3^i +A_3^f) +N_c c_8 A_3^i \right],
 \nonumber \\
 b_{\rm 4,EW}^p &=& {C_F \over N_c^2} \left[ c_{10}A _1^i +c_8 A_2^i \right].
\end{eqnarray}
The subscripts 1,2,3 of $A_n^{i,f}$ denote the annihilation amplitudes induced from $(V-A)(V-A)$,
$(V-A)(V+A)$ and $(S-P)(S+P)$ operators, respectively, and the superscripts $i$ and $f$ refer to gluon
emission from the initial and final-state quarks, respectively. We choose the convention that $M_1$
contains an antiquark from the weak vertex and $M_2$ contains a quark from the weak vertex, as in
Ref.~\cite{Beneke:2003zv}.

For the explicit expressions of vertex, hard spectator corrections and annihilation contributions, we
refer to \cite{Beneke:1999br, Beneke:2003zv, Beneke:2006hg} for details.
In practice, it is more convenient to express the decay amplitudes in terms of the flavor operators
$\alpha_i^p$ and the annihilation operators  $\beta_j^p$ which are related to the coefficients
$a_i^p$ and $b_j^p$ by
\begin{eqnarray}
 \alpha_1 (M_1 M_2) &=& a_1 (M_1 M_2) ~,  \qquad  \qquad
  \alpha_2 (M_1 M_2) = a_2 (M_1 M_2) ~,  \nonumber \\
 \alpha_3^p (M_1 M_2) &=& \left\{
   \begin{array}{cl}
     a_3^p (M_1 M_2) - a_5^p (M_1 M_2) & \quad \mbox{for~} M_1 M_2 = PP, ~VP ~, \\
     a_3^p (M_1 M_2) + a_5^p (M_1 M_2) & \quad \mbox{for~} M_1 M_2 = PV ~,
   \end{array}\right. \nonumber \\
 \alpha_4^p (M_1 M_2) &=& \left\{
   \begin{array}{cl}
     a_4^p (M_1 M_2) + r_{\chi}^{M_2}~ a_6^p (M_1 M_2)
       & \quad \mbox{for~} M_1 M_2 = PP, ~PV ~, \\
     a_4^p (M_1 M_2) - r_{\chi}^{M_2}~ a_6^p (M_1 M_2)
       & \quad \mbox{for~} M_1 M_2 = VP ~,
   \end{array} \right. \\
 \alpha_{3,\rm EW}^p (M_1 M_2) &=& \left\{
   \begin{array}{cl}
     a_9^p (M_1 M_2) - a_7^p (M_1 M_2) & \quad \mbox{for~} M_1 M_2 = PP, ~VP ~, \\
     a_9^p (M_1 M_2) + a_7^p (M_1 M_2) & \quad \mbox{for~} M_1 M_2 = PV  ~,
   \end{array} \right. \nonumber\\
 \alpha_{4,\rm EW}^p (M_1 M_2) &=& \left\{
   \begin{array}{cl}
     a_{10}^p (M_1 M_2) + r_{\chi}^{M_2}~ a_8^p (M_1 M_2)
       & \quad \mbox{for~} M_1 M_2 = PP, ~PV ~, \\
     a_{10}^p (M_1 M_2) - r_{\chi}^{M_2}~ a_8^p (M_1 M_2)
       & \quad \mbox{for~} M_1 M_2 = VP ~,
     \end{array} \right. \nonumber
\end{eqnarray}
and
\begin{eqnarray}
 \beta_j^p (M_1 M_2) = \frac{i f_B f_{M_1} f_{M_2}}{X^{(\bar B M_1, M_2)}}~ b_j^p (M_1 M_2) ~,
\end{eqnarray}
where $f_M$ is the decay constant of a meson $M$ and the chiral factors $r_\chi^{M_2}$ are given by
\begin{eqnarray}
 r_\chi^P(\mu) = {2m_P^2 \over m_b(\mu)(m_2+m_1)(\mu)},  \qquad
 r_\chi^V(\mu) = \frac{2m_V}{m_b(\mu)} ~\frac{f_V^\perp (\mu)}{f_V} ~,
\end{eqnarray}
with $f_V^\perp (\mu)$ being the scale-dependent transverse decay constant of the vector meson $V$.
The relevant factorizable matrix elements are
\begin{eqnarray}
 X^{(\bar B P_1, P_2)} &\equiv& \langle P_2| J^{\mu} |0 \rangle
  \langle P_1| J'_{\mu} |\overline B \rangle
  =i f_{P_2} (m_{B}^2 -m^2_{P_1}) ~F_0^{B P_1} (m_{P_2}^2) ~,  \nonumber \\
 X^{(\bar BP, V)} &\equiv & \langle V| J^{\mu} |0 \rangle
  \langle P| J'_{\mu} |\overline B \rangle
  =2 f_V ~m_B p_c ~F_1^{BP} (m_{V}^2) ~,  \nonumber \\
 X^{(\bar B V,P)} &\equiv & \langle P| J^{\mu} |0 \rangle
  \langle V| J'_{\mu} |\overline B \rangle
  = 2 f_P ~m_B p_c ~A_0^{BV} (m_{P}^2) ~,
\end{eqnarray}
with $p_c$ being the c.m. momentum.  Here we have followed the conventional Bauer-Stech-Wirbel
definition for form factors $F_{0,1}^{BP}$ and $A_0^{BV}$~\cite{Bauer:1986bm}.

%%%%%%%%%%%%%%%%%%%%%%%%%%%%%%%%%%%%%%%%%%%%%%%%%%%%%%%%%%%%%%%%%%
%%%%%%%%%%%%%%%%%%%%%%%%%%%%%%%%%%%%%%%%%%%%%%%%%%%%%%%%%%%%%%%%%%
\subsection{SU(3)$_{\rm F}$ decomposition of decay amplitudes}

The decay amplitudes of  $\bar B_{u,d} (\bar B_s) \to M_1 M_2$ modes can be analyzed by the relevant
quark diagrams and written in terms of the SU(3)$_{\rm F}$ amplitudes.  The decomposition of the
decay amplitudes of these modes is displayed in Tables~\ref{B_PP_DelS0_1}$-$\ref{Bs_PV_DelS1_3}.
In these tables, the subscript $M_1$ (or $M_2$) of the amplitudes $T_{M_1 ~[M_2]}^{(\zeta)}$, $\cdots$, etc.,
represents the case that the meson $M_1$ (or $M_2$) contains the spectator quark in the final state.
The superscript $\zeta$ of the amplitudes is only applied to the case involving an $\eta^{(\prime)}$ or
an $\omega / \phi$, or both $\eta^{(\prime)}$ and $\omega / \phi$ in the final state and denotes the
quark content ($\zeta = q, s, c$) of $\eta^{(\prime)}$ and $\omega / \phi$ with $q = u$ or $d$.
The value of $\zeta$ is shown in the parenthesis as $(q)$, $(s)$ or $(c)$.
For $\bar B^0 (\bar B_s) \to \eta^{(\prime)} \eta^{(\prime)}$, two values of $\zeta$ are shown in one
parenthesis: e.g., $(q, s)$, where $q$ and $s$ denote the quark content of the first and second
$\eta^{(\prime)}$, respectively. A similar rule is also applied to the case of
$\bar B^0 (\bar B_s) \to \eta^{(\prime)} ~\omega/\phi$.
On the other hand, to distinguish the decays with $|\Delta S| = 1$ from those with $\Delta S = 0$, we will
put the prime to all the SU(3)$_{\rm F}$ amplitudes for the former, for example
$T^{\prime (\zeta)}_{M_1 ~[M_2]}$.
The SU(3)$_{\rm F}$-singlet amplitudes $S^{(\prime)}_{M_1 [M_2]}$ are involved only when the
SU(3)$_{\rm F}$-singlet meson(s) ($\eta, ~\eta', ~\omega, ~\phi$) appear(s) in the final state.

We will give some examples for illustration.  The decay amplitude of $B^- \to \pi^- \bar K^0$ which is
a $\bar B_u \to PP$ mode with $|\Delta S| = 1$ can be written, from
Tables~\ref{B_PP_DelS1_1}$-$\ref{B_PP_DelS1_2}, as
\begin{eqnarray}
 {\cal A}_{B^- \to \pi^- \bar K^0}
 &=& P'_{\pi} -\frac{1}{3} P_{{\rm EW}, \pi}^{C \prime} +A'_{\pi} +PE'_{\pi}
   +\frac{2}{3} PE'_{{\rm EW}, \pi} ~.
\label{B_pi1K0}
\end{eqnarray}
The decay amplitude of $\bar B^0 \to \eta^{(\prime)} \bar K^{*0}$ which is a $\bar B_d \to PV$ mode with
$|\Delta S| = 1$ can be recast, from Tables~\ref{B_PV_DelS1_1}$-$\ref{B_PV_DelS1_3}, to
\begin{eqnarray}
 && \mbox{} \hspace{-0.6cm}
   \sqrt{2}~ {\cal A}_{\bar B^0 \to \eta^{(\prime)} \bar K^{*0}}  \nonumber \\
 &=& \Big( C_{K^*}^{\prime (q)} +2 S_{K^*}^{\prime (q)}
   + \frac{1}{3} P_{{\rm EW}, ~K^*}^{\prime (q)}
   +2 SPE_{K^*}^{\prime (q)}
   - \frac{2}{3} SPE_{{\rm EW}, ~K^*}^{\prime (q)} \Big)  \nonumber \\
 && + \sqrt{2}~ \Big( S_{K^*}^{\prime (s)} + P_{K^*}^{\prime (s)}
   - \frac{1}{3} P_{{\rm EW}, ~K^*}^{\prime (s)} - \frac{1}{3} P_{{\rm EW}, ~K^*}^{C \prime, ~(s)}
   + PE_{K^*}^{\prime (s)} - \frac{1}{3} PE_{{\rm EW}, ~K^*}^{\prime (s)}  \nonumber \\
 && + SPE_{K^*}^{\prime (s)}
   - \frac{1}{3} SPE_{{\rm EW}, ~K^*}^{\prime (s)} \Big) \nonumber \\
 && + \sqrt{2}~ \Big( C_{K^*}^{\prime (c)} + S_{K^*}^{\prime (c)} \Big)  \nonumber \\
 && + \Big( P_{\eta^{(\prime)}}^{\prime (q)}
   - \frac{1}{3} P_{{\rm EW}, ~\eta^{(\prime)}}^{C \prime, ~(q)}
   + PE_{\eta^{(\prime)}}^{\prime (q)}
   - \frac{1}{3} PE_{{\rm EW}, ~\eta^{(\prime)}}^{\prime (q)} \Big) ~,
\label{B_etaKst0}
\end{eqnarray}
where the superscripts $(q)$, $(s)$ and $(c)$ represent the quark contents of $\eta^{(\prime)}$,
such as $\eta^{(\prime)}_q$, $\eta^{(\prime)}_s$ and $\eta^{(\prime)}_c$, respectively.
Likewise, from Tables~\ref{B_PV_DelS0_1}$-$\ref{B_PV_DelS0_3}, the decay amplitude of
$\bar B^0 \to \eta^{(\prime)} ~\omega / \phi$ which is a $\bar B_d \to PV$ mode with $\Delta S = 0$ is
given by
\begin{eqnarray}
 2~ {\cal A}_{\bar B^0 \to \eta^{(\prime)} ~\omega / \phi}
 &=& \Big( C_{\eta^{(\prime)}}^{(q, ~q)} +2 S_{\eta^{(\prime)}}^{(q, ~q)}
   + P_{\eta^{(\prime)}}^{(q, ~q)} + \frac{1}{3} P_{{\rm EW}, ~\eta^{(\prime)}}^{(q, ~q)}
   - \frac{1}{3} P_{{\rm EW}, ~\eta^{(\prime)}}^{C, ~(q, ~q)}  \nonumber \\
 && + E_{\eta^{(\prime)}}^{(q, ~q)} + PE_{\eta^{(\prime)}}^{(q, ~q)}
   +2 PA_{\eta^{(\prime)}}^{(q, ~q)}
   - \frac{1}{3} PE_{{\rm EW}, ~\eta^{(\prime)}}^{(q, ~q)}
   + \frac{1}{3} PA_{{\rm EW}, ~\eta^{(\prime)}}^{(q, ~q)}  \nonumber \\
 && +2 SE_{\eta^{(\prime)}}^{(q, ~q)} +2 SPE_{\eta^{(\prime)}}^{(q, ~q)}
   +4 SPA_{\eta^{(\prime)}}^{(q, ~q)}
   - \frac{2}{3} SPE_{{\rm EW}, ~\eta^{(\prime)}}^{(q, ~q)}
   + \frac{2}{3} SPA_{{\rm EW}, ~\eta^{(\prime)}}^{(q, ~q)} \Big)
   \nonumber \\
 &+& \sqrt{2}~ \Big( S_{\eta^{(\prime)}}^{(q, ~s)}
   - \frac{1}{3} P_{{\rm EW}, ~\eta^{(\prime)}}^{(q, ~s)}
   + SE_{\eta^{(\prime)}}^{(q, ~s)} + SPE_{\eta^{(\prime)}}^{(q, ~s)}
   +2 SPA_{\eta^{(\prime)}}^{(q, ~s)}  \nonumber \\
 && - \frac{1}{3} SPE_{{\rm EW}, ~\eta^{(\prime)}}^{(q, ~s)}
   + \frac{1}{3} SPA_{{\rm EW}, ~\eta^{(\prime)}}^{(q, ~s)} \Big)  \nonumber \\
 &+& \sqrt{2}~ \Big( 2 SPA_{\eta^{(\prime)}}^{(s, ~q)}
   - \frac{2}{3} SPA_{{\rm EW}, ~\eta^{(\prime)}}^{(s, ~q)} \Big)  \nonumber \\
 &+& 2~ \Big( PA_{\eta^{(\prime)}}^{(s, ~s)}
   - \frac{1}{3} PA_{{\rm EW}, ~\eta^{(\prime)}}^{(s, ~s)}
   + SPA_{\eta^{(\prime)}}^{(s, ~s)}
   - \frac{1}{3} SPA_{{\rm EW}, ~\eta^{(\prime)}}^{(s, ~s)} \Big)  \nonumber \\
 &+& \Big( C_{\omega / \phi}^{(q, ~q)} +2 S_{\omega / \phi}^{(q, ~q)}
   + P_{\omega / \phi}^{(q, ~q)} + \frac{1}{3} P_{{\rm EW}, ~\omega / \phi}^{(q, ~q)}
   - \frac{1}{3} P_{{\rm EW}, ~\omega / \phi}^{C, ~(q, ~q)}  \nonumber \\
 && + E_{\omega / \phi}^{(q, ~q)} + PE_{\omega / \phi}^{(q, ~q)}
   +2 PA_{\omega / \phi}^{(q, ~q)}
   - \frac{1}{3} PE_{{\rm EW}, ~\omega / \phi}^{(q, ~q)}
   + \frac{1}{3} PA_{{\rm EW}, ~\omega / \phi}^{(q, ~q)}  \nonumber \\
 && +2 SE_{\omega / \phi}^{(q, ~q)} +2 SPE_{\omega / \phi}^{(q, ~q)}
   +4 SPA_{\omega / \phi}^{(q, ~q)}
   - \frac{2}{3} SPE_{{\rm EW}, ~\omega / \phi}^{(q, ~q)}
   + \frac{2}{3} SPA_{{\rm EW}, ~\omega / \phi}^{(q, ~q)} \Big)
   \nonumber \\
 &+& \sqrt{2}~ \Big( S_{\omega / \phi}^{(q, ~s)}
   - \frac{1}{3} P_{{\rm EW}, ~\omega / \phi}^{(q, ~s)}
   + SE_{\omega / \phi}^{(q, ~s)} + SPE_{\omega / \phi}^{(q, ~s)}
   +2 SPA_{\omega / \phi}^{(q, ~s)}  \nonumber \\
 && - \frac{1}{3} SPE_{{\rm EW}, ~\omega / \phi}^{(q, ~s)}
   + \frac{1}{3} SPA_{{\rm EW}, ~\omega / \phi}^{(q, ~s)} \Big)  \nonumber \\
 &+& \sqrt{2}~ \Big( C_{\omega / \phi}^{(q, ~c)} +S_{\omega / \phi}^{(q, ~c)} \Big)  \nonumber \\
 &+& \sqrt{2}~ \Big( 2 SPA_{\omega / \phi}^{(s, ~q)}
   - \frac{2}{3} SPA_{{\rm EW}, ~\omega / \phi}^{(s, ~q)} \Big)  \nonumber \\
 &+& 2~ \Big( PA_{\omega / \phi}^{(s, ~s)}
   - \frac{1}{3} PA_{{\rm EW}, ~\omega / \phi}^{(s, ~s)}
   + SPA_{\omega / \phi}^{(s, ~s)}
   - \frac{1}{3} SPA_{{\rm EW}, ~\omega / \phi}^{(s, ~s)} \Big) ~,
\label{B_etaomega}
\end{eqnarray}
where the superscripts $(q^{\prime}, ~q^{\prime \prime})$ with $q^{\prime}, q^{\prime \prime}
= q, s$ denote the quark contents of $(\eta^{(\prime)}, ~\omega / \phi)$, such as
$(\eta^{(\prime)}_q, ~\omega_q / \phi_q)$, $(\eta^{(\prime)}_q, ~\omega_s / \phi_s)$, etc.
When ideal mixing for $\omega$ and $\phi$ is assumed, $\omega_s$ and $\phi_q$ terms vanish: i.e.,
the amplitudes with the superscripts $(q, ~s)$ or $(s, ~s)$ for $B \to \eta^{(\prime)} \omega$ and
the superscripts $(q, ~q)$ or $(s, ~q)$ for $B \to \eta^{(\prime)} \phi$ are set to be zero.

%%%%%%%%%%%%%%%%%%%%%%%%%%%%%%%%%%%%%%%%%%%%%%%%%%%%%%%%%%%%%%%%%%
%%%%%%%%%%%%%%%%%%%%%%%%%%%%%%%%%%%%%%%%%%%%%%%%%%%%%%%%%%%%%%%%%%
\subsection{The SU(3)$_{\rm F}$ amplitudes and QCD factorization}

The SU(3)$_{\rm F}$ amplitudes for $\bar B_{u,d} (\bar B_s) \to M_1 M_2$ (with
$M_1 M_2 = P_1 P_2, ~PV, ~VP$) decays can be expressed in terms of the quantities calculated in the
framework of QCD factorization as follows:~\footnote{The factorizable amplitude $X^{(\bar BM_1,~M_2)}$
is denoted by $A_{M_1\!M_2}$ in~\cite{Beneke:2003zv}.}
\begin{eqnarray}
 T^{(\zeta)}_{M_1 [M_2]}
   &=& {G_F \over \sqrt{2}} ~~\lambda_u^r ~\alpha_1 (M_1M_2)
   ~X^{(\bar B M_1, ~M_2)} ~[X^{(\bar B M_2, ~M_1)}]~,  \nonumber \\
 C^{(\zeta)}_{M_1 [M_2]}
   &=& {G_F \over \sqrt{2}} ~~\lambda_u^r ~\alpha_2 (M_1M_2)
   ~X^{(\bar B M_1, ~M_2)} ~[X^{(\bar B M_2, ~M_1)}]~,  \nonumber \\
 S^{(\zeta)}_{M_1 [M_2]}
   &=& {G_F \over \sqrt{2}} ~\sum_{p=u,c} ~\lambda_p^r ~\alpha_3^p (M_1M_2)
   ~X^{(\bar B M_1, ~M_2)} ~[X^{(\bar B M_2, ~M_1)}]  ~,  \nonumber \\
 P^{(\zeta)}_{M_1 [M_2]}
   &=& {G_F \over \sqrt{2}} ~\sum_{p=u,c} ~\lambda_p^r ~\alpha_4^p (M_1M_2)
   ~X^{(\bar B M_1, ~M_2)} ~[X^{(\bar B M_2, ~M_1)}]  ~,  \nonumber \\
 P_{{\rm EW}, M_1 [M_2]} ^{(\zeta)}
   &=& {G_F \over \sqrt{2}} ~\sum_{p=u,c} ~\lambda_p^r
   ~{3 \over 2} \alpha_{3, {\rm EW}}^p (M_1M_2)
   ~X^{(\bar B M_1, ~M_2)} ~[X^{(\bar B M_2, ~M_1)}]  ~, \nonumber \\
 P_{{\rm EW}, M_1 [M_2]} ^{C, ~(\zeta)}
   &=& {G_F \over \sqrt{2}} ~\sum_{p=u,c} ~\lambda_p^r
   ~{3 \over 2} \alpha_{4, {\rm EW}}^p (M_1M_2)
   ~X^{(\bar B M_1, ~M_2)} ~[X^{(\bar B M_2, ~M_1)}]  ~,
\label{SU3_T_Pew}
\end{eqnarray}
where the superscript $\zeta = q, s, c$, which is only applied to the case when $M_1$ (or $M_2$)
$= \eta^{(\prime)}$ or $\omega / \phi$, or $M_1 M_2 = \eta^{(\prime)} \eta^{(\prime)}$ or
$\eta^{(\prime)} \omega / \phi$.
As mentioned before, for $|\Delta S| = 1$ decays, we will put the prime to all the SU(3)$_{\rm F}$
amplitudes.
The {\it unprimed} and {\it primed} amplitudes have the CKM factor $\lambda_p^r \equiv V_{pb} V^*_{pr}$
with $r = d$ and $r = s$, respectively.
The weak annihilation amplitudes are given by
\begin{eqnarray}
 E^{(\zeta)}_{M_1 [M_2]}
 &=& {G_F \over \sqrt{2}}~ \lambda_u^r ~(i f_B f_{M_1} f_{M_2})
   ~\left[ b_1 \right]_{M_1 M_2 ~[M_2 M_1]}~, \nonumber \\
 A^{(\zeta)}_{M_1 [M_2]}
 &=& {G_F \over \sqrt{2}}~ \lambda_u^r ~(i f_B f_{M_1} f_{M_2})
   ~\left[ b_2 \right]_{M_1 M_2 ~[M_2 M_1]}~, \nonumber \\
 PE^{(\zeta)}_{M_1 [M_2]}
 &=& {G_F \over \sqrt{2}}~ \sum_{p=u,c} \lambda_p^r
   ~(i f_B f_{M_1} f_{M_2})~ \left[ b_3^p \right]_{M_1 M_2 ~[M_2 M_1]} ~, \nonumber \\
 PA^{(\zeta)}_{M_1 [M_2]}
 &=& {G_F \over \sqrt{2}}~ \sum_{p=u,c} \lambda_p^r
   ~(i f_B f_{M_1} f_{M_2})~ \left[ b_4^p \right]_{M_1 M_2 ~[M_2 M_1]} ~, \nonumber \\
 PE^{(\zeta)}_{{\rm EW}, M_1 [M_2]}
 &=& {G_F \over \sqrt{2}}~ \sum_{p=u,c} \lambda_p^r
   ~(i f_B f_{M_1} f_{M_2})~ \left[ \frac{3}{2} b_{3, {\rm EW}}^p
   \right]_{M_1 M_2 ~[M_2 M_1]} ~, \nonumber \\
 PA^{(\zeta)}_{{\rm EW}, M_1 [M_2]}
 &=& {G_F \over \sqrt{2}}~ \sum_{p=u,c} \lambda_p^r
   ~(i f_B f_{M_1} f_{M_2})~ \left[ \frac{3}{2} b_{4, {\rm EW}}^p
   \right]_{M_1 M_2 ~[M_2 M_1]} ~,
\label{SU3_E_PAew}
\end{eqnarray}
and the singlet weak annihilation amplitudes by
\begin{eqnarray}
 SE^{(\zeta)}_{M_1 [M_2]}
 &=& {G_F \over \sqrt{2}}~ \lambda_u^r ~(i f_B f_{M_1} f_{M_2})
   ~\left[ b_{S1} \right]_{M_1 M_2 ~[M_2 M_1]}~, \nonumber \\
 SA^{(\zeta)}_{M_1 [M_2]}
 &=& {G_F \over \sqrt{2}}~ \lambda_u^r ~(i f_B f_{M_1} f_{M_2})
   ~\left[ b_{S2} \right]_{M_1 M_2 ~[M_2 M_1]}~, \nonumber \\
 SPE^{(\zeta)}_{M_1 [M_2]}
 &=& {G_F \over \sqrt{2}}~ \sum_{p=u,c} \lambda_p^r
   ~(i f_B f_{M_1} f_{M_2})~ \left[ b_{S3}^p \right]_{M_1 M_2 ~[M_2 M_1]} ~, \nonumber \\
 SPA^{(\zeta)}_{M_1 [M_2]}
 &=& {G_F \over \sqrt{2}}~ \sum_{p=u,c} \lambda_p^r
   ~(i f_B f_{M_1} f_{M_2})~ \left[ b_{S4}^p \right]_{M_1 M_2 ~[M_2 M_1]} ~, \nonumber \\
 SPE^{(\zeta)}_{{\rm EW}, M_1 [M_2]}
 &=& {G_F \over \sqrt{2}}~ \sum_{p=u,c} \lambda_p^r
   ~(i f_B f_{M_1} f_{M_2})~ \left[ \frac{3}{2} b_{S3, {\rm EW}}^p
   \right]_{M_1 M_2 ~[M_2 M_1]} ~, \nonumber \\
 SPA^{(\zeta)}_{{\rm EW}, M_1 [M_2]}
 &=& {G_F \over \sqrt{2}}~ \sum_{p=u,c} \lambda_p^r
   ~(i f_B f_{M_1} f_{M_2})~ \left[ \frac{3}{2} b_{S4, {\rm EW}}^p
   \right]_{M_1 M_2 ~[M_2 M_1]} ~,
\label{SU3_SE_SPAew}
\end{eqnarray}
where we have used the notation $[b_j^p]_{M_1 M_2} \equiv b_j^p (M_1, M_2)$.
Note that the weak annihilation contributions in the QCDF approach given in Eq.~(\ref{weak_ann})
include all the above SU(3)$_{\rm F}$ amplitudes given in Eqs.~(\ref{SU3_E_PAew}) and
(\ref{SU3_SE_SPAew}), such as $E_{M_i}^{(\prime)}$, $A_{M_i}^{(\prime)}$, \ldots ,
$SE_{M_i}^{(\prime)}$, $SA_{M_i}^{(\prime)}$, $\cdots$, etc.

Using the above relations, one can easily translate the decay amplitude expressed in terms of the
SU(3)$_{\rm F}$ amplitudes as shown in Tables~\ref{B_PP_DelS0_1}$-$\ref{Bs_PV_DelS1_3} into
that expressed in terms of the quantities calculated in the framework of QCDF.
For example, the decay amplitude of $B^- \to \pi^- \bar K^0$ given in Eq.~(\ref{B_pi1K0}) can be
rewritten in terms of the quantities calculated in QCDF:
\begin{eqnarray}
 {\cal A}_{B^- \to \pi^- \bar K^0}
  = {G_F \over \sqrt{2}} ~\sum_{p=u,c} \lambda_p^s
   \Big[ \delta_{pu} ~\beta_2 +\alpha_4^p -\frac{1}{2} \alpha_{4, {\rm EW}}^p
   +\beta_3^p +\beta_{3, {\rm EW}}^p \Big] ~X^{(\bar B \pi , ~\bar K)} ~.
\end{eqnarray}
Likewise, the decay amplitude of $\bar B^0 \to \eta^{(\prime)} \bar K^{*0}$ in Eq.~(\ref{B_etaKst0})
now reads
\begin{eqnarray}
 && \mbox{} \hspace{-0.6cm}
   \sqrt{2}~ {\cal A}_{\bar B^0 \to \eta^{(\prime)} \bar K^{*0}}  \nonumber \\
 &=& {G_F \over \sqrt{2}} ~\sum_{p=u,c} \lambda_p^s \Bigg\{
   \Big[ \delta_{pu} ~\alpha_2 +2 \alpha_3^p
   +\frac{1}{2} \alpha_{3, {\rm EW}}^p +\beta_3^p +\beta_{3, {\rm EW}}^p \Big]
   ~X^{(\bar B K^*, ~\eta^{(\prime)}_q)}  \nonumber \\
 && + \sqrt{2}~ \Big[ \alpha_3^p +\alpha_4^p
   -\frac{1}{2} \alpha_{3, {\rm EW}}^p -\frac{1}{2} \alpha_{4, {\rm EW}}^p
   +\beta_3^p -\frac{1}{2} \beta_{3, {\rm EW}}^p
   +\beta_{S3}^p -\frac{1}{2} \beta_{S3, {\rm EW}}^p \Big]
   ~X^{(\bar B K^*, ~\eta^{(\prime)}_s)}  \nonumber \\
 && +\sqrt{2}~ \Big[ \delta_{pc}~ \alpha_2 +\alpha_3^p \Big]
   ~X^{(\bar B K^*, ~\eta^{(\prime)}_c)}  \nonumber \\
 && + \Big[ \alpha_4^p -\frac{1}{2} \alpha_{4, {\rm EW}}^p
   +\beta_3^p -\frac{1}{2} \beta_{3, {\rm EW}}^p \Big]
   ~X^{(\bar B \eta^{(\prime)}_q , ~K^*)} \Bigg\} ~.
\end{eqnarray}
Finally, the decay amplitude of $\bar B^0 \to \eta^{(\prime)} ~\omega / \phi$ in Eq.~(\ref{B_etaomega})
is recast to
\begin{eqnarray}
 && \mbox{} \hspace{-0.4cm}
   2~ {\cal A}_{\bar B^0 \to \eta^{(\prime)} ~\omega / \phi}  \nonumber \\
 &=& {G_F \over \sqrt{2}} ~\sum_{p=u,c} \lambda_p^d \Bigg\{
   \Big[ \delta_{pu} ~( \alpha_2 +\beta_1 +2 \beta_{S1} )
    +2 \alpha_3^p +\alpha_4^p +\frac{1}{2} \alpha_{3,{\rm EW}}^p -\frac{1}{2} \alpha_{4,{\rm EW}}^p
    \nonumber \\
 && \hspace{1.4cm}
    +\beta_3^p +2 \beta_4^p -\frac{1}{2} \beta_{3, {\rm EW}}^p +\frac{1}{2} \beta_{4, {\rm EW}}^p
    +2 \beta_{S3}^p +4 \beta_{S4}^p -\beta_{S3, {\rm EW}}^p +\beta_{S4, {\rm EW}}^p \Big]
   ~ X^{(\bar B \eta^{(\prime)}_q, ~\omega_q /\phi_q)}  \nonumber \\
 && +\sqrt{2} ~\Big[ \delta_{pu} ~\beta_{S1} +\alpha_3^p -\frac{1}{2} \alpha_{3,{\rm EW}}^p
    +\beta_{S3} +2 \beta_{S4}^p -\frac{1}{2} \beta_{S3, {\rm EW}}^p
    +\frac{1}{2} \beta_{S4, {\rm EW}}^p \Big]
    ~ X^{(\bar B \eta^{(\prime)}_q, ~\omega_s /\phi_s)}  \nonumber \\
 && +\sqrt{2} ~(-i f_B f_{\eta^{(\prime)}} f_{\omega/\phi})
    ~\Bigg[ 2 b_{S4}^p -b_{S4, {\rm EW}}^p \Bigg]_{\eta^{(\prime)}_s ~\omega_q /\phi_q}  \nonumber \\
 && +2 ~(-i f_B f_{\eta^{(\prime)}} f_{\omega/\phi})
    ~\Bigg[ b_4^p -\frac{1}{2} b_{4, {\rm EW}}^p +b_{S4}^p -\frac{1}{2} b_{S4, {\rm EW}}^p
    \Bigg]_{\eta^{(\prime)}_s ~\omega_s /\phi_s}  \nonumber \\
 && +\Big[ \delta_{pu} ~( \alpha_2 +\beta_1 +2 \beta_{S1} )
    +2 \alpha_3^p +\alpha_4^p +\frac{1}{2} \alpha_{3,{\rm EW}}^p -\frac{1}{2} \alpha_{4,{\rm EW}}^p
    \nonumber \\
 && \hspace{1.4cm}
    +\beta_3^p +2 \beta_4^p -\frac{1}{2} \beta_{3, {\rm EW}}^p +\frac{1}{2} \beta_{4, {\rm EW}}^p
    +2 \beta_{S3}^p +4 \beta_{S4}^p -\beta_{S3, {\rm EW}}^p +\beta_{S4, {\rm EW}}^p \Big]
   ~ X^{(\bar B ~\omega_q /\phi_q , ~\eta^{(\prime)}_q)}  \nonumber \\
 && +\sqrt{2} ~\Big[ \delta_{pu} ~\beta_{S1} +\alpha_3^p -\frac{1}{2} \alpha_{3,{\rm EW}}^p
    +\beta_{S3} +2 \beta_{S4}^p -\frac{1}{2} \beta_{S3, {\rm EW}}^p
    +\frac{1}{2} \beta_{S4, {\rm EW}}^p \Big]
    ~ X^{(\bar B ~\omega_q /\phi_q , ~\eta^{(\prime)}_s)}  \nonumber \\
 && +\sqrt{2} ~\Big[ \delta_{pc} ~\alpha_2 +\alpha_3^p \Big]
    ~ X^{(\bar B ~\omega_q /\phi_q , ~\eta^{(\prime)}_c)}  \nonumber \\
 && +\sqrt{2} ~(-i f_B f_{\eta^{(\prime)}} f_{\omega/\phi})
    ~\Bigg[ 2 b_{S4}^p -b_{S4, {\rm EW}}^p \Bigg]_{\omega_s /\phi_s , ~\eta^{(\prime)}_q}  \nonumber \\
 && +2 ~(-i f_B f_{\eta^{(\prime)}} f_{\omega/\phi})
    ~\Bigg[ b_4^p -\frac{1}{2} b_{4, {\rm EW}}^p +b_{S4}^p -\frac{1}{2} b_{S4, {\rm EW}}^p
    \Bigg]_{\omega_s /\phi_s , ~\eta^{(\prime)}_s} \Bigg\} ~.
\end{eqnarray}

All the decay amplitudes of $\bar B\to PP,VP$ in QCD factorization shown in Appendix A of
Ref.~\cite{Beneke:2003zv} can be obtained from the $SU(3)_{\rm F}$ amplitudes displayed in
Tables~\ref{B_PP_DelS0_1}$-$\ref{Bs_PV_DelS1_3}.

%%%%%%%%%%%%%%%%%%%%%%%%%%%%%%%%%%%%%%%%%%%%%%%%%%%%%%%%%%%%%%%%%%%%%%%%%%%%%%%%%%%%%%%%%%
%%%%%%%%%%%%%%%%%%%%%%%%%%%%%%%%%%%%%%%%%%%%%%%%%%%%%%%%%%%%%%%%%%%%%%%%%%%%%%%%%%%%%%%%%%
\section{Numerical analysis of flavor SU(3) amplitudes in QCDF}

In this section we estimate the magnitude of each SU(3)$_{\rm F}$ amplitude in the framework of QCDF
by using the relations given in Eqs.~(\ref{SU3_T_Pew}), (\ref{SU3_E_PAew}) and (\ref{SU3_SE_SPAew}).
For the numerical analysis, we shall use the same input values for the relevant parameters as those in
Ref.~\cite{Cheng:2009cn}.
Specifically we use the values of the form factors for $B_{u,d} \to P$ and $B_{u,d} \to V$ transitions
obtained in the light-cone QCD sum rules~\cite{Ball:2004ye} and those for $B_s \to V$ transitions
obtained in the covariant light-front quark model~\cite{Cheng:2003sm} with some modifications :
\begin{eqnarray}
 && F_0^{B \pi} (0) = 0.25 ~, \qquad~~ F_0^{B K} (0) = 0.35 ~, \qquad \quad
  F_0^{B \eta_q} (0) = 0.296 ~,  \nonumber \\
 && F_0^{B_s K} (0) = 0.24 ~, \qquad~ F_0^{B_s \eta_s} (0) = 0.28 ~,  \nonumber \\
 && A_0^{B \rho} (0) = 0.303 ~, \qquad~ A_0^{B K^*} (0) = 0.374 ~, \qquad
  A_0^{B \omega} (0) = 0.281 ~, \nonumber \\
 && A_0^{B_s K^*} (0) = 0.30 ~, \qquad A_0^{B_s \phi} (0) = 0.32 ~.
\end{eqnarray}
Here for $\eta^{(\prime)}$ we have used the flavor states
$q \bar q \equiv (u \bar u +d \bar d)/\sqrt{2}$, $s \bar s$ and $c \bar c$ labeled by the $\eta_q$,
$\eta_s$ and $\eta_{c}$, respectively, and the form factors for $B \to \eta^{(\prime)}$ are given
by
\begin{eqnarray}
 && F^{B \eta} = F^{B \eta_q} \cos\theta ~, \qquad \quad~ F^{B \eta'} = F^{B \eta_q} \sin\theta ~,
  \nonumber \\
 && F^{B_s \eta} = -F^{B_s \eta_s} \sin\theta ~, \qquad F^{B_s \eta'} = F^{B_s \eta_s} \cos\theta ~,
\end{eqnarray}
where the small mixing with $\eta_c$ is neglected and the $\eta_q$-$\eta_s$ mixing angle $\theta$ defined by
\begin{eqnarray}
|\eta\ra &=& \cos\theta|\eta_q\ra-\sin\theta|\eta_s\ra, \non \\
|\eta'\ra &=& \sin\theta|\eta_q\ra+\cos\theta|\eta_s\ra,
\end{eqnarray}
is
$(39.3 \pm 1.0)^\circ$ in the Feldmann-Kroll-Stech mixing scheme~\cite{Feldmann:1998vh}.
For the decay constants, we use the values (in units of MeV)~\cite{Ball:2006eu,Feldmann:1998vh}
\begin{eqnarray}
 && f_{\pi} = 132 ~, \qquad ~ f_K = 160 ~, \qquad f_{B_{u,d}} = 210 ~, \quad ~ f_{B_s} = 230 ~,
  \nonumber \\
 && f_{\eta^{\prime}}^q = 107~, \qquad f_{\eta}^q = 89 ~,
  \qquad ~~  f_{\eta^{\prime}}^s = -112~, \quad~~ f_{\eta}^s =137 ~,  \nonumber \\
 && f_{\rho} = 216 ~, \qquad ~ f_{K^*} = 220 ~, \quad ~~
  f_{\omega} = 187 ~, \qquad f_{\phi} = 215 ~.
\end{eqnarray}

It is known that although physics behind nonleptonic $B$ decays is extremely complicated, it is greatly
simplified in the heavy quark limit $m_b\to\infty$ as the decay amplitude becomes factorizable and can
be expressed in terms of decay constants and form factors. However, this simple approach encounters
three major difficulties:
(i) the predicted branching fractions for penguin-dominated $\bar B\to PP, ~VP, ~VV$ decays are
systematically below the measurements \cite{Beneke:2003zv}   (ii) direct {\it CP}-violating asymmetries
for $\bar B^0\to K^-\pi^+$, $\bar B^0\to K^{*-}\pi^+$, $B^-\to K^-\rho^0$, $\bar B^0 \to \pi^+\pi^-$ and
$\bar B_s^0\to K^+\pi^-$ disagree with experiment in signs \cite{Cheng:2009eg}, and (iii) the transverse
polarization fraction in penguin-dominated charmless $\bar B\to VV$ decays is predicted to be very small,
while experimentally it is comparable to the longitudinal polarization one. All these indicate the
necessity of going beyond zeroth $1/m_b$ power expansion. In the QCDF approach one considers the power
correction to penguin amplitudes due to weak penguin annihilation characterized by the parameter
$\beta_3^p$ or $b_3^p$. However, QCD-penguin exchange amplitudes involve troublesome endpoint
divergences and hence they can be studied only in a phenomenological way.
We shall follow~\cite{Beneke:1999br} to model the endpoint divergence $X\equiv\int^1_0 dx/\bar x$ in
the annihilation diagrams as
 \be \label{eq:XA}
 X_A=\ln\left({m_B\over \Lambda_h}\right)(1+\rho_A e^{i\phi_A}),
 \en
where $\Lambda_h$ is a typical scale of order 500 MeV, and $\rho_A$ and $\phi_A$ are the unknown real
parameters.

By adjusting the magnitude $\rho_A$ and the phase $\phi_A$ in this scenario, all the above-mentioned
difficulties can be resolved. However, a scrutiny of the QCDF predictions reveals more puzzles with
respect to direct {\it CP} violation, as pointed out in \cite{Cheng:2009eg,Cheng:2009cn}. While the
signs of {\it CP} asymmetries in $K^-\pi^+,~ K^-\rho^0$ modes are flipped to the right ones in the
presence of power corrections from penguin annihilation, the signs of $A_{CP}$ in
$B^-\to K^-\pi^0,~K^-\eta,~\pi^-\eta$ and $\bar B^0\to\pi^0\pi^0,~\bar K^{*0}\eta$ will also get
reversed in such a way that they disagree with experiment.
This indicates that it is necessary to consider subleading power corrections other than penguin
annihilation. It turns out that an additional subleading $1/m_b$ power correction to color-suppressed
tree amplitudes is crucial for resolving the aforementioned  {\it CP} puzzles and explaining the decay
rates for the color-suppressed tree-dominated $\pi^0\pi^0$, $\rho^0\pi^0$
modes~\cite{Cheng:2009eg,Cheng:2009cn}. A solution to the $B\to K\pi$ {\it CP}-puzzle related to the
difference of {\it CP} asymmetries of $B^-\to K^-\pi^0$ and $\bar B^0\to K^-\pi^+$ requires  a large
complex color-suppressed tree amplitude and/or a large complex electroweak penguin. These two
possibilities can be discriminated in tree-dominated $B$ decays. The {\it CP} puzzles with $\pi^-\eta$,
$\pi^0\pi^0$ and the rate deficit problems with $\pi^0\pi^0,~\rho^0\pi^0$ can only be resolved by
having a large complex color-suppressed tree topology $C$. While the New Physics solution to the
$B\to K\pi$ {\it CP} puzzle is interesting, it is most likely irrelevant for tree-dominated decays.

We shall use the fitted values of the parameters $\rho_A$ and $\phi_A$ given in \cite{Cheng:2009cn}:
\begin{eqnarray}
 && {\rm For}~ \bar B_{u,d} (\bar B_s) \to PP ~, \qquad \rho_A = 1.10 ~(1.00) ~,
  \quad \phi_A = -50^{\circ} ~(-55^{\circ}) ~,  \nonumber \\
 && {\rm For}~ \bar B_{u,d} (\bar B_s) \to VP ~, \qquad \rho_A = 1.07 ~(0.90) ~,
  \quad \phi_A = -70^{\circ} ~(-65^{\circ}) ~,  \nonumber \\
 && {\rm For}~ \bar B_{u,d} (\bar B_s) \to PV ~, \qquad \rho_A = 0.87 ~(0.85) ~,
  \quad \phi_A = -30^{\circ} ~(-30^{\circ}) ~.
\end{eqnarray}
Following \cite{Cheng:2009eg}, power corrections to the color-suppressed topology are parametrized as
\be \label{eq:a2}
a_2 \to a_2(1+\rho_C e^{i\phi_C}),
\en
with the unknown parameters $\rho_C$ and $\phi_C$ to be inferred from experiment. We shall
use~\cite{Cheng:2009eg}
\be
 \rho_C\approx 1.3\,,~0.8\,,~0, \qquad
 \phi_C\approx -70^\circ\,,-80^\circ\,,~0,
\en
for $\bar B\to PP, ~VP, ~VV$ decays, respectively.

%%%%%%%%%%%%%%%%%%%%%%%%%%%%%%%%%%%%%%%%%%%%%%%%%%%%%%%%%%%%%%%%%%
%%%%%%%%%%%%%%%%%%%%%%%%%%%%%%%%%%%%%%%%%%%%%%%%%%%%%%%%%%%%%%%%%%
\subsection{Magnitudes and strong phases of the SU(3)$_{\rm F}$ amplitudes}

\mbox{} From Eqs.~(\ref{SU3_T_Pew})$-$(\ref{SU3_SE_SPAew}), it is obvious that the SU(3)$_{\rm F}$
amplitudes for $\bar B_{u,d} (B_s) \to M_1 M_2$ depend on the specific final states $M_1$ and $M_2$.
Thus, the magnitudes of these SU(3)$_F$ amplitudes are process-dependent in general, though numerically
the dependence turns out to be moderate.
In order to find typical magnitudes of these SU(3)$_{\rm F}$ ones, we choose typical decay processes
as explained below, and use only the central values of the input parameters.
>From now on, we shall use a notation for the relevant strong and weak phases as follows: e.g., the
color-favored tree amplitude for $\Delta S = 0$ $(|\Delta S| = 1)$ decays is denoted as
$T^{(\prime)}_P \equiv |T^{(\prime)}_P| ~e^{i \big( \delta_P^{T (\prime)} +\theta^{T(\prime)} \big) }$
with the strong phase $\delta_P^{T (\prime)}$ and the weak phase
$\theta^{T (\prime)} = {\rm arg} \big( V_{ub} V_{ud (s)}^* \big)$.

The numerical estimates of the SU(3)$_{\rm F}$ amplitudes are displayed in
Tables~\ref{Abs_B_PP}$-$\ref{Abs_Bs_PV}.  In the case of $\bar B_{u,d} (\bar B_s) \to P_1 P_2$ decays
with $\Delta S = 0$, the modes $\bar B_{u,d} (\bar B_s) \to \pi \pi ~(\pi K)$ are used to numerically
compute the relevant SU(3)$_{\rm F}$ amplitudes such as $T_P$, $C_P$, $P_P$ and so on, except for
$S_P^{(q,s,c)}$ which $\bar B_{u,d} (\bar B_s) \to \pi \eta ~(K \eta)$ are used for.
Similarly, for $|\Delta S| = 1$ decays the processes $\bar B_{u,d} (\bar B_s) \to \pi K ~(K K)$ are
used to estimate $T'_P$, $C'_P$, $P'_P$ and so on, except for $S_P^{\prime (q,s,c)}$ which
$\bar B_{u,d} (\bar B_s) \to K \eta ~(\eta \eta)$ are used for.
In the case of $\Delta S = 0$ $\bar B_{u,d} (\bar B_s) \to P V$ decays, the modes
$\bar B_{u,d} (\bar B_s) \to \pi \rho ~(\pi K^*, ~K \rho)$ are used for the numerical calculation of the
amplitudes $T_{P,V}$, $C_{P,V}$, $P_{P,V}$ and so on, except for $S_P^{(q,s,c)}$ and $S_V^{(q,s,c)}$
for which $\bar B_{u,d} (\bar B_s) \to \pi \omega /\phi ~(K \omega /\phi)$ and $\eta \rho ~(\eta K^*)$,
respectively, are used.
For $|\Delta S| = 1$ decays the processes $\bar B_{u,d} (\bar B_s) \to \pi K^*, ~K \rho ~(K K^*)$ are
used to estimate $T'_{P,V}$, $C'_{P,V}$, $P'_{P,V}$ and so on, except for $S_P^{\prime (q,s,c)}$ and
$S_V^{\prime (q,s,c)}$ for which $\bar B_{u,d} (\bar B_s) \to K \omega /\phi ~(\eta \omega /\phi)$
and $\eta K^* ~(\eta \phi)$, respectively, are used.
The strong phases of the SU(3)$_{\rm F}$ amplitudes are generated from the flavor operators
$\alpha_i^p$ and $b_j^p$. As shown in Eqs.~(\ref{SU3_T_Pew})$-$(\ref{SU3_SE_SPAew}), except the
amplitudes $T^{(\prime)}, ~C^{(\prime)}, ~E^{(\prime)}, ~A^{(\prime)}, ~SE^{(\prime)}$ and
$SA^{(\prime)}$, all the other amplitudes including penguin ones are the sum of two terms each of which
is proportional to $\lambda_p^{r} \alpha_i^p$ or $\lambda_p^{r} b_j^p$ with $p = u,c$.
Because the CKM factors $\lambda_u^{r}$ and $\lambda_c^{r}$ involve different weak
phases from each other, to exhibit the strong phase of each amplitude in the tables, we shall use the
approximations, $\alpha_{3,4 ({\rm EW})}^u \simeq \alpha_{3,4 ({\rm EW})}^c$ and
$b_{3,4 ({\rm EW})}^u \simeq b_{3,4 ({\rm EW})}^c$, where the former (latter) relation holds roughly
(very well) in QCDF.
\footnote{ In QCDF the value of $\alpha_i^c$ $(i = 3, 4, (3, {\rm EW}), (4, {\rm EW}))$
differ from that of $\alpha_i^u$ by about $(25 - 30)$\%, while the value of $b_i^c$ is  the
same as that of $b_i^u$.
It should be emphasized that using Eqs.~(\ref{SU3_T_Pew})$-$(\ref{SU3_SE_SPAew}), one can compute both
the magnitude and strong phase of each SU(3)$_{\rm F}$ amplitude without invoking these approximations on
$\alpha_i^{u,c}$ and $b_i^{u,c}$.
In our numerical analysis, we use these approximations only for expressing the strong phases as shown
in Tables~\ref{Abs_B_PP}$-$\ref{Abs_Bs_PV}. All the magnitudes of the amplitudes are obtained without
using these approximations. }
Thus, for instance, the QCD-penguin amplitude can be understood as
$P^{(\prime)} \equiv -|P^{(\prime)}| e^{i \delta^{P^{(\prime)}}} e^{i \theta^{P^{(\prime)}}}$ with
the strong phase $\delta^{P^{(\prime)}}$ and the weak phase
$\theta^{P^{(\prime)}} = {\rm arg}(V_{tb} V_{td (s)}^*)$.

\mbox{} From Tables~\ref{Abs_B_PP}$-$\ref{Abs_Bs_PV},  hierarchies among the SU(3)$_{\rm F}$
amplitudes are numerically found as follows.
For $\Delta S = 0$ $\bar B_{u,d} (\bar B_s) \to PP$ decays, the hierarchical relation is
\begin{eqnarray}
 |T_P| &>& |C_P| > |P_P| \gtrsim |PE_P| > |E_P| > |S_P^{(s)}| \sim |S_P^{(q)}| \sim |P_{{\rm EW}, ~P}|
  \sim |A_P| \sim |PA_P|  \nonumber \\
 &>& |P_{{\rm EW}, ~P}^C| > |PE_{{\rm EW}, ~P}| \sim |PA_{{\rm EW}, ~P}| \sim |S_P^{(c)}| ~,
\end{eqnarray}
and for $\bar B_{u,d} (\bar B_s) \to PV$,
\begin{eqnarray}
 |T_{P,V}| &>& |C_{P,V}| > |PE_{P,V}| \gtrsim |P_{P,V}| \sim |E_{P,V}| >  |P_{{\rm EW}, ~{P,V}}|
  \gtrsim |A_{P,V}| \sim |S_{P,V}^{(s)}| \sim |S_{P,V}^{(q)}|  \nonumber \\
 &\gtrsim& |P_{{\rm EW}, ~{P,V}}^C| > |PA_{P,V}| \gtrsim |PE_{{\rm EW}, ~{P,V}}|
  \gtrsim |PA_{{\rm EW}, ~{P,V}}| \sim |S_V^{(c)}| ~.
\end{eqnarray}
Likewise, for $\Delta S = 1$ $\bar B_{u,d} (\bar B_s) \to PP$ decays, the hierarchical relation
is found to be
\begin{eqnarray}
 |P'_P| &\gtrsim& |PE'_P| > |T'_P| \gtrsim |P'_{{\rm EW}, ~P}| \gtrsim |C'_P|
  \gtrsim |S_P^{\prime (s)}| \sim |S_P^{\prime (q)}| \sim |PA'_P|  \nonumber \\
 &>& |P_{{\rm EW}, ~P}^{C \prime}| > |E'_P| > |A'_P| \sim |PE'_{{\rm EW}, ~P}|
  \sim |PA'_{{\rm EW}, ~P}| \gtrsim |S_P^{\prime (c)}| ~,
\end{eqnarray}
and for $\bar B_{u,d} (\bar B_s) \to PV$,
\begin{eqnarray}
 && |PE'_{P,V}| \gtrsim |P'_{P,V}| > |T'_{P,V}| \gtrsim |P'_{{\rm EW}, ~P,V}| > |C'_{P,V}|
  \sim |S_{P,V}^{\prime (s)}| \sim |S_{P,V}^{\prime (q)}| \gtrsim |P_{{\rm EW}, ~P,V}^{C \prime}|
  \nonumber \\
 && > |PE'_{{\rm EW}, ~V}| \gtrsim |PA'_{P,V}| \sim |E'_{P,V}| > |A'_{P,V}| > |PE'_{{\rm EW}, ~P}|
  \sim |PA'_{{\rm EW}, ~P,V}| \sim |S_V^{\prime (c)}| ~.
\end{eqnarray}

Several remarks are in order:

\begin{enumerate}
\item
It is well known that  the penguin contributions are dominant in $|\Delta S| = 1$ decays due to the
CKM enhancement $|V_{cs} V^*_{cb}| \approx |V_{ts} V^*_{tb}| \gg |V_{us} V^*_{ub}|$ and the large top
quark mass.
Especially, it is interesting to note that in addition to the QCD-penguin contributions $P'_{P,V}$,
the QCD-penguin exchange ones $PE'_{P,V}$ are large for $\bar B_{u,d} (\bar B_s) \to PP$ and $PV$
decays. Since the strong phase of $PE'_{P (V)}$ is comparable to that of
$P'_{P (V)}$ in magnitude with the {\it same} sign ({\it i.e.},
$\delta_{P (V)}^{PE \prime} \sim \delta_{P (V)}^{P \prime}$), the effects from $PE'_{P (V)}$ and
$P'_{P (V)}$ are strongly constructive to each other.
It has been shown~\cite{Cheng:2009cn} that in order to accommodate the data including the
branching fractions and {\it CP} asymmetries of those decays, the QCD-penguin exchange contributions
($PE'_{P,V} \propto b_3^p$) are important and play a crucial role.
For example, for penguin-dominated $\bar B_{u,d} \to PP$ decays, the effects of the QCD-penguin
exchange dictated by the values of $\rho_A$ and $\phi_A$  paly a key role in resolving the problems
of the smallness of predicted decay rates and of the wrong sign of the predicted direct {\it CP} asymmetry
$A_{CP} (\pi^+ K^-)$. Also, for $B_{u,d} \to K \rho$ and $\pi K^*$ decays, the QCD-penguin exchange
contributions will enhance the rates by $(15 \sim 100)\%$ for $K \rho$ modes and by a factor of
$2 \sim 3$ for $\pi K^*$ ones.

\item
The SU(3)$_{\rm F}$-singlet contributions $S^{(\prime)}_{P,V}$ are involved in the decay modes including
$\eta^{(\prime)}, ~\omega, ~\phi$ in the final state, such as $\bar B \to \pi \eta^{(\prime)}$,
$K \eta^{(\prime)}$, $\pi \omega/\phi$, $K \omega/\phi$, $\cdots$, etc.
They are expected to be small because of the Okubo-Zweig-Iizuka (OZI) suppression rule which favors
connected quark diagrams.  Indeed they are found to be :
$|S^{(\prime)}_P / P^{(\prime)}_P| \approx (10 \sim 24)\%$~ for $\bar B_{u,d} (\bar B_s) \to PP$~ and
$|S^{(\prime)}_{P,V} / P^{(\prime)}_{P,V}| \approx (11 \sim 27)\%$~ for
$\bar B_{u,d} (\bar B_s) \to PV$.
\footnote{  When the effects from
$PE^{(\prime)}_{P,V}$ which are comparable to $P^{(\prime)}_{P,V}$ are taken into account, the ratio
~$|S^{(\prime)}_{P,V} / (P^{(\prime)}_{P,V} +PE^{(\prime)}_{P,V})|$ becomes $\lesssim 10\%$~
for $\bar B_{u,d} (\bar B_s) \to PP$ and $PV$.
}
In contrast, in the framework of generalized factorization, the SU(3)$_{\rm F}$-singlet contribution
depend strongly on the parameter $\xi \equiv 1 / N_c$ ($N_c$ being the effective number of color)
and could be large, particularly for $\bar B_{u,d} \to VV$ decays~\cite{Oh:1998wa} : e.g., up to 77\%
of the dominant QCD-penguin contribution. In the flavor SU(3) analyses with a  global fit of the
SU(3)$_{\rm F}$ amplitudes to the data, a large effect from $S'_P$ is also needed for explaining the
large BFs of the $B \to \eta' K$ modes~\cite{large_S}:
e.g., $|S'_P / P'_P| \approx 38\%$~\cite{Chiang06}.

\vspace{0.2cm}
Among the two-body $B$ decays, $B\to K\eta'$ has the largest branching fraction, of order
$70\times 10^{-6}$, while ${\cal B}(B\to\eta K)$ is only $(1 \sim 3)\times 10^{-6}$. This can be
qualitatively  understood as  the interference between the $B\to K\eta_q$ amplitude induced by the
$b \to sq \bar q$ penguin and the $B\to K\eta_s$ amplitude induced by $b\to ss\bar s$, which is
constructive for $B\to K\eta'$ and destructive for $B \to \eta K$~\cite{Lipkin}. This explains the
large rate of the former and the suppression of the latter.
As stressed in~\cite{Beneke:2002jn,Beneke:2003zv}, the observed large $B\to K\eta'$ rates are naturally
explained in QCDF without invoking large flavor-singlet contributions.

\item
In $\Delta S = 0$ decays, as expected, the tree contributions $T_{P,V}$ dominate and the
color-suppressed tree amplitudes $C_{P,V}$ are larger than the penguin ones.
Among the penguin contributions, the QCD-penguin ones $P_{P,V}$ and the QCD-penguin exchange ones
$PE_{P,V}$ are comparable.
Large strong phases in the decay amplitudes are needed to generate sizable direct {\it CP} violation in
$\bar B$ decay processes.
For tree-dominated $\bar B_{u,d} (\bar B_s) \to PP$ decays we have
$C_P/T_P \approx 0.63 \,e^{-i 56^\circ} ~(0.83 \,e^{-i 53^\circ})$ which is larger than the naive
expectation of $C_P/T_P \sim 1/3$ in both magnitude and phase. Recall that a large complex
color-suppressed tree topology $C$ is needed to solve the rate deficit problems with $\pi^0 \pi^0$ and
$\pi^0 \rho^0$ and give the correct sign for direct {\it CP} violation in the decays $K^-\pi^0$, $K^-\eta$,
$\bar K^{*0}\eta$, $\pi^0\pi^0$ and $\pi^-\eta$~\cite{Cheng:2009cn}.

\item
The $W$-exchange $E_{P,V}^{(\prime)}$, $W$-annihilation $A_{P,V}^{(\prime)}$ and QCD-penguin
annihilation $PA_{P,V}^{(\prime)}$ contributions are small, as expected because of a helicity
suppression factor of $f_B / m_B \approx 5\%$ arising from the smallness of the $B$ meson wave function
at the origin~\cite{Gronau:1994rj}; they are at most only a few~\% (or up to 12\% in the case of $PA'_A$)
of the dominant contributions $T_{P,V}$ or $P'_{P,V}$.
\end{enumerate}

Finally let us compare the numerical values of the SU(3)$_F$ amplitudes computed in QCDF with those
obtained from global fits to charmless $\bar B_{u,d} \to PP$ and $\bar B_{u,d} \to PV$ decays.
The ratios of
the SU(3)$_F$ amplitudes extracted from global fits to charmless $\bar B_{u,d} \to PP$
modes~\cite{Chiang06} are~\footnote{We only show the cases of ``Scheme 4'' in~\cite{Chiang06} for
$\bar B_{u,d} \to PP$ and of ``Scheme B2'' in~\cite{ChiangPV} for $\bar B_{u,d} \to PV$ below, since
these cases take into account the largest set of SU(3) breaking effects among the four schemes presented
in~\cite{Chiang06} and~\cite{ChiangPV}.
For comparison to our results, only the central values are shown.}
\begin{eqnarray}
 && \left| \frac{C^{(\prime)}_P}{T^{(\prime)}_P} \right| = 0.67 ~(0.67) , \qquad \qquad \quad~
   \left| \frac{P^{(\prime)}_P /\lambda_t^{d (s)}}{T^{(\prime)}_P /\lambda_u^{d (s)}} \right|
   = 0.17 ~(0.14) ,  \nonumber \\
 && \left| \frac{S^{(\prime)}_P /\lambda_t^{d (s)}}{T^{(\prime)}_P /\lambda_u^{d (s)}} \right|
   = 0.065 ~(0.053) , \qquad
   \left| \frac{P^{(\prime)}_{{\rm EW}, P} /\lambda_t^{d (s)}}{T^{(\prime)}_P /\lambda_u^{d (s)}}
   \right| = 0.020 ~(0.016) ,
\label{ratio_SU3amp_PP}
\end{eqnarray}
with $\lambda_q^r \equiv V_{qb} V_{qr}^*$ ($q = u, t$ and $r = d, s$),
and the relative strong phases are
\begin{eqnarray}
 && \delta^{C (\prime)}_P - \delta^{T (\prime)}_P = -68.3^{\circ} ~, \qquad
   \delta^{P (\prime)}_P - \delta^{T (\prime)}_P = -15.9^{\circ}  ~,  \nonumber \\
 && \delta^{S (\prime)}_P - \delta^{T (\prime)}_P = -42.9^{\circ}  ~, \qquad
   \delta^{P_{\rm EW} (\prime)}_P - \delta^{T (\prime)}_P = -57.6^{\circ}  ~.
\label{strong_phase_PP}
\end{eqnarray}
Likewise, for charmless $\bar B_{u,d} \to PV$ modes, the ratios of the SU(3)$_F$ amplitudes extracted
from global fits~\cite{ChiangPV} are
\begin{eqnarray}
 && \left| \frac{C^{(\prime)}_P}{T^{(\prime)}_P} \right| = 0.15 ~(0.15) ,
      \qquad \qquad \quad~
   \left| \frac{C^{(\prime)}_V}{T^{(\prime)}_V} \right| = 0.76 ~(0.76) ,  \nonumber \\
 &&  \left| \frac{P^{(\prime)}_P /\lambda_t^{d (s)}}{T^{(\prime)}_P /\lambda_u^{d (s)}}
   \right| = 0.11 ~(0.11) ,  \qquad~~~
   \left| \frac{P^{(\prime)}_V /\lambda_t^{d (s)}}{T^{(\prime)}_V /\lambda_u^{d (s)}}
   \right| = 0.056 ~(0.046) , \nonumber \\
 && \left| \frac{S^{(\prime)}_P /\lambda_t^{d (s)}}{T^{(\prime)}_P /\lambda_u^{d (s)}} \right|
      = 0.018 ~(0.018) , \quad~~~
    \left| \frac{S^{(\prime)}_V /\lambda_t^{d (s)}}{T^{(\prime)}_V /\lambda_u^{d (s)}} \right|
      = 0.041 ~(0.034) , \nonumber \\
 && \left| \frac{P^{(\prime)}_{{\rm EW}, P} /\lambda_t^{d (s)}}
      {T^{(\prime)}_P /\lambda_u^{d (s)}} \right| = 0.039 ~(0.039) , \quad
   \left| \frac{P^{(\prime)}_{{\rm EW}, V} /\lambda_t^{d (s)}}
     {T^{(\prime)}_V /\lambda_u^{d (s)}} \right| = 0.074 ~(0.061) ,
\label{ratio_SU3amp_PV}
\end{eqnarray}
and the relative strong phases are
\begin{eqnarray}
 && \delta^{C (\prime)}_P - \delta^{T (\prime)}_P = 149.0^{\circ}  ~, \qquad~~
   \delta^{T (\prime)}_V - \delta^{T (\prime)}_P = 0.6^{\circ}  ~, \qquad~
   \delta^{C (\prime)}_V - \delta^{T (\prime)}_P = -75.9^{\circ}  ~, \nonumber \\
 && \delta^{P (\prime)}_P - \delta^{T (\prime)}_P = -2.6^{\circ}  ~, \qquad~~
   \delta^{P (\prime)}_V - \delta^{T (\prime)}_P = 172.5^{\circ}  ~, \nonumber \\
 && \delta^{S (\prime)}_P - \delta^{T (\prime)}_P = -139.8^{\circ}  ~, \quad~~
   \delta^{S (\prime)}_V - \delta^{T (\prime)}_P = -47.7^{\circ}  ~, \nonumber \\
 && \delta^{P_{\rm EW} (\prime)}_P - \delta^{T (\prime)}_P = 59.0^{\circ} ~, \qquad
   \delta^{P_{\rm EW} (\prime)}_V - \delta^{T (\prime)}_P = -111.0^{\circ} ~.
\label{strong_phase_PV}
\end{eqnarray}
In the above the numerical values outside (inside) parentheses correspond to $\Delta S = 0$
($|\Delta S| = 1$) decays.  In the case of $\bar B_{u,d} \to PP$ with $|\Delta S| = 1$, the
{\it primed} amplitudes were obtained by including the SU(3) breaking factor $f_K /f_{\pi}$ for both
$|T'_P|$ and $|C'_P|$ and a universal SU(3) breaking factor $\xi = 1.04$ for all the amplitudes except
$P'_{\rm EW, P}$.
But, in $\bar B_{u,d} \to PV$, the {\it primed} amplitudes were extracted by imposing
partial SU(3) breaking factors on $T$ and $C$ only: {\it i.e.}, including $f_{K^*} /f_{\rho}$ for
$|T'_P|$ and $|C'_P|$, and $f_K /f_{\pi}$ for $|T'_V|$ and $|C'_V|$.
Also, for both $\bar B_{u,d} \to PP$ and $PV$, the top penguin dominance was assumed, which is
equivalent to the assumption that $\alpha_{3,4 ({\rm EW})}^u \simeq \alpha_{3,4 ({\rm EW})}^c$ in QCDF.
For the strong phases, exact flavor SU(3) symmetry was assumed in the fits so that
$\delta^T_P =\delta^{T \prime}_P$, $\delta^C_P =\delta^{C \prime}_P$, etc.
In $\bar B_{u,d} \to PV$, all the relative strong phases were found relative to the strong phase of
$T_P$ ({\it i.e.}, $\delta^T_P$).

On the other hand, from Table~\ref{Abs_B_PP}, the ratios of the SU(3)$_F$ amplitudes for
$\bar B_{u,d} \to PP$ are given by
\begin{eqnarray}
 && \left| \frac{C^{(\prime)}_P}{T^{(\prime)}_P} \right| = 0.63 ~(0.64) , \qquad \qquad \quad~~~
   \left| \frac{P^{(\prime)}_P /\lambda_t^{d (s)}}{T^{(\prime)}_P /\lambda_u^{d (s)}} \right|
   = 0.091 ~(0.097) ,  \nonumber \\
 && \left| \frac{S^{(\prime) (q)}_P /\lambda_t^{d (s)}}{T^{(\prime)}_P /\lambda_u^{d (s)}}
   \right| = 0.014 ~(0.012) , \qquad
   \left| \frac{P^{(\prime)}_{{\rm EW}, P} /\lambda_t^{d (s)}}{T^{(\prime)}_P /\lambda_u^{d (s)}}
   \right| = 0.013 ~(0.016) , \nonumber \\
 && \left| \frac{PE^{(\prime)}_P /\lambda_t^{d (s)}}{T^{(\prime)}_P /\lambda_u^{d (s)})}
   \right| = 0.061 ~(0.060) ,
\label{ratio_SU3amp_PP2}
\end{eqnarray}
and the relative strong phases are
\begin{eqnarray}
 && \delta^{C (\prime)}_P - \delta^{T (\prime)}_P = -55.7^{\circ} ~(-57.5^{\circ}) ~, \qquad
   \delta^{P (\prime)}_P - \delta^{T (\prime)}_P = -158.6^{\circ} ~(-158.3^{\circ}) ~,  \nonumber \\
 && \delta^{S^{(\prime) (q)}}_P - \delta^{T (\prime)}_P = 158.2^{\circ} ~(149.4^{\circ}) ~, \qquad
   \delta^{P_{\rm EW} (\prime)}_P - \delta^{T (\prime)}_P = -179.8^{\circ} ~(-179.8^{\circ}) ~,
   \nonumber \\
 && \delta^{PE (\prime)}_P - \delta^{T (\prime)}_P = -147.1^{\circ} ~(-147.1^{\circ}) ~.
\label{strong_phase_PP2}
\end{eqnarray}
Likewise, for $\bar B_{u,d} \to PV$, we obtain
\begin{eqnarray}
 && \left| \frac{C^{(\prime)}_P}{T^{(\prime)}_P} \right| = 0.30 ~(0.35) , \qquad \qquad \quad~~
   \left| \frac{C^{(\prime)}_V}{T^{(\prime)}_V} \right| = 0.39 ~(0.33) ,  \nonumber \\
 &&  \left| \frac{P^{(\prime)}_P /\lambda_t^{d (s)}}{T^{(\prime)}_P /\lambda_u^{d (s)}}
   \right| = 0.030 ~(0.036) ,  \qquad~
   \left| \frac{P^{(\prime)}_V /\lambda_t^{d (s)}}{T^{(\prime)}_V /\lambda_u^{d (s)}}
   \right| = 0.041 ~(0.039) ,  \nonumber \\
 && \left| \frac{S^{(\prime)}_P /\lambda_t^{d (s)}}{T^{(\prime)}_P /\lambda_u^{d (s)}} \right|
      = 0.005 ~(0.006) , ~~~~~~~
    \left| \frac{S^{(\prime)}_V /\lambda_t^{d (s)}}{T^{(\prime)}_V /\lambda_u^{d (s)}} \right|
      = 0.010 ~(0.007) , \nonumber \\
 && \left| \frac{P^{(\prime)}_{{\rm EW}, P} /\lambda_t^{d (s)}}
     {T^{(\prime)}_P /\lambda_u^{d (s)}} \right| = 0.014 ~(0.019) , ~~~
   \left| \frac{P^{(\prime)}_{{\rm EW}, V} /\lambda_t^{d (s)}}
     {T^{(\prime)}_V /\lambda_u^{d (s)}} \right| = 0.013 ~(0.014) , \nonumber \\
 && \left| \frac{PE^{(\prime)}_P /\lambda_t^{d (s)}}
     {T^{(\prime)}_P /\lambda_u^{d (s)}} \right| = 0.037 ~(0.040) , \quad~
   \left| \frac{PE^{(\prime)}_V /\lambda_t^{d (s)}}
     {T^{(\prime)}_V /\lambda_u^{d (s)}} \right| = 0.051 ~(0.049) ,
\label{ratio_SU3amp_PV2}
\end{eqnarray}
and the relative strong phases are
\begin{eqnarray}
 && \delta^{C (\prime)}_P - \delta^{T (\prime)}_P = -16.8^{\circ} ~(-19.8^{\circ}) ~, \qquad \quad
   \delta^{C (\prime)}_V - \delta^{T (\prime)}_P = -52.5^{\circ} ~(-56.2^{\circ}) ~, \nonumber \\
 && \delta^{P (\prime)}_P - \delta^{T (\prime)}_P = -145.5^{\circ} ~(-145.2^{\circ}) ~, \qquad
   \delta^{P (\prime)}_V - \delta^{T (\prime)}_P = 7.6^{\circ} ~(7.1^{\circ}) ~, \nonumber \\
 && \delta^{S (\prime)}_P - \delta^{T (\prime)}_P = -5.5^{\circ} ~(-6.4^{\circ}) ~, \qquad \qquad
   \delta^{S (\prime)}_V - \delta^{T (\prime)}_P = 150.7^{\circ} ~(133.4^{\circ}) ~, \nonumber \\
 && \delta^{P_{\rm EW} (\prime)}_P - \delta^{T (\prime)}_P = 179.8^{\circ} ~(179.8^{\circ}) ~, \qquad~~
   \delta^{P_{\rm EW} (\prime)}_V - \delta^{T (\prime)}_P = -179.7^{\circ} ~(-179.7^{\circ}) ~,
   \nonumber \\
 && \delta^{PE (\prime)}_P - \delta^{T (\prime)}_P = -124.4^{\circ} ~(-125.0^{\circ}) ~, \qquad~~~
   \delta^{PE (\prime)}_V - \delta^{T (\prime)}_P = 4.5^{\circ} ~(4.2^{\circ}) ~,
\label{strong_phase_PV2}
\end{eqnarray}
where the numerical values outside (inside) parentheses correspond to $\Delta S = 0$ ($|\Delta S| = 1$)
decays. In our case, ~$\delta^{T}_P =\delta^{T}_V =\delta^{T \prime}_P =\delta^{T \prime}_V$.

In comparison of Eqs.~(\ref{ratio_SU3amp_PP})$-$(\ref{strong_phase_PV}) [{\it ``fitting case''}] with
Eqs.~(\ref{ratio_SU3amp_PP2})$-$(\ref{strong_phase_PV2}) [{\it ``QCDF case''}], it is found that the
values of $|C_P^{(\prime)} /T_P^{(\prime)}|$ for $\bar B_{u,d} \to PP$ in the {\it fitting case} are
very similar to those of our {\it QCDF case}: both results show the large magnitudes of $C_P^{(\prime)}$
together with large strong phases, as discussed in the above ``{\it remark} 3''.
But, for $\bar B_{u,d} \to PV$, the values of $|C_{P,V}^{(\prime)} /T_{P,V}^{(\prime)}|$ from both cases
are different: in the {\it fitting case}, the ratios $|C_V^{(\prime)} /T_V^{(\prime)}|$ are significantly
larger than $|C_P^{(\prime)} /T_P^{(\prime)}|$, though the values of $|C_P^{(\prime)}|$ include large
errors in~\cite{ChiangPV}, while in the {\it QCDF case} $|C_V^{(\prime)} /T_V^{(\prime)}| \sim
|C_P^{(\prime)} /T_P^{(\prime)}|$.
For the penguin amplitudes, the results from both cases are also different.  The values of
$|(P^{(\prime)}_P /\lambda_t^{d (s)}) /(T^{(\prime)}_P /\lambda_u^{d (s)})|$ for both $\bar B_{u,d} \to PP$
and $PV$ in the {\it fitting case} are larger than those in the {\it QCDF case}.  In the latter case,
the effects from $PE_{P}^{(\prime)}$ are comparable to and contribute constructively to those of
$P_{P}^{(\prime)}$, as discussed in the above ``{\it remark} 1''.
Interestingly it is found that the combined effects from $P_{P}^{(\prime)}$ and $PE_{P}^{(\prime)}$
obtained in the {\it QCDF case} are comparable to that of $P_{P}^{(\prime)}$ determined in the
{\it fitting case}. In contrast, the combined effects from $P_{V}^{(\prime)}$ and $PE_{V}^{(\prime)}$
found in the {\it QCDF case} are (roughly two times) larger than that of $P_{P}^{(\prime)}$ obtained
in the {\it fitting case}.
Also, for $\bar B_{u,d} \to PV$ decays, the ratio
$|(P^{(\prime)}_P /\lambda_t^{d (s)}) /(T^{(\prime)}_P /\lambda_u^{d (s)})| \sim
|(P^{(\prime)}_V /\lambda_t^{d (s)}) /(T^{(\prime)}_V /\lambda_u^{d (s)})|$ in the {\it QCDF case},
while $|(P^{(\prime)}_P /\lambda_t^{d (s)}) /(T^{(\prime)}_P /\lambda_u^{d (s)})| \sim
2 |(P^{(\prime)}_V /\lambda_t^{d (s)}) /(T^{(\prime)}_V /\lambda_u^{d (s)})|$ in the {\it fitting case}.
For the SU(3)$_{\rm F}$-singlet contributions, as discussed in the above ``{\it remark} 2'',
$S^{(\prime)}_{P,V}$ obtained in the {\it fitting case} are much larger than those found in the
{\it QCDF case}: e.g., for $\bar B_{u,d} \to PV$, the ratio
$|S^{(\prime)}_V /P^{(\prime)}_V| \approx 73\%$ in the {\it fitting case}, in contrast to
$|S^{(\prime)}_V /(P^{(\prime)}_V +PE^{(\prime)}_V)| \lesssim 10\%$~
[or $|S^{(\prime)}_V /P^{(\prime)}_V| \approx (18 - 24)\%$] in the {\it QCDF case}.

%%%%%%%%%%%%%%%%%%%%%%%%%%%%%%%%%%%%%%%%%%%%%%%%%%%%%%%%%%%%%%%%%%
%%%%%%%%%%%%%%%%%%%%%%%%%%%%%%%%%%%%%%%%%%%%%%%%%%%%%%%%%%%%%%%%%%
\subsection{Estimates of decay amplitudes, SU(3)$_{\rm F}$ breaking effects and SU(3)$_{\rm F}$ relations}

Using Tables~\ref{Abs_B_PP}$-$\ref{Abs_Bs_PV}, one can easily estimate the decay amplitudes of
$\bar B_{u,d} (\bar B_s) \to PP, ~PV$ numerically. For example, the decay amplitude of
$B^- \to \pi^- \pi^0$ is obtained as
\begin{eqnarray}
 \sqrt{2} {\cal A}_{B^- \to \pi^- \pi^0}
 &=& T_{\pi} +C_{\pi} +P_{{\rm EW}, \pi} +P^C_{{\rm EW}, \pi}  \nonumber \\
 &=& (1.52 -i ~24.94) \times 10^{-9} ~{\rm GeV} ~,
\label{B_pi1pi0}
\end{eqnarray}
and the decay amplitude of $B^- \to \pi^- \bar K^0$ given in Eq.~(\ref{B_pi1K0}) is estimated as
\begin{eqnarray}
 {\cal A}_{B^- \to \pi^- \bar K^0}  = (-49.71 -i ~24.77) \times 10^{-9} ~{\rm GeV} ~.
\end{eqnarray}
Likewise, the decay amplitude of $\bar B_s \to K^0 \pi^0$ is found to be
\begin{eqnarray}
 {\cal A}_{\bar B_s \to K^0 \pi^0}
 &=& C_K -P_K +P_{{\rm EW}, K} +\frac{1}{3} P_{{\rm EW}, K}^C -PE_K +\frac{1}{3} PE_{{\rm EW}, K}
   \nonumber \\
 &=& (-16.32 -i ~17.91) \times 10^{-9} ~{\rm GeV} ~,
\label{Bs_K0pi0}
\end{eqnarray}
and the decay amplitude of $\bar B_s \to K^+ K^{*-}$ is given by
\begin{eqnarray}
 {\cal A}_{\bar B_s \to K^+ K^{*-}}
 &=& T'_K +P'_K +\frac{2}{3} P_{{\rm EW}, K}^{C \prime} +E'_{K^*} +PE'_K +PA'_K +PA'_{K^*} \nonumber \\
 && -\frac{1}{3} PE'_{{\rm EW}, K} -\frac{1}{3} PA'_{{\rm EW}, K} +\frac{2}{3} PE'_{{\rm EW}, K^*}
   \nonumber \\
 &=& (-30.20 -i ~4.97) \times 10^{-9} ~{\rm GeV} ~.
\label{Bs_KpKstm}
\end{eqnarray}
In the above, the color-suppressed and color-favored tree amplitudes are, for example,
$C_K \equiv |C_K| ~e^{i \big( \delta_K^C +\theta^C \big) }$ and
$T'_K \equiv |T'_K| ~e^{i \big( \delta_K^{T \prime} +\theta^{T \prime} \big) }$, respectively,
with the strong phases $\delta_K^C$ and $\delta_K^{T \prime}$ and the weak phases
$\theta^C = {\rm arg} \big( V_{ub} V_{ud}^* \big)$ and
$\theta^{T \prime} = {\rm arg} \big( V_{ub} V_{us}^* \big)$.
The QCD- and EW-penguin and weak annihilation amplitudes have the similar form, such as
$P_K^{(\prime)} \equiv |P_K^{(\prime)}| ~e^{i \big( \delta_K^{P (\prime)} +\theta^{P (\prime)} \big) }$
and $E_{K^*} \equiv |E'_{K^*}| ~e^{i \big( \delta_{K^*}^{E \prime} +\theta^{E \prime} \big) }$, etc,
where the strong phases $\delta_K^P \neq \delta_K^{P \prime} \neq \delta_{K^*}^{E \prime}$ in general
and the weak phases $\theta^P = {\rm arg} \big( -V_{tb} V_{td}^* \big)$ and
$\theta^{P \prime} =\theta^{E \prime} = {\rm arg} \big( -V_{tb} V_{ts}^* \big)$.
By using Eqs.~(\ref{B_pi1pi0})$-$(\ref{Bs_KpKstm}), and noting that each SU(3)$_{\rm F}$ amplitude
and its {\it CP}-conjugate one are the same except for having the weak phase with opposite sign to each
other (e.g., the {\it CP}-conjugate amplitude to $C_K$ is $|C_K| ~e^{i \big( \delta_K^C -\theta^C \big) }$),
the estimation of direct {\it CP} asymmetries as well as the decay rates can be easily obtained.

The SU(3)$_{\rm F}$ breaking effects in the amplitudes arise from the decay constants, masses of the
mesons and the form factors in addition to the CKM matrix elements.  For example, taking into account
the effects of SU(3)$_{\rm F}$ breaking in $\bar B_{u,d} \to PP$, the ratio of $T'_P$ and $T_P$ is
estimated by $|T'_P / T_P| \approx [ |V_{us}| ~f_K (m_B^2 -m_K^2) F^{B K}_0 ] /
[ |V_{ud}| ~f_{\pi} (m_B^2 -m_{\pi}^2) F^{B \pi}_0 ]$. From Tables~\ref{Abs_B_PP}$-$\ref{Abs_Bs_PV},
the numerical estimates of the SU(3)$_{\rm F}$ breaking effects in the amplitudes can be obtained.
For both $\bar B_{u,d} (\bar B_s) \to PP$ decays with $\Delta S = 0$ and $|\Delta S| = 1$,
\begin{eqnarray}
 && \left| \frac{V_{ud}~ T'_P}{V_{us}~ T_P} \right| = 1.22 ~(1.21) ,
    \quad  \left| \frac{V_{ud}~ C'_P}{V_{us}~ C_P} \right| = 1.25 ~(1.26) ,  \nonumber \\
 && \left| \frac{V_{td}~ P'_P}{V_{ts}~ P_P} \right| = 1.18 ~(1.18) ,
    \quad  \left| \frac{V_{td}~ PE'_P}{V_{ts}~ PE_P} \right| = 1.19 ~(1.19) .
\end{eqnarray}
Likewise, for $\bar B_{u,d} \to PV$ decays,
\begin{eqnarray}
 && \left| \frac{V_{ud}~ T'_{P(V)}}{V_{us}~ T_{P(V)}} \right| = 1.02 ~(1.23),
    \qquad  \left| \frac{V_{ud}~ C'_{P(V)}}{V_{us}~ C_{P(V)}} \right| = 1.19 ~(1.02),  \nonumber \\
 && \left| \frac{V_{td}~ P'_{P(V)}}{V_{ts}~ P_{P(V)}} \right| = 1.07 ~(1.12),
    \qquad \left| \frac{V_{td}~ PE'_{P(V)}}{V_{ts}~ PE_{P(V)}} \right| = 1.10 ~(1.18),
\end{eqnarray}
and for $\bar B_s \to PV$ decays,
\begin{eqnarray}
 && \left| \frac{V_{ud}~ T'_{P(V)}}{V_{us}~ T_{P(V)}} \right| = 1.02 ~(1.22),
    \qquad  \left| \frac{V_{ud}~ C'_{P(V)}}{V_{us}~ C_{P(V)}} \right| = 1.00 ~(1.28),  \nonumber \\
 && \left| \frac{V_{td}~ P'_{P(V)}}{V_{ts}~ P_{P(V)}} \right| = 1.07 ~(1.15),
    \qquad  \left| \frac{V_{td}~ PE'_{P(V)}}{V_{ts}~ PE_{P(V)}} \right| = 1.11 ~(1.18).
\end{eqnarray}
In the above, we have factored out the relevant CKM matrix element from each SU(3)$_{\rm F}$ amplitude.
The results show that the SU(3)$_{\rm F}$ breaking is up to 28\% for the tree and color-suppressed tree
amplitudes, and 19\% for the QCD-penguin and QCD-penguin exchange amplitudes.

In Refs.~\cite{Gronau:1994rj, Oh:1998wa}, a number of SU(3)$_{\rm F}$ linear relations among various decay
amplitudes were presented. These relations were suggested to be used in testing various assumptions
made in the SU(3)$_{\rm F}$ analysis and extracting {\it CP} phases and strong final-state phases and so on.
In the previous studies, certain diagrams, such as the QCD-penguin exchange $PE'$ and the EW-penguin
exchange $PE'_{\rm EW}$, were ignored. As we have discussed in the previous subsection, the contribution
from the $PE'$ diagram turns out to be important in $|\Delta S| = 1$ decay processes.
Because of its topology, the QCD-penguin exchange amplitude $PE'$ always appears in the decay amplitude
together with the QCD-penguin one $P'$.  Thus, all the SU(3)$_{\rm F}$ linear relations obtained
in~\cite{Gronau:1994rj, Oh:1998wa} still hold after replacing $P'$ by $(P' +PE')$.
However, due to this replacement, the relevant strong phase of $P'$ should be changed as follows:
\begin{eqnarray}
\mbox{} \hspace{-0.3cm}
 P' = |P'| e^{i \delta^{P'}} e^{i \theta^{P'}} ~~~ \to ~~~
 P' +PE' = |P'| e^{i \delta^{P'}} e^{i \theta^{P'}} +|PE'| e^{i \delta^{PE'}} e^{i \theta^{PE'}}
   \equiv |\tilde P'| e^{i \delta^{\tilde P'}} e^{i \theta^{\tilde P'}},
\end{eqnarray}
where the weak phases $\theta^{P'} =\theta^{PE'} =\theta^{\tilde P'}$ under the assumption that the
top quark dominates the penguin amplitudes in the relevant processes.  Apparently, the strong phase
$\delta^{\tilde P'}$ is generally not the same as $\delta^{P'}$,
although they differ not much  because roughly $|P'| \sim |PE'|$ and
$\delta^{P'} \sim \delta^{PE'}$, as shown in Eqs.~(\ref{ratio_SU3amp_PP2})$-$(\ref{strong_phase_PV2}).
In fact, $\delta^{\tilde P'}$ arises from the different flavor operators $\alpha_4^{u,c}$ and
$b_3^{u,c}$ in QCDF.

For completeness, we present some useful SU(3)$_{\rm F}$ relations among the decay amplitudes of
$\bar B_d (\bar B_s) \to PV$ which are not given in~\cite{Gronau:1994rj}.
>From Tables~\ref{B_PV_DelS0_1}$-$\ref{B_PV_DelS1_3} and \ref{Bs_PV_DelS0_1}$-$\ref{Bs_PV_DelS1_3},
we find
\begin{eqnarray}
\label{SU3relation_1}
 && {\cal A}_{\bar B_s \to \pi^- K^{*+}}
   = T_{K^*} +P_{K^*} +\frac{2}{3} P_{{\rm EW}, K^*}^C +PE_{K^*} -\frac{1}{3} PE_{{\rm EW}, K^*}~,
   \nonumber \\
 && {\cal A}_{\bar B_d \to \pi^- \rho^+}
   = T_{\rho} +P_{\rho} +\frac{2}{3} P_{{\rm EW}, \rho}^C + E_{\pi} +PE_{\rho} +PA_{\pi} +PA_{\rho}
     -\frac{1}{3} PE_{{\rm EW}, \rho}  \nonumber \\
 && \hspace{2.3cm} +\frac{2}{3} PA_{{\rm EW}, \pi} -\frac{1}{3} PA_{{\rm EW}, \rho}~,   \nonumber \\
 && {\cal A}_{\bar B_s \to K^+ \rho^-}
   = T_{K} +P_{K} +\frac{2}{3} P_{{\rm EW}, K}^C +PE_{K} -\frac{1}{3} PE_{{\rm EW}, K}~,  \nonumber \\
 && {\cal A}_{\bar B_d \to \pi^+ \rho^-}
   = T_{\pi} +P_{\pi} +\frac{2}{3} P_{{\rm EW}, \pi}^C + E_{\rho} +PE_{\pi} +PA_{\rho} +PA_{\pi}
     -\frac{1}{3} PE_{{\rm EW}, \pi}  \nonumber \\
 && \hspace{2.3cm} +\frac{2}{3} PA_{{\rm EW}, \rho} -\frac{1}{3} PA_{{\rm EW}, \pi}~,   \nonumber \\
 && {\cal A}_{\bar B_s \to K^- K^{*+}}
   = T'_{K^*} +P'_{K^*} +\frac{2}{3} P_{{\rm EW}, K^*}^{C \prime} +E'_K +PE'_{K^*}
     +PA'_{K^*} +PA'_K \nonumber \\
 && \hspace{2.3cm} -\frac{1}{3} PE'_{{\rm EW}, K^*} -\frac{1}{3} PA'_{{\rm EW}, K^*}
     +\frac{2}{3} PA'_{{\rm EW}, K} ~, \nonumber \\
 && {\cal A}_{\bar B_d \to K^- \rho^+}
   = T'_{\rho} +P'_{\rho} +\frac{2}{3} P_{{\rm EW}, \rho}^{C \prime} +PE'_{\rho}
     -\frac{1}{3} PE'_{{\rm EW}, \rho} ~, \\
 && {\cal A}_{\bar B_d \to \pi^+ K^{*-}}
   = T'_{\pi} +P'_{\pi} +\frac{2}{3} P_{{\rm EW}, \pi}^{C \prime} +PE'_{\pi}
     -\frac{1}{3} PE'_{{\rm EW}, \pi} ~.   \nonumber
\end{eqnarray}
Also, from Tables~\ref{Abs_B_PV} and \ref{Abs_Bs_PV}, we see that
$|E_{\pi, \rho, K^{(*)}}^{(\prime)}|, |PA_{\pi, \rho, K^{(*)}}^{(\prime)}|,
|PA_{{\rm EW}, {\pi, \rho, K^{(*)}}}^{(\prime)}| \ll |T_{\pi, \rho, K^{(*)}}^{(\prime)}|$ and the
dominant contributions $|T_{K^* (K)}| \simeq |T_{\rho (\pi)}|$, $|P'_{K^* (K)}| \simeq |P'_{\rho (\pi)}|$
and $|PE'_{K^* (K)}| \simeq |PE'_{\rho (\pi)}|$.  Thus, it is expected from Eqs.~(\ref{Bs_KpKstm})
and (\ref{SU3relation_1}) that
\begin{eqnarray}
 && {\cal A}_{\bar B_s \to \pi^- K^{*+}} \simeq {\cal A}_{\bar B_d \to \pi^- \rho^+}~,  \qquad \qquad
   {\cal A}_{\bar B_s \to K^+ \rho^-} \simeq {\cal A}_{\bar B_d \to \pi^+ \rho^-}~, \nonumber \\
 && {\cal A}_{\bar B_s \to K^- K^{*+}} \simeq {\cal A}_{\bar B_d \to K^- \rho^+}~,  \qquad \quad ~
   {\cal A}_{\bar B_s \to K^+ K^{*-}} \simeq {\cal A}_{\bar B_d \to \pi^+ K^{*-}}~.
\end{eqnarray}
Consequently, we obtain the relations for the BFs and the direct {\it CP} asymmetries :
\begin{eqnarray}
 && {\cal B}(\bar B_s \to \pi^- K^{*+}) \simeq {\cal B}(\bar B_d \to \pi^- \rho^+) ~, \qquad \quad ~~
   {\cal B}(\bar B_s \to K^+ \rho^-) \simeq {\cal B}(\bar B_d \to \pi^+ \rho^-) ~, \nonumber \\
 && {\cal B}(\bar B_s \to K^- K^{*+}) \simeq {\cal B}(\bar B_d \to K^- \rho^+) ~, \qquad \quad
   {\cal B}(\bar B_s \to K^+ K^{*-}) \simeq {\cal B}(\bar B_d \to \pi^+ K^{*-}) ~,  \\
 && A_{CP}(\bar B_s \to \pi^- K^{*+}) \simeq A_{CP}(\bar B_d \to \pi^- \rho^+) ~,  \quad ~
   A_{CP}(\bar B_s \to K^+ \rho^-) \simeq A_{CP}(\bar B_d \to \pi^+ \rho^-) ~, \nonumber \\
 && A_{CP}(\bar B_s \to K^- K^{*+}) \simeq A_{CP}(\bar B_d \to K^- \rho^+) ~,  ~~
   A_{CP}(\bar B_s \to K^+ K^{*-}) \simeq A_{CP}(\bar B_d \to \pi^+ K^{*-}) ~. \nonumber
\end{eqnarray}
Numerically the above SU(3)$_{\rm F}$ relations are generally respected~\cite{Cheng:2009cn}.

%%%%%%%%%%%%%%%%%%%%%%%%%%%%%%%%%%%%%%%%%%%%%%%%%%%%%%%%%%%%%%%%%%%%%%%%%%%%%%%%%%%%%%%%%%
%%%%%%%%%%%%%%%%%%%%%%%%%%%%%%%%%%%%%%%%%%%%%%%%%%%%%%%%%%%%%%%%%%%%%%%%%%%%%%%%%%%%%%%%%%
\section{Conclusion}

Based on flavor SU(3) symmetry, we have presented a model-independent analysis of
$\bar B_{u,d} (\bar B_s) \to PP, ~PV$ decays.  Based on the topological diagrams, all the decay
amplitudes of interest have been expressed in terms of the the SU(3)$_{\rm F}$ amplitudes.
In order to bridge the topological-diagram approach (or the flavor SU(3) analysis) and the QCDF approach,
we have explicitly shown how to translate each SU(3)$_{\rm F}$ amplitude involved in these decay modes
into the corresponding terms in the framework of QCDF.
This is practically a way to easily find the rather {\it sophisticated} results of the relevant decay
amplitudes calculated in QCDF by taking into account the simpler and more {\it intuitive}
topological diagrams of relevance.
For further quantitative discussions, we have numerically computed each SU(3)$_{\rm F}$ amplitude in
QCDF and shown its magnitude and strong phase.

In our analysis, we have included the presumably subleading diagrams, such as the QCD- and EW-penguin
exchange ones ($PE$ and $PE_{\rm EW}$) and flavor-singlet weak annihilation ones ($SE$, $SA$, $SPE$,
$SPA$, $SPE_{\rm EW}$, $SPA_{\rm EW}$).  Among them, the contribution from the QCD-penguin exchange
diagram plays a crucial role in understanding the branching fractions and direct {\it CP} asymmetries
for penguin-dominant decays with $|\Delta S| = 1$, such as $\bar B_{u,d} \to \pi^+ K^-$,
$K \rho, ~\pi K^*$ decays.
Numerically the SU(3)$_{\rm F}$-singlet amplitudes $S^{(\prime)}_{P,V}$ involved in
$\bar B_{u,d} (\bar B_s) \to \pi \eta^{(\prime)}$, $K \eta^{(\prime)}$, $\pi\,\omega/\phi$,
$K \,\omega/\phi$, etc, are found to be small as expected from the OZI suppression rule.
On the other hand, the color-suppressed tree amplitude $C$ is found to be large and complex : e.g.,
for tree-dominated $\bar B_{u,d} (\bar B_s) \to PP$ decays,
$C_P/T_P \approx 0.63 \,e^{-i 56^\circ} ~(0.83 \,e^{-i 53^\circ})$ which is larger than the naive
expectation of $C_P/T_P \sim 1/3$ in phase and magnitude.  This large complex $C$ is needed to
understand the experimental data for the branching fractions of $\bar B_d \to \pi^0 \pi^0$,
$\pi^0 \rho^0$ and the direct {\it CP} asymmetries in $\bar B_{u,d} \to K^-\pi^0$, $K^-\eta$,
$\bar K^{*0}\eta$, $\pi^0\pi^0$, $\pi^-\eta$ modes.
We have also compared our results with those obtained from global fits to $\bar B_{u,d} \to PP, ~PV$
decays. Certain results, such as the effects of $C^{(\prime)}_P$ for $\bar B_{u,d} \to PP$, are
consistent with each other, but some other results, such as the contributions of $P^{(\prime)}_{P,V}$
and $S^{(\prime)}_{P,V}$ for $\bar B_{u,d} \to PP, ~PV$, are different from each other. These
differences stem mainly from the different ways of explaining the current data of $\bar B_{u,d} \to PP$,
$PV$ in these two approaches, depending on which SU(3)$_{\rm F}$ amplitudes become more important in
a particular mode.

As an example of the applications, we have discussed the SU(3)$_{\rm F}$ breaking effects.
Our results show that the SU(3)$_{\rm F}$ breaking is up to 28\% for the tree and color-suppressed tree
amplitudes and 19\% for the QCD-penguin and QCD-penguin exchange ones.  Using the SU(3)$_{\rm F}$
amplitudes, we have also derived some useful relations among the decay amplitudes of $\bar B_s \to PV$
and $\bar B_d \to PV$.
These SU(3)$_{\rm F}$ relations are expected to be tested in future experiments such as the upcoming
LHCb one.

%%%%%%%%%%%%%%%%%%%%%%%%%%%%%%%%%%%%%%%%%%%%%%%%%%%%%%%%%%%%%%%%%%%%%%%%%%%%%%%%%%%%%%%%%%
%%%%%%%%%%%%%%%%%%%%%%%%%%%%%%%%%%%%%%%%%%%%%%%%%%%%%%%%%%%%%%%%%%%%%%%%%%%%%%%%%%%%%%%%%%
\acknowledgments{
This work was supported in part by the National Science Council of R.O.C. under Grants Numbers:
NSC-97-2112-M-001-004-MY3 and NSC-99-2811-M-001-038. }

%%%%%%%%%%%%%%%%%%%%%%%%%%%%%%%%%%%%%%%%%%%%%%%%%%%%%%%%%%%%%%%%%%%%%%%%%%%%%%%%%%%%%%%%%%
%%%%%%%%%%%%%%%%%%%%%%%%%%%%%%%%%%%%%%%%%%%%%%%%%%%%%%%%%%%%%%%%%%%%%%%%%%%%%%%%%%%%%%%%%%

%==============================================================================
%                               Tables
%==============================================================================

\clearpage
%%%%%%%%%%%%%%%%%%%%%%%%%%%%%%%%%%%%%%%%%%%%%%%%%%%%%%%%%%%%%%%%%%  Table 1
\begin{table}
\caption{Coefficients of SU(3)$_{\rm F}$ amplitudes in $\bar B \to P_1 P_2$ ( $\Delta S = 0$ ).}
\begin{tabular}{c||c|c|c|c|c|c|c}
\hline
$\bar B \to P_1 P_2$ & factor & {\footnotesize $T_{P_1 ~[P_2]}^{(\zeta)}$}
  & {\footnotesize $C_{P_1 ~[P_2]}^{(\zeta)}$}
  & {\footnotesize $S_{P_1 ~[P_2]}^{(\zeta)}$}
  & {\footnotesize $P_{P_1 ~[P_2]}^{(\zeta)}$}
  & {\footnotesize $P_{{\rm EW}, ~P_1 ~[P_2]}^{(\zeta)}$}
  & {\footnotesize $P_{{\rm EW}, ~P_1 ~[P_2]}^{C, ~(\zeta)}$}
\\ \hline
$B^- \to \pi^- \pi^0$ & $\frac{1}{\sqrt{2}}$ & 0 & 1 & 0 & $-1$ & 1
  & $\frac{1}{3}$ \\
  &  & [1] & [0] & [0] & [1] & [0] & $[\frac{2}{3}]$ \\
$\bar B^0 \to \pi^- \pi^+$ & 1 & 1 & 0 & 0 & 1 & 0 & $\frac{2}{3}$  \\
  &  & [0] & [0] & [0] & [0] & [0] & [0] \\
$\bar B^0 \to \pi^0 \pi^0$ & $- \frac{1}{2}$ & 0 & 1 & 0 & $-1$ & 1 & $\frac{1}{3}$ \\
  &  & [0] & [1] & [0] & $[-1]$ & [1] & $[\frac{1}{3}]$ \\
$B^- \to K^- K^0$ & 1 & 0 & 0 & 0 & 1 & 0 & $- \frac{1}{3}$  \\
  &  & [0] & [0] & [0] & [0] & [0] & [0] \\
$\bar B^0 \to K^- K^+$ & 1 & 0 & 0 & 0 & 0 & 0 & 0  \\
  &  & [0] & [0] & [0] & [0] & [0] & [0] \\
$\bar B^0 \to \bar K^0 K^0$ & 1 & 0 & 0 & 0 & 1 & 0 & $- \frac{1}{3}$  \\
  &  & [0] & [0] & [0] & [0] & [0] & [0] \\
$B^- \to \pi^- \eta^{(\prime)}$ & $\frac{1}{\sqrt{2}}$ & 0 & {\footnotesize $1 (q) +\sqrt{2} (c)$}
  & {\footnotesize $2 (q) +\sqrt{2} (s) +\sqrt{2} (c)$} & $1 (q)$
  & {\footnotesize $\frac{1}{3} (q) - \frac{\sqrt{2}}{3} (s)$} & $- \frac{1}{3} (q)$  \\
  &  & $[1 (q)]$ & [0] & [0] & $[1 (q)]$ & [0] & $[\frac{2}{3} (q)]$  \\
$\bar B^0 \to \pi^0 \eta^{(\prime)}$ & $- \frac{1}{2}$ & 0 & {\footnotesize $1 (q) +\sqrt{2} (c)$}
  & {\footnotesize $2 (q) +\sqrt{2} (s) +\sqrt{2} (c)$} & $1 (q)$
  & {\footnotesize $\frac{1}{3} (q) - \frac{\sqrt{2}}{3} (s)$} & $- \frac{1}{3} (q)$  \\
  &  & [0] & $[-1 (q)]$ & [0] & $[1 (q)]$ & $[-1 (q)]$ & $[- \frac{1}{3} (q)]$  \\
$\bar B^0 \to \eta^{(\prime)} \eta^{(\prime)}$ & $\frac{1}{2}$ & 0 & {\footnotesize $1 (q,q)$}
  & {\footnotesize $2 (q,q) +\sqrt{2} (q,s)$} & {\footnotesize $1 (q,q)$}
  & {\footnotesize $\frac{1}{3} (q,q)$} & {\footnotesize $- \frac{1}{3} (q,q)$}  \\
  &  &  & $+\sqrt{2} (q,c)$ & $+\sqrt{2} (q,c)$ & & $- \frac{\sqrt{2}}{3} (q,s)$ &  \\
  &  & [0] & {\footnotesize $[1 (q,q)$} & {\footnotesize $[2 (q,q) +\sqrt{2} (q,s)$}
  & {\footnotesize $[1 (q,q)]$} & {\footnotesize $[\frac{1}{3} (q,q)$}
  & {\footnotesize $[- \frac{1}{3} (q,q)]$}  \\
  &  &  & {\footnotesize $+\sqrt{2} (q,c)]$} & {\footnotesize $+\sqrt{2} (q,c)]$}
  & & {\footnotesize $- \frac{\sqrt{2}}{3} (q,s)]$} &
\\ \hline
\end{tabular}
\label{B_PP_DelS0_1}
\end{table}
%%%%%%%%%%%%%%%%%%%%%%%%%%%%%%%%%%%%%%%%%%%%%%%%%%%%%%%%%%%%%%%%%%

%%%%%%%%%%%%%%%%%%%%%%%%%%%%%%%%%%%%%%%%%%%%%%%%%%%%%%%%%%%%%%%%%%  Table 2
\begin{table}
\caption{({\it Continued from Table~\ref{B_PP_DelS0_1}})~ Weak annihilation contributions.}
\begin{tabular}{c||c|c|c|c|c|c|c}
\hline
$\bar B \to P_1 P_2$ & factor & $E_{P_1 ~[P_2]}^{(\zeta)}$ & $A_{P_1 ~[P_2]}^{(\zeta)}$
  & $PE_{P_1 ~[P_2]}^{(\zeta)}$ & $PA_{P_1 ~[P_2]}^{(\zeta)}$
  & $PE_{{\rm EW}, ~P_1 ~[P_2]}^{(\zeta)}$ & $PA_{{\rm EW}, ~P_1 ~[P_2]}^{(\zeta)}$
\\ \hline
$B^- \to \pi^- \pi^0$ & $\frac{1}{\sqrt{2}}$ & 0 & $-1$ & $-1$ & 0 & $- \frac{2}{3}$ & 0  \\
  &  & [0] & [1] & [1] & [0] & $[\frac{2}{3}]$ & [0]  \\
$\bar B^0 \to \pi^- \pi^+$ & 1 & 0 & 0 & 1 & 1 & $- \frac{1}{3}$ & $- \frac{1}{3}$  \\
  &  & [1] & [0] & [0] & [1] & [0] & $[\frac{2}{3}]$   \\
$\bar B^0 \to \pi^0 \pi^0$ & $- \frac{1}{2}$ & $-1$ & 0 & $-1$
  & $-2$ & $\frac{1}{3}$ & $- \frac{1}{3}$  \\
  &  & $[-1]$ & [0] & $[-1]$ & $[-2]$ & $[\frac{1}{3}]$ & $[- \frac{1}{3}]$  \\
$B^- \to K^- K^0$ & 1 & 0 & 1 & 1 & 0 & $\frac{2}{3}$ & 0  \\
  &  & [0] & [0] & [0] & [0] & [0] & [0] \\
$\bar B^0 \to K^- K^+$ & 1 & 1 & 0 & 0 & 1 & 0 & $\frac{2}{3}$ \\
  &  & [0] & [0] & [0] & [1] & [0] & $[- \frac{1}{3}]$  \\
$\bar B^0 \to \bar K^0 K^0$ & 1 & 0 & 0 & 1 & 1 & $- \frac{1}{3}$ & $- \frac{1}{3}$  \\
  &  & [0] & [0] & [0] & [1] & [0] & $[- \frac{1}{3}]$ \\
$B^- \to \pi^- \eta^{(\prime)}$ & $\frac{1}{\sqrt{2}}$ & 0 & $1 (q)$
  & $1 (q)$ & 0 & $\frac{2}{3} (q)$ & 0  \\
  &  & [0] & $[1 (q)]$ & $[1 (q)]$ & [0] & $[\frac{2}{3} (q)]$ & [0]  \\
$\bar B^0 \to \pi^0 \eta^{(\prime)}$ & $- \frac{1}{2}$ & $-1 (q)$ & 0
  & $1 (q)$ & 0 & $- \frac{1}{3} (q)$ & $-1 (q)$  \\
  &  & $[-1 (q)]$ & [0] & $[1 (q)]$ & [0] & $[- \frac{1}{3} (q)]$ & $[-1 (q)]$  \\
$\bar B^0 \to \eta^{(\prime)} \eta^{(\prime)}$ & $\frac{1}{2}$ & $1 (q,q)$ & 0
  & $1 (q,q)$ & {\footnotesize $2 (q,q) +2 (s,s)$} & $- \frac{1}{3} (q,q)$
  & {\footnotesize $\frac{1}{3} (q,q) -\frac{2}{3} (s,s)$}  \\
  &  & $[1 (q,q)]$ & [0] & $[1 (q,q)]$ & {\footnotesize $[2 (q,q) +2 (s,s)]$}
  & $[- \frac{1}{3} (q,q)]$ & {\footnotesize $[\frac{1}{3} (q,q) -\frac{2}{3} (s,s)]$}
\\ \hline
\end{tabular}
\label{B_PP_DelS0_2}
\end{table}
%%%%%%%%%%%%%%%%%%%%%%%%%%%%%%%%%%%%%%%%%%%%%%%%%%%%%%%%%%%%%%%%%%

\clearpage
%%%%%%%%%%%%%%%%%%%%%%%%%%%%%%%%%%%%%%%%%%%%%%%%%%%%%%%%%%%%%%%%%%  Table 3
{\squeezetable
\begin{table}
\caption{({\it Continued from Table~\ref{B_PP_DelS0_2}})~
 Singlet weak annihilation contributions.}
\begin{tabular}{c||c|c|c|c|c|c|c}
\hline
$\bar B \to P_1 P_2$ & factor & $SE_{P_1 ~[P_2]}^{(\zeta)}$ & $SA_{P_1 ~[P_2]}^{(\zeta)}$
  & $SPE_{P_1 ~[P_2]}^{(\zeta)}$ & $SPA_{P_1 ~[P_2]}^{(\zeta)}$
  & $SPE_{{\rm EW}, ~P_1 ~[P_2]}^{(\zeta)}$ & $SPA_{{\rm EW}, ~P_1 ~[P_2]}^{(\zeta)}$
\\ \hline
$B^- \to \pi^- \eta^{(\prime)}$ & $\frac{1}{\sqrt{2}}$ & 0 & $2 (q) +\sqrt{2} (s)$
  & $2 (q) +\sqrt{2} (s)$ & 0 & $\frac{4}{3} (q) +\frac{2 \sqrt{2}}{3} (s)$ & 0  \\
  &  & [0] & [0] & [0] & [0] & [0] & [0] \\
$\bar B^0 \to \pi^0 \eta^{(\prime)}$ & $- \frac{1}{2}$ & $-2 (q) -\sqrt{2} (s)$
  & 0 & $2 (q) +\sqrt{2} (s)$ & 0 & $- \frac{2}{3} (q) - \frac{\sqrt{2}}{3} (s)$
  & $-2 (q) - \sqrt{2} (s)$  \\
  &  & [0] & [0] & [0] & [0] & [0] & [0] \\
$\bar B^0 \to \eta^{(\prime)} \eta^{(\prime)}$ & $\frac{1}{2}$ & $2 (q,q)$
  & 0 & $2 (q,q)$ & $4 (q,q) +2 \sqrt{2} (q,s)$
  & $- \frac{2}{3} (q,q)$ & $\frac{2}{3} (q,q) +\frac{\sqrt{2}}{3} (q,s)$  \\
  &  & $+\sqrt{2} (q,s)$ &  & $+\sqrt{2} (q,s)$ &  $+2 \sqrt{2} (s,q) +2 (s,s)$
  & $- \frac{\sqrt{2}}{3} (q,s)$ & $- \frac{2 \sqrt{2}}{3} (s,q) - \frac{2}{3} (s,s)$  \\
  &  & $[2 (q,q)$ & [0] & $[2 (q,q)$ & $[4 (q,q) +2 \sqrt{2} (q,s)$
  & $[- \frac{2}{3} (q,q)$ & $[\frac{2}{3} (q,q) +\frac{\sqrt{2}}{3} (q,s)$  \\
  &  & $+\sqrt{2} (q,s)]$ &  & $+\sqrt{2} (q,s)]$ &  $+2 \sqrt{2} (s,q) +2 (s,s)]$
  & $- \frac{\sqrt{2}}{3} (q,s)]$ & $- \frac{2 \sqrt{2}}{3} (s,q) - \frac{2}{3} (s,s)]$
\\ \hline
\end{tabular}
\label{B_PP_DelS0_3}
\end{table}
}
%%%%%%%%%%%%%%%%%%%%%%%%%%%%%%%%%%%%%%%%%%%%%%%%%%%%%%%%%%%%%%%%%%

%%%%%%%%%%%%%%%%%%%%%%%%%%%%%%%%%%%%%%%%%%%%%%%%%%%%%%%%%%%%%%%%%%  Table 4
\begin{table}
\caption{Coefficients of SU(3)$_{\rm F}$ amplitudes in $\bar B \to P_1 P_2$ ( $|\Delta S| = 1$ ).}
\begin{tabular}{c||c|c|c|c|c|c|c}
\hline
$\bar B \to P_1 P_2$ & factor & {\footnotesize $T_{P_1 ~[P_2]}^{\prime (\zeta)}$}
  & {\footnotesize $C_{P_1 ~[P_2]}^{\prime (\zeta)}$}
  & {\footnotesize $S_{P_1 ~[P_2]}^{\prime (\zeta)}$}
  & {\footnotesize $P_{P_1 ~[P_2]}^{\prime (\zeta)}$}
  & {\footnotesize $P_{{\rm EW}, ~P_1 ~[P_2]}^{\prime (\zeta)}$}
  & {\footnotesize $P_{{\rm EW}, ~P_1 ~[P_2]}^{C \prime , ~(\zeta)}$}
\\ \hline
$B^- \to \pi^- \bar K^0$ & 1 & 0 & 0 & 0 & 1 & 0 & $- \frac{1}{3}$ \\
    &  & [0] & [0] & [0] & [0] & [0] & [0] \\
$B^- \to \pi^0 K^-$ & $\frac{1}{\sqrt{2}}$ & 1 & 0 & 0 & 1 & 0 & $\frac{2}{3}$ \\
    &  & [0] & [1] & [0] & [0] & [1] & [0] \\
$\bar B^0 \to \pi^+ K^-$ & 1 & 1 & 0 & 0 & 1 & 0 & $\frac{2}{3}$  \\
    &  & [0] & [0] & [0] & [0] & [0] & [0] \\
$\bar B^0 \to \pi^0 \bar K^0$ & $\frac{1}{\sqrt{2}}$ & 0 & 0 & 0 & $-1$ & 0 & $\frac{1}{3}$ \\
    &  & [0] & [1] & [0] & [0] & [1] & [0] \\
$B^- \to K^- \eta^{(\prime)}$ & $\frac{1}{\sqrt{2}}$ & 0 & {\footnotesize $1 (q) +\sqrt{2} (c)$}
  & {\footnotesize $2 (q) +\sqrt{2} (s) +\sqrt{2} (c)$} & $\sqrt{2} (s)$
  & {\footnotesize $\frac{1}{3} (q) - \frac{\sqrt{2}}{3} (s)$} & $- \frac{\sqrt{2}}{3} (s)$  \\
  &  & $[1 (q)]$ & [0] & [0] & $[1 (q)]$ & [0] & $[\frac{2}{3} (q)]$  \\
$\bar B^0 \to \bar K^0 \eta^{(\prime)}$ & $\frac{1}{\sqrt{2}}$ & 0 & {\footnotesize $1 (q) +\sqrt{2} (c)$}
  & {\footnotesize $2 (q) +\sqrt{2} (s) +\sqrt{2} (c)$} & $\sqrt{2} (s)$
  & {\footnotesize $\frac{1}{3} (q) - \frac{\sqrt{2}}{3} (s)$} & $- \frac{\sqrt{2}}{3} (s)$  \\
  &  & $[0]$ & [0] & [0] & $[1 (q)]$ & [0] & $[- \frac{1}{3} (q)]$
\\ \hline
\end{tabular}
\label{B_PP_DelS1_1}
\end{table}
%%%%%%%%%%%%%%%%%%%%%%%%%%%%%%%%%%%%%%%%%%%%%%%%%%%%%%%%%%%%%%%%%%

\clearpage
%%%%%%%%%%%%%%%%%%%%%%%%%%%%%%%%%%%%%%%%%%%%%%%%%%%%%%%%%%%%%%%%%%  Table 5
\begin{table}
\caption{({\it Continued from Table~\ref{B_PP_DelS1_1}})~ Weak annihilation contributions.}
\begin{tabular}{c||c|c|c|c|c|c|c}
\hline
$\bar B \to P_1 P_2$ & factor & {\footnotesize $E_{P_1 ~[P_2]}^{\prime (\zeta)}$}
  & {\footnotesize $A_{P_1 ~[P_2]}^{\prime (\zeta)}$}
  & {\footnotesize $PE_{P_1 ~[P_2]}^{\prime (\zeta)}$}
  & {\footnotesize $PA_{P_1 ~[P_2]}^{\prime (\zeta)}$}
  & {\footnotesize $PE_{{\rm EW}, ~P_1 ~[P_2]}^{\prime (\zeta)}$}
  & {\footnotesize $PA_{{\rm EW}, ~P_1 ~[P_2]}^{\prime (\zeta)}$}
\\ \hline
$B^- \to \pi^- \bar K^0$ & 1 & 0 & 1 & 1 & 0 & $\frac{2}{3}$ & 0 \\
  &  & [0] & [0] & [0] & [0] & [0] & [0] \\
$B^- \to \pi^0 K^-$ & $\frac{1}{\sqrt{2}}$ & 0 & 1 & 1 & 0 & $\frac{2}{3}$ & 0 \\
  &  & [0] & [0] & [0] & [0] & [0] & [0] \\
$\bar B^0 \to \pi^+ K^-$ & 1 & 0 & 0 & 1 & 0 & $- \frac{1}{3}$ & 0  \\
  &  & [0] & [0] & [0] & [0] & [0] & [0] \\
$\bar B^0 \to \pi^0 \bar K^0$ & $\frac{1}{\sqrt{2}}$ & 0 & 0 & $-1$ & 0 & $\frac{1}{3}$ & 0 \\
  &  & [0] & [0] & [0] & [0] & [0] & [0] \\
$B^- \to K^- \eta^{(\prime)}$ & $\frac{1}{\sqrt{2}}$ & 0 & $\sqrt{2} (s)$
  & $\sqrt{2} (s)$ & 0 & $\frac{2 \sqrt{2}}{3} (s)$ & 0  \\
  &  & [0] & $[1 (q)]$ & $[1 (q)]$ & [0] & $[\frac{2}{3} (q)]$ & [0]  \\
$\bar B^0 \to \bar K^0 \eta^{(\prime)}$ & $\frac{1}{\sqrt{2}}$ & 0 & 0
  & $\sqrt{2} (s)$ & 0 & $- \frac{\sqrt{2}}{3} (s)$ & 0  \\
  &  & [0] & [0] & $[1 (q)]$ & [0] & $[- \frac{1}{3} (q)]$ & [0]
\\ \hline
\end{tabular}
\label{B_PP_DelS1_2}
\end{table}
%%%%%%%%%%%%%%%%%%%%%%%%%%%%%%%%%%%%%%%%%%%%%%%%%%%%%%%%%%%%%%%%%%

%%%%%%%%%%%%%%%%%%%%%%%%%%%%%%%%%%%%%%%%%%%%%%%%%%%%%%%%%%%%%%%%%%  Table 6
\begin{table}
\caption{({\it Continued from Table~\ref{B_PP_DelS1_2}})~ Singlet weak annihilation contributions.}
\begin{tabular}{c||c|c|c|c|c|c|c}
\hline
$\bar B \to P_1 P_2$ & factor & {\footnotesize $SE_{P_1 ~[P_2]}^{\prime (\zeta)}$}
  & {\footnotesize $SA_{P_1 ~[P_2]}^{\prime (\zeta)}$}
  & {\footnotesize $SPE_{P_1 ~[P_2]}^{\prime (\zeta)}$}
  & {\footnotesize $SPA_{P_1 ~[P_2]}^{\prime (\zeta)}$}
  & {\footnotesize $SPE_{{\rm EW}, ~P_1 ~[P_2]}^{\prime (\zeta)}$}
  & {\footnotesize $SPA_{{\rm EW}, ~P_1 ~[P_2]}^{\prime (\zeta)}$}
\\ \hline
$B^- \to K^- \eta^{(\prime)}$ & $\frac{1}{\sqrt{2}}$ & 0
  & {\footnotesize $2 (q) +\sqrt{2} (s)$}
  & {\footnotesize $2 (q) +\sqrt{2} (s)$} & 0
  & {\footnotesize $\frac{4}{3} (q) +\frac{2 \sqrt{2}}{3} (s)$} & 0  \\
  &  & [0] & [0] & [0] & [0] & [0] & [0] \\
$\bar B^0 \to \bar K^0 \eta^{(\prime)}$ & $\frac{1}{\sqrt{2}}$ & 0
  & 0 & {\footnotesize $2 (q) +\sqrt{2} (s)$} & 0
  & {\footnotesize $- \frac{2}{3} (q) - \frac{\sqrt{2}}{3} (s)$} & 0  \\
  &  & [0] & [0] & [0] & [0] & [0] & [0]
\\ \hline
\end{tabular}
\label{B_PP_DelS1_3}
\end{table}
%%%%%%%%%%%%%%%%%%%%%%%%%%%%%%%%%%%%%%%%%%%%%%%%%%%%%%%%%%%%%%%%%%

\clearpage
%%%%%%%%%%%%%%%%%%%%%%%%%%%%%%%%%%%%%%%%%%%%%%%%%%%%%%%%%%%%%%%%%%  Table 7
{\squeezetable
\begin{table}
\caption{Coefficients of SU(3)$_{\rm F}$ amplitudes in $\bar B \to PV$ ( $\Delta S = 0$ ).
When ideal mixing for $\omega$ and $\phi$ is assumed, i) for $B^- \to \pi^- \omega ~(\pi^- \phi)$ and
$\bar B^0 \to \pi^0 \omega ~(\pi^0 \phi)$, set the coefficients of SU(3)$_{\rm F}$ amplitudes with the
subscript $\pi$ and the superscript $\zeta = s ~(q)$ to zero: {\it i.e.}, for $\bar B \to \pi \omega$,
$S_{\pi}^{(s)} = P_{{\rm EW}, \pi}^{(s)} = 0$, and for $\bar B \to \pi \phi$,
$C_{\pi}^{(q)} = S_{\pi}^{(q)} = P_{\pi}^{(q)} = \cdots = 0$, and
ii) for $\bar B^0 \to \eta^{(\prime)} \omega ~[\eta^{(\prime)} \phi]$, set the coefficients of
SU(3)$_{\rm F}$ amplitudes with the subscript $\eta^{(\prime)}$ and the superscript
$\zeta = (q,s) ~{\rm or}~ (s,s)$ $[(q,q) ~{\rm or}~ (s,q)]$ to zero: {\it i.e.}, for
$\bar B^0 \to \eta^{(\prime)} \omega$,
$S_{\eta^{(\prime)}}^{(q,s)} = P_{{\rm EW}, \eta^{(\prime)}}^{(q,s)} = 0$, and
for $\bar B^0 \to \eta^{(\prime)} \phi$, $C_{\eta^{(\prime)}}^{(q,q)} = S_{\eta^{(\prime)}}^{(q,q)}
= P_{\eta^{(\prime)}}^{(q,q)} = \cdots = 0$. }
\begin{tabular}{c||c|c|c|c|c|c|c}
\hline
$\bar B \to P V$ & factor & {\footnotesize $T_{P ~[V]}^{(\zeta)}$}
  & {\footnotesize $C_{P ~[V]}^{(\zeta)}$}
  & {\footnotesize $S_{P ~[V]}^{(\zeta)}$}
  & {\footnotesize $P_{P ~[V]}^{(\zeta)}$}
  & {\footnotesize $P_{{\rm EW}, ~P ~[V]}^{(\zeta)}$}
  & {\footnotesize $P_{{\rm EW}, ~P ~[V]}^{C, ~(\zeta)}$}
\\ \hline
$B^- \to \pi^- \rho^0$ & $\frac{1}{\sqrt{2}}$ & 0 & 1 & 0 & $-1$ & 1
  & $\frac{1}{3}$ \\
  &  & [1] & [0] & [0] & $[1]$ & [0] & $[\frac{2}{3}]$ \\
$B^- \to \pi^0 \rho^-$ & $\frac{1}{\sqrt{2}}$ & 1 & 0 & 0 & 1 & 0
  & $\frac{2}{3}$ \\
  &  & [0] & [1] & [0] & $[-1]$ & [1] & $[\frac{1}{3}]$ \\
$\bar B^0 \to \pi^+ \rho^-$ & 1 & 1 & 0 & 0 & 1 & 0 & $\frac{2}{3}$  \\
  &  & [0] & [0] & [0] & [0] & [0] & [0] \\
$\bar B^0 \to \pi^- \rho^+$ & 1 & 0 & 0 & 0 & 0 & 0 & 0  \\
  &  & [1] & [0] & [0] & [1] & [0] & $[\frac{2}{3}]$ \\
$\bar B^0 \to \pi^0 \rho^0$ & $- \frac{1}{2}$ & 0 & 1 & 0 & $-1$ & 1 & $\frac{1}{3}$ \\
  &  & [0] & [1] & [0] & $[-1]$ & [1] & $[\frac{1}{3}]$ \\
$B^- \to K^- K^{*0}$ & 1 & 0 & 0 & 0 & 1 & 0 & $- \frac{1}{3}$  \\
  &  & [0] & [0] & [0] & [0] & [0] & [0] \\
$B^- \to K^0 K^{*-}$ & 1 & 0 & 0 & 0 & 0 & 0 & 0  \\
  &  & [0] & [0] & [0] & [1] & [0] & $[- \frac{1}{3}]$ \\
$\bar B^0 \to K^- K^{*+}$ & 1 & 0 & 0 & 0 & 0 & 0 & 0  \\
  &  & [0] & [0] & [0] & [0] & [0] & [0] \\
$\bar B^0 \to K^+ K^{*-}$ & 1 & 0 & 0 & 0 & 0 & 0 & 0  \\
  &  & [0] & [0] & [0] & [0] & [0] & [0] \\
$\bar B^0 \to \bar K^0 K^{*0}$ & 1 & 0 & 0 & 0 & 1 & 0 & $- \frac{1}{3}$  \\
  &  & [0] & [0] & [0] & [0] & [0] & [0] \\
$\bar B^0 \to K^0 \bar K^{*0}$ & 1 & 0 & 0 & 0 & 0 & 0 & 0  \\
  &  & [0] & [0] & [0] & [1] & [0] & $[- \frac{1}{3}]$ \\
$B^- \to \eta^{(\prime)} \rho^-$ & $\frac{1}{\sqrt{2}}$ & $1 (q)$ & 0 & 0 & $1 (q)$
  & 0 & $\frac{2}{3} (q)$  \\
  & & [0] & $[1 (q) +\sqrt{2} (c)]$ & $[2 (q) +\sqrt{2} (s) +\sqrt{2} (c)]$ & $[1 (q)]$
  & $[\frac{1}{3} (q) - \frac{\sqrt{2}}{3} (s)]$ & $[- \frac{1}{3} (q)]$  \\
$\bar B^0 \to \eta^{(\prime)} \rho^0$ & $- \frac{1}{2}$ & 0 & $-1 (q)$ & 0 & $1 (q)$
  & $-1 (q)$ & $- \frac{1}{3} (q)$  \\
  & & [0] & $[1 (q) +\sqrt{2} (c)]$ & $[2 (q) +\sqrt{2} (s) +\sqrt{2} (c)]$ & $[1 (q)]$
  & $[\frac{1}{3} (q) - \frac{\sqrt{2}}{3} (s)]$ & $[- \frac{1}{3} (q)]$  \\
$B^- \to \pi^- ~\omega/\phi$ & $\frac{1}{\sqrt{2}}$ & 0 & $1 (q)$
  & $2 (q) +\sqrt{2} (s)$ & $1 (q)$
  & $\frac{1}{3} (q) - \frac{\sqrt{2}}{3} (s)$ & $- \frac{1}{3} (q)$  \\
  &  & $[1 (q)]$ & [0] & [0] & $[1 (q)]$ & [0] & $[\frac{2}{3} (q)]$  \\
$\bar B^0 \to \pi^0 ~\omega/\phi$ & $- \frac{1}{2}$ & 0 & $1 (q)$ & $2 (q) +\sqrt{2} (s)$ & $1 (q)$
  & $\frac{1}{3} (q) - \frac{\sqrt{2}}{3} (s)$ & $- \frac{1}{3} (q)$  \\
  &  & [0] & $[-1 (q)]$ & [0] & $[1 (q)]$ & $[-1 (q)]$ & $[- \frac{1}{3} (q)]$  \\
$\bar B^0 \to \eta^{(\prime)} ~\omega/\phi$ & $\frac{1}{2}$ & 0 & $1 (q,q)$
  & $2 (q,q) +\sqrt{2} (q,s)$ & $1 (q,q)$
  & $\frac{1}{3} (q,q) - \frac{\sqrt{2}}{3} (q,s)$ & $- \frac{1}{3} (q,q)$  \\
  &  & [0] & $[1 (q,q)  +\sqrt{2} (q,c)]]$ & $[2 (q,q) +\sqrt{2} (q,s)$
  & $[1 (q,q)$ & $[\frac{1}{3} (q,q) - \frac{\sqrt{2}}{3} (q,s)]$ & $[- \frac{1}{3} (q,q)]$ \\
  &  &  &  & $+\sqrt{2} (q,c)]]$ &  &  &
\\ \hline
\end{tabular}
\label{B_PV_DelS0_1}
\end{table}
}
%%%%%%%%%%%%%%%%%%%%%%%%%%%%%%%%%%%%%%%%%%%%%%%%%%%%%%%%%%%%%%%%%%

%%%%%%%%%%%%%%%%%%%%%%%%%%%%%%%%%%%%%%%%%%%%%%%%%%%%%%%%%%%%%%%%%%  Table 8
{\squeezetable
\begin{table}
\caption{({\it Continued from Table~\ref{B_PV_DelS0_1}})~ Weak annihilation contributions.
When ideal mixing for $\omega$ and $\phi$ is assumed, the same rules as used in
{\it Table~\ref{B_PV_DelS0_1}} are applied. }
\begin{tabular}{c||c|c|c|c|c|c|c}
\hline
$\bar B \to P V$ & factor & $E_{P ~[V]}^{(\zeta)}$ & $A_{P ~[V]}^{(\zeta)}$
  & $PE_{P ~[V]}^{(\zeta)}$ & $PA_{P ~[V]}^{(\zeta)}$
  & $PE_{{\rm EW}, ~P ~[V]}^{(\zeta)}$ & $PA_{{\rm EW}, ~P ~[V]}^{(\zeta)}$
\\ \hline
$B^- \to \pi^- \rho^0$ & $\frac{1}{\sqrt{2}}$ & 0 & $-1$ & $-1$ & 0 & $- \frac{2}{3}$ & 0  \\
  &  & [0] & [1] & [1] & [0] & $[\frac{2}{3}]$ & [0]  \\
$B^- \to \pi^0 \rho^-$ & $\frac{1}{\sqrt{2}}$ & 0 & 1 & 1 & 0 & $\frac{2}{3}$ & 0  \\
  &  & [0] & $[-1]$ & $[-1]$ & [0] & $[- \frac{2}{3}]$ & [0]  \\
$\bar B^0 \to \pi^+ \rho^-$ & 1 & 0 & 0 & 1 & 1 & $- \frac{1}{3}$ & $- \frac{1}{3}$  \\
  &  & [1] & [0] & [0] & [1] & [0] & $[\frac{2}{3}]$   \\
$\bar B^0 \to \pi^- \rho^+$ & 1 & 1 & 0 & 0 & 1 & 0 & $\frac{2}{3}$   \\
  &  & [0] & [0] & [1] & [1] & $[- \frac{1}{3}]$ & $[- \frac{1}{3}]$  \\
$\bar B^0 \to \pi^0 \rho^0$ & $- \frac{1}{2}$ & $-1$ & 0 & $-1$
  & $-2$ & $\frac{1}{3}$ & $- \frac{1}{3}$  \\
  &  & $[-1]$ & [0] & $[-1]$ & $[-2]$ & $[\frac{1}{3}]$ & $[- \frac{1}{3}]$  \\
$B^- \to K^- K^{*0}$ & 1 & 0 & 1 & 1 & 0 & $\frac{2}{3}$ & 0  \\
  &  & [0] & [0] & [0] & [0] & [0] & [0] \\
$B^- \to K^0 K^{*-}$ & 1 & 0 & 0 & 0 & 0 & 0 & 0 \\
  &  & [0] & [1] & [1] & [0] & $[\frac{2}{3}]$ & [0]  \\
$\bar B^0 \to K^- K^{*+}$ & 1 & 1 & 0 & 0 & 1 & 0 & $\frac{2}{3}$ \\
  &  & [0] & [0] & [0] & [1] & [0] & $[- \frac{1}{3}]$  \\
$\bar B^0 \to K^+ K^{*-}$ & 1 & 0 & 0 & 0 & 1 & 0 & $- \frac{1}{3}$  \\
  &  & [1] & [0] & [0] & [1] & [0] & $[\frac{2}{3}]$ \\
$\bar B^0 \to \bar K^0 K^{*0}$ & 1 & 0 & 0 & 1 & 1 & $- \frac{1}{3}$ & $- \frac{1}{3}$  \\
  &  & [0] & [0] & [0] & [1] & [0] & $[- \frac{1}{3}]$ \\
$\bar B^0 \to K^0 \bar K^{*0}$ & 1 & 0 & 0 & 0 & 1 & 0 & $- \frac{1}{3}$ \\
  &  & [0] & [0] & [1] & [1] & $[- \frac{1}{3}]$ & $[- \frac{1}{3}]$  \\
$B^- \to \eta^{(\prime)} \rho^-$ & $\frac{1}{\sqrt{2}}$ & 0 & $1 (q)$
  & $1 (q)$ & 0 & $\frac{2}{3} (q)$ & 0  \\
  &  & [0] & $[1 (q)]$ & $[1 (q)]$ & [0] & $[\frac{2}{3} (q)]$ & [0]  \\
$\bar B^0 \to \eta^{(\prime)} \rho^0$ & $- \frac{1}{2}$ & $-1 (q)$ & 0
  & $1 (q)$ & 0 & $- \frac{1}{3} (q)$ & $-1 (q)$  \\
  &  & $[-1 (q)]$ & [0] & $[1 (q)]$ & [0] & $[- \frac{1}{3} (q)]$ & $[-1 (q)]$  \\
$B^- \to \pi^- ~\omega/\phi$ & $\frac{1}{\sqrt{2}}$ & 0 & $1 (q)$
  & $1 (q)$ & 0 & $\frac{2}{3} (q)$ & 0  \\
  &  & [0] & $[1 (q)]$ & $[1 (q)]$ & [0] & $[\frac{2}{3} (q)]$ & [0]  \\
$\bar B^0 \to \pi^0 ~\omega/\phi$ & $- \frac{1}{2}$ & $-1 (q)$ & 0
  & $1 (q)$ & 0 & $- \frac{1}{3} (q)$ & $-1 (q)$  \\
  &  & $[-1 (q)]$ & [0] & $[1 (q)]$ & [0] & $[- \frac{1}{3} (q)]$ & $[-1 (q)]$  \\
$\bar B^0 \to \eta^{(\prime)} ~\omega/\phi$ & $\frac{1}{2}$ & $1 (q,q)$ & 0
  & $1 (q,q)$ & $2 (q,q) +2 (s,s)$ & $- \frac{1}{3} (q,q)$
  & $\frac{1}{3} (q,q) -\frac{2}{3} (s,s)$  \\
  &  & $[1 (q,q)]$ & [0] & $[1 (q,q)]$ & $[2 (q,q) +2 (s,s)]$
  & $[- \frac{1}{3} (q,q)]$ & $[\frac{1}{3} (q,q) -\frac{2}{3} (s,s)]$
\\ \hline
\end{tabular}
\label{B_PV_DelS0_2}
\end{table}
}
%%%%%%%%%%%%%%%%%%%%%%%%%%%%%%%%%%%%%%%%%%%%%%%%%%%%%%%%%%%%%%%%%%

%%%%%%%%%%%%%%%%%%%%%%%%%%%%%%%%%%%%%%%%%%%%%%%%%%%%%%%%%%%%%%%%%%  Table 9
{\squeezetable
\begin{table}
\caption{({\it Continued from Table~\ref{B_PV_DelS0_2}})~
Singlet weak annihilation contributions.
When ideal mixing for $\omega$ and $\phi$ is assumed, the same rules as used in
{\it Table~\ref{B_PV_DelS0_1}} are applied. }
\begin{tabular}{c||c|c|c|c|c|c|c}
\hline
$\bar B \to P V$ & factor & $SE_{P ~[V]}^{(\zeta)}$ & $SA_{P ~[V]}^{(\zeta)}$
  & $SPE_{P ~[V]}^{(\zeta)}$ & $SPA_{P ~[V]}^{(\zeta)}$
  & $SPE_{{\rm EW}, ~P ~[V]}^{(\zeta)}$ & $SPA_{{\rm EW}, ~P ~[V]}^{(\zeta)}$
\\ \hline
$B^- \to \eta^{(\prime)} \rho^-$ & $\frac{1}{\sqrt{2}}$ & 0 & 0 & 0 & 0 & 0 & 0 \\
  &  & [0] & $[2 (q) +\sqrt{2} (s)]$ & $[2 (q) +\sqrt{2} (s)]$ & [0]
  & $[\frac{4}{3} (q) +\frac{2 \sqrt{2}}{3} (s)]$ & [0]  \\
$\bar B^0 \to \eta^{(\prime)} \rho^0$ & $- \frac{1}{2}$ & 0 & 0 & 0 & 0 & 0 & 0 \\
  &  & $[-2 (q) -\sqrt{2} (s)]$ & [0] & $[2 (q) +\sqrt{2} (s)]$ & [0]
  & $[- \frac{2}{3} (q) - \frac{\sqrt{2}}{3} (s)]$ & $[-2 (q) - \sqrt{2} (s)]$  \\
$B^- \to \pi^- ~\omega/\phi$ & $\frac{1}{\sqrt{2}}$ & 0 & $2 (q) +\sqrt{2} (s)$
  & $2 (q) +\sqrt{2} (s)$ & 0 & $\frac{4}{3} (q) +\frac{2 \sqrt{2}}{3} (s)$ & 0  \\
  &  & [0] & [0] & [0] & [0] & [0] & [0] \\
$\bar B^0 \to \pi^0 ~\omega/\phi$ & $- \frac{1}{2}$ & $-2 (q) -\sqrt{2} (s)$
  & 0 & $2 (q) +\sqrt{2} (s)$ & 0 & $- \frac{2}{3} (q) - \frac{\sqrt{2}}{3} (s)$
  & $-2 (q) - \sqrt{2} (s)$  \\
  &  & [0] & [0] & [0] & [0] & [0] & [0] \\
$\bar B^0 \to \eta^{(\prime)} ~\omega/\phi$ & $\frac{1}{2}$ & $2 (q,q)$
  & 0 & $2 (q,q)$ & $4 (q,q) +2 \sqrt{2} (q,s)$
  & $- \frac{2}{3} (q,q)$ & $\frac{2}{3} (q,q) +\frac{\sqrt{2}}{3} (q,s)$  \\
  &  & $+\sqrt{2} (q,s)$ &  & $+\sqrt{2} (q,s)$ &  $+2 \sqrt{2} (s,q) +2 (s,s)$
  & $- \frac{\sqrt{2}}{3} (q,s)$ & $- \frac{2 \sqrt{2}}{3} (s,q) - \frac{2}{3} (s,s)$  \\
  &  & $[2 (q,q)$ & [0] & $[2 (q,q)$ & $[4 (q,q) +2 \sqrt{2} (q,s)$
  & $[- \frac{2}{3} (q,q)$ & $[\frac{2}{3} (q,q) +\frac{\sqrt{2}}{3} (q,s)$  \\
  &  & $+\sqrt{2} (q,s)]$ &  & $+\sqrt{2} (q,s)]$ &  $+2 \sqrt{2} (s,q) +2 (s,s)]$
  & $- \frac{\sqrt{2}}{3} (q,s)]$ & $- \frac{2 \sqrt{2}}{3} (s,q) - \frac{2}{3} (s,s)]$
\\ \hline
\end{tabular}
\label{B_PV_DelS0_3}
\end{table}
}
%%%%%%%%%%%%%%%%%%%%%%%%%%%%%%%%%%%%%%%%%%%%%%%%%%%%%%%%%%%%%%%%%%

%%%%%%%%%%%%%%%%%%%%%%%%%%%%%%%%%%%%%%%%%%%%%%%%%%%%%%%%%%%%%%%%%%  Table 10
\begin{table}
\caption{Coefficients of SU(3)$_{\rm F}$ amplitudes in $\bar B \to P V$ ( $|\Delta S| = 1$ ).
When ideal mixing for $\omega$ and $\phi$ is assumed, for $B^- \to K^- \omega ~(K^- \phi)$ and
$\bar B^0 \to K^0 \omega ~(K^0 \phi)$, set the coefficients of SU(3)$_{\rm F}$ amplitudes with the
subscript $K$ and the superscript $\zeta = s ~(q)$ to zero: {\it i.e.}, for $\bar B \to K \omega$,
$S_K^{\prime (s)} = P_K^{\prime (s)} = \cdots = 0$, and for $\bar B \to K \phi$,
$C_K^{\prime (q)} = S_K^{\prime (q)} = P_{{\rm EW}, K}^{\prime (q)} = 0$.}
\begin{tabular}{c||c|c|c|c|c|c|c}
\hline
$\bar B \to P V$ & factor & {\footnotesize $T_{P ~[V]}^{\prime (\zeta)}$}
  & {\footnotesize $C_{P ~[V]}^{\prime (\zeta)}$}
  & {\footnotesize $S_{P ~[V]}^{\prime (\zeta)}$}
  & {\footnotesize $P_{P ~[V]}^{\prime (\zeta)}$}
  & {\footnotesize $P_{{\rm EW}, ~P ~[V]}^{\prime (\zeta)}$}
  & {\footnotesize $P_{{\rm EW}, ~P ~[V]}^{C \prime , ~(\zeta)}$}
\\ \hline
$B^- \to \pi^- \bar K^{*0}$ & 1 & 0 & 0 & 0 & 1 & 0 & $- \frac{1}{3}$ \\
  &  & [0] & [0] & [0] & [0] & [0] & [0] \\
$B^- \to \pi^0 K^{*-}$ & $\frac{1}{\sqrt{2}}$ & 1 & 0 & 0 & 1 & 0 & $\frac{2}{3}$ \\
  &  & [0] & [1] & [0] & [0] & [1] & [0] \\
$\bar B^0 \to \pi^+ K^{*-}$ & 1 & 1 & 0 & 0 & 1 & 0 & $\frac{2}{3}$  \\
  &  & [0] & [0] & [0] & [0] & [0] & [0] \\
$\bar B^0 \to \pi^0 \bar K^{*0}$ & $\frac{1}{\sqrt{2}}$ & 0 & 0 & 0 & $-1$ & 0 & $\frac{1}{3}$ \\
  &  & [0] & [1] & [0] & [0] & [1] & [0] \\
$B^- \to \bar K^0 \rho^-$ & 1 & 0 & 0 & 0 & 0 & 0 \\
  &  & [0] & [0] & [0] & [1] & [0] & $[- \frac{1}{3}]$ \\
$B^- \to K^- \rho^0$ & $\frac{1}{\sqrt{2}}$ & 0 & 1 & 0 & 0 & 1 & 0 \\
  &  & [1] & [0] & [0] & [1] & [0] & $[\frac{2}{3}]$ \\
$\bar B^0 \to K^- \rho^+$ & 1 & 0 & 0 & 0 & 0 & 0 & 0 \\
  &  & [1] & [0] & [0] & [1] & [0] & $[\frac{2}{3}]$  \\
$\bar B^0 \to \bar K^0 \rho^0$ & $\frac{1}{\sqrt{2}}$ & 0 & 1 & 0 & 0 & 1 & 0 \\
  &  & [0] & [0] & [0] & $[-1]$ & [0] & $[\frac{1}{3}]$ \\
$B^- \to \eta^{(\prime)} K^{*-}$ & $\frac{1}{\sqrt{2}}$ & $1 (q)$ & 0 & 0 & $1 (q)$ & 0
  & $\frac{2}{3} (q)$  \\
  &  & [0] & {\footnotesize $[1 (q) +\sqrt{2} (c)]$}
  & {\footnotesize $[2 (q) +\sqrt{2} (s) +\sqrt{2} (c)]$} & {\footnotesize $[\sqrt{2} (s)]$}
  & {\footnotesize $[\frac{1}{3} (q) - \frac{\sqrt{2}}{3} (s)]$}
  & {\footnotesize $[- \frac{\sqrt{2}}{3} (s)]$}  \\
$\bar B^0 \to \eta^{(\prime)} \bar K^{*0}$ & $\frac{1}{\sqrt{2}}$ & 0 & 0 & 0 & $1 (q)$ & 0
  & $- \frac{1}{3} (q)$  \\
  &  & [0] & {\footnotesize $[1 (q) +\sqrt{2} (c)]$}
  & {\footnotesize $[2 (q) +\sqrt{2} (s) +\sqrt{2} (c)]$} & {\footnotesize $[\sqrt{2} (s)]$}
  & {\footnotesize $[\frac{1}{3} (q) - \frac{\sqrt{2}}{3} (s)]$}
  & {\footnotesize $[- \frac{\sqrt{2}}{3} (s)]$}  \\
$B^- \to K^- ~\omega/\phi$ & $\frac{1}{\sqrt{2}}$ & 0 & {\footnotesize $1 (q)$}
  & {\footnotesize $2 (q) +\sqrt{2} (s)$} & {\footnotesize $\sqrt{2} (s)$}
  & {\footnotesize $\frac{1}{3} (q) - \frac{\sqrt{2}}{3} (s)$}
  & {\footnotesize $- \frac{\sqrt{2}}{3} (s)$}  \\
  &  & $[1 (q)]$ & [0] & [0] & $[1 (q)]$ & [0] & $[\frac{2}{3} (q)]$  \\
$\bar B^0 \to \bar K^0 ~\omega/\phi$ & $\frac{1}{\sqrt{2}}$ & 0 & {\footnotesize $1 (q)$}
  & {\footnotesize $2 (q) +\sqrt{2} (s)$} & {\footnotesize $\sqrt{2} (s)$}
  & {\footnotesize $\frac{1}{3} (q) - \frac{\sqrt{2}}{3} (s)$}
  & {\footnotesize $- \frac{\sqrt{2}}{3} (s)$}  \\
  &  & $[0]$ & [0] & [0] & $[1 (q)]$ & [0] & $[- \frac{1}{3} (q)]$
\\ \hline
\end{tabular}
\label{B_PV_DelS1_1}
\end{table}
%%%%%%%%%%%%%%%%%%%%%%%%%%%%%%%%%%%%%%%%%%%%%%%%%%%%%%%%%%%%%%%%%%

%%%%%%%%%%%%%%%%%%%%%%%%%%%%%%%%%%%%%%%%%%%%%%%%%%%%%%%%%%%%%%%%%%  Table 11
\begin{table}
\caption{({\it Continued from Table~\ref{B_PV_DelS1_1}})~ Weak annihilation contributions.
When ideal mixing for $\omega$ and $\phi$ is assumed, the same rules as used in
{\it Table~\ref{B_PV_DelS1_1}} are applied. }
\begin{tabular}{c||c|c|c|c|c|c|c}
\hline
$\bar B \to P V$ & factor & {\footnotesize $E_{P ~[V]}^{\prime (\zeta)}$}
  & {\footnotesize $A_{P ~[V]}^{\prime (\zeta)}$}
  & {\footnotesize $PE_{P ~[V]}^{\prime (\zeta)}$}
  & {\footnotesize $PA_{P ~[V]}^{\prime (\zeta)}$}
  & {\footnotesize $PE_{{\rm EW}, ~P ~[V]}^{\prime (\zeta)}$}
  & {\footnotesize $PA_{{\rm EW}, ~P ~[V]}^{\prime (\zeta)}$}
\\ \hline
$B^- \to \pi^- \bar K^{*0}$ & 1 & 0 & 1 & 1 & 0 & $\frac{2}{3}$ & 0 \\
  &  & [0] & [0] & [0] & [0] & [0] & [0] \\
$B^- \to \pi^0 K^{*-}$ & $\frac{1}{\sqrt{2}}$ & 0 & 1 & 1 & 0 & $\frac{2}{3}$ & 0 \\
  &  & [0] & [0] & [0] & [0] & [0] & [0] \\
$\bar B^0 \to \pi^+ K^{*-}$ & 1 & 0 & 0 & 1 & 0 & $- \frac{1}{3}$ & 0  \\
  &  & [0] & [0] & [0] & [0] & [0] & [0] \\
$\bar B^0 \to \pi^0 \bar K^{*0}$ & $\frac{1}{\sqrt{2}}$ & 0 & 0 & $-1$ & 0 & $\frac{1}{3}$ & 0 \\
  &  & [0] & [0] & [0] & [0] & [0] & [0] \\
$B^- \to \bar K^0 \rho^-$ & 1 & 0 & 0 & 0 & 0 & 0 & 0 \\
  &  & [0] & [1] & [1] & [0] & $[\frac{2}{3}]$ & [0] \\
$B^- \to K^- \rho^0$ & $\frac{1}{\sqrt{2}}$ & 0 & 0 & 0 & 0 & 0 & 0 \\
  &  & [0] & [1] & [1] & [0] & $[\frac{2}{3}]$ & [0] \\
$\bar B^0 \to K^- \rho^+$ & 1 & 0 & 0 & 0 & 0 & 0 & 0 \\
  &  & [0] & [0] & [1] & [0] & $[- \frac{1}{3}]$ & [0]  \\
$\bar B^0 \to \bar K^{*0} \rho^0$ & $\frac{1}{\sqrt{2}}$ & 0 & 0 & 0 & 0 & 0 & 0 \\
  &  & [0] & [0] & $[-1]$ & [0] & $[\frac{1}{3}]$ & [0] \\
$B^- \to \eta^{(\prime)} K^{*-}$ & $\frac{1}{\sqrt{2}}$ & 0 & $1 (q)$ & $1 (q)$ & 0
  & $\frac{2}{3} (q)$ & 0  \\
  &  & [0] & $[\sqrt{2} (s)]$ & $[\sqrt{2} (s)]$ & [0] & $[\frac{2 \sqrt{2}}{3} (s)]$ & [0]  \\
$\bar B^0 \to \eta^{(\prime)} \bar K^{*0}$ & $\frac{1}{\sqrt{2}}$ & 0 & 0 & $1 (q)$ & 0
  & $- \frac{1}{3} (q)$ & 0  \\
  &  & [0] & [0] & $[\sqrt{2} (s)]$ & [0] & $[- \frac{\sqrt{2}}{3} (s)]$ & [0]  \\
$B^- \to K^- ~\omega/\phi$ & $\frac{1}{\sqrt{2}}$ & 0 & $\sqrt{2} (s)$
  & $\sqrt{2} (s)$ & 0 & $\frac{2 \sqrt{2}}{3} (s)$ & 0  \\
  &  & [0] & $[1 (q)]$ & $[1 (q)]$ & [0] & $[\frac{2}{3} (q)]$ & [0]  \\
$\bar B^0 \to \bar K^0 ~\omega/\phi$ & $\frac{1}{\sqrt{2}}$ & 0 & 0
  & $\sqrt{2} (s)$ & 0 & $- \frac{\sqrt{2}}{3} (s)$ & 0  \\
  &  & [0] & [0] & $[1 (q)]$ & [0] & $[- \frac{1}{3} (q)]$ & [0]
\\ \hline
\end{tabular}
\label{B_PV_DelS1_2}
\end{table}
%%%%%%%%%%%%%%%%%%%%%%%%%%%%%%%%%%%%%%%%%%%%%%%%%%%%%%%%%%%%%%%%%%

%%%%%%%%%%%%%%%%%%%%%%%%%%%%%%%%%%%%%%%%%%%%%%%%%%%%%%%%%%%%%%%%%%  Table 12
\begin{table}
\caption{({\it Continued from Table~\ref{B_PV_DelS1_2}})~ Singlet weak annihilation contributions.
When ideal mixing for $\omega$ and $\phi$ is assumed, the same rules as used in
{\it Table~\ref{B_PV_DelS1_1}} are applied. }
\begin{tabular}{c||c|c|c|c|c|c|c}
\hline
$\bar B \to P V$ & factor & {\footnotesize $SE_{P ~[V]}^{\prime (\zeta)}$}
  & {\footnotesize $SA_{P ~[V]}^{\prime (\zeta)}$}
  & {\footnotesize $SPE_{P ~[V]}^{\prime (\zeta)}$}
  & {\footnotesize $SPA_{P ~[V]}^{\prime (\zeta)}$}
  & {\footnotesize $SPE_{{\rm EW}, ~P ~[V]}^{\prime (\zeta)}$}
  & {\footnotesize $SPA_{{\rm EW}, ~P ~[V]}^{\prime (\zeta)}$}
\\ \hline
$B^- \to \eta^{(\prime)} K^{*-}$ & $\frac{1}{\sqrt{2}}$ & 0 & 0 & 0 & 0 & 0 & 0 \\
  &  & [0] & {\footnotesize $[2 (q) +\sqrt{2} (s)]$}
  & {\footnotesize $[2 (q) +\sqrt{2} (s)]$} & [0]
  & {\footnotesize $[\frac{4}{3} (q) +\frac{2 \sqrt{2}}{3} (s)]$} & [0]  \\
$\bar B^0 \to \eta^{(\prime)} \bar K^{*0}$ & $\frac{1}{\sqrt{2}}$ & 0 & 0 & 0 & 0 & 0 & 0 \\
  &  & [0] & [0] & {\footnotesize $[2 (q) +\sqrt{2} (s)]$} & [0]
  & {\footnotesize $[- \frac{2}{3} (q) - \frac{\sqrt{2}}{3} (s)]$} & [0]  \\
$B^- \to K^- ~\omega/\phi$ & $\frac{1}{\sqrt{2}}$ & 0
  & {\footnotesize $2 (q) +\sqrt{2} (s)$}
  & {\footnotesize $2 (q) +\sqrt{2} (s)$} & 0
  & {\footnotesize $\frac{4}{3} (q) +\frac{2 \sqrt{2}}{3} (s)$} & 0  \\
  &  & [0] & [0] & [0] & [0] & [0] & [0] \\
$\bar B^0 \to \bar K^0 ~\omega/\phi$ & $\frac{1}{\sqrt{2}}$ & 0
  & 0 & {\footnotesize $2 (q) +\sqrt{2} (s)$} & 0
  & {\footnotesize $- \frac{2}{3} (q) - \frac{\sqrt{2}}{3} (s)$} & 0  \\
  &  & [0] & [0] & [0] & [0] & [0] & [0]
\\ \hline
\end{tabular}
\label{B_PV_DelS1_3}
\end{table}
%%%%%%%%%%%%%%%%%%%%%%%%%%%%%%%%%%%%%%%%%%%%%%%%%%%%%%%%%%%%%%%%%%

%%%%%%%%%%%%%%%%%%%%%%%%%%%%%%%%%%%%%%%%%%%%%%%%%%%%%%%%%%%%%%%%%%  Table 13
\begin{table}
\caption{Coefficients of SU(3)$_{\rm F}$ amplitudes in $\bar B_s \to P_1 P_2$ ( $\Delta S = 0$ ).}
\begin{tabular}{c||c|c|c|c|c|c|c}
\hline
$\bar B_s \to P_1 P_2$ & factor & {\footnotesize $T_{P_1 ~[P_2]}^{(\zeta)}$}
  & {\footnotesize $C_{P_1 ~[P_2]}^{(\zeta)}$} & {\footnotesize $S_{P_1 ~[P_2]}^{(\zeta)}$}
  & {\footnotesize $P_{P_1 ~[P_2]}^{(\zeta)}$}
  & {\footnotesize $P_{{\rm EW}, ~P_1 ~[P_2]}^{(\zeta)}$}
  & {\footnotesize $P_{{\rm EW}, ~P_1 ~[P_2]}^{C, ~(\zeta)}$}
\\ \hline
$\bar B_s \to K^+ \pi^-$ & 1 & 1 & 0 & 0 & 1 & 0 & $\frac{2}{3}$ \\
  &  & [0] & [0] & [0] & [0] & [0] & [0] \\
$\bar B_s \to K^0 \pi^0$ & $\frac{1}{\sqrt{2}}$ & 0 & 1 & 0 & $-1$ & 1 & $\frac{1}{3}$ \\
  &  & [0] & [0] & [0] & $[0]$ & [0] & [0] \\
$\bar B_s \to K^0 \eta^{(\prime)}$ & $\frac{1}{\sqrt{2}}$ & 0
  & {\footnotesize $1 (q) +\sqrt{2} (c)$}
  & {\footnotesize $2 (q) +\sqrt{2} (s) +\sqrt{2} (c)$} & $1 (q)$
  & {\footnotesize $\frac{1}{3} (q) - \frac{\sqrt{2}}{3} (s)$} & $- \frac{1}{3} (q)$  \\
  &  & [0] & [0] & [0] & $[\sqrt{2} (s)]$ & [0] & $[-\frac{\sqrt{2}}{3} (s)]$
\\ \hline
\end{tabular}
\label{Bs_PP_DelS0_1}
\end{table}
%%%%%%%%%%%%%%%%%%%%%%%%%%%%%%%%%%%%%%%%%%%%%%%%%%%%%%%%%%%%%%%%%%

\clearpage
%%%%%%%%%%%%%%%%%%%%%%%%%%%%%%%%%%%%%%%%%%%%%%%%%%%%%%%%%%%%%%%%%%  Table 14
\begin{table}
\caption{({\it Continued from Table~\ref{Bs_PP_DelS0_1}})~ Weak annihilation contributions.}
\begin{tabular}{c||c|c|c|c|c|c|c}
\hline
$\bar B_s \to P_1 P_2$ & factor & $E_{P_1 ~[P_2]}^{(\zeta)}$ & $A_{P_1 ~[P_2]}^{(\zeta)}$
  & $PE_{P_1 ~[P_2]}^{(\zeta)}$ & $PA_{P_1 ~[P_2]}^{(\zeta)}$
  & $PE_{{\rm EW}, ~P_1 ~[P_2]}^{(\zeta)}$ & $PA_{{\rm EW}, ~P_1 ~[P_2]}^{(\zeta)}$
\\ \hline
$\bar B_s \to K^+ \pi^-$ & 1 & 0 & 0 & 1 & 0 & $-\frac{1}{3}$ & 0 \\
  &  & [0] & [0] & [0] & [0] & [0] & [0] \\
$\bar B_s \to K^0 \pi^0$ & $\frac{1}{\sqrt{2}}$ & 0 & 0 & $-1$ & 0 & $\frac{1}{3}$ & 0 \\
  &  & [0] & [0] & [0] & $[0]$ & [0] & [0] \\
$\bar B_s \to K^0 \eta^{(\prime)}$ & $\frac{1}{\sqrt{2}}$ & 0
  & 0 & $1 (q)$ & 0 & {\footnotesize $-\frac{1}{3} (q)$} & 0  \\
  &  & [0] & [0] & $[\sqrt{2} (s)]$ & [0] & $[-\frac{\sqrt{2}}{3} (s)]$ & [0]
\\ \hline
\end{tabular}
\label{Bs_PP_DelS0_2}
\end{table}
%%%%%%%%%%%%%%%%%%%%%%%%%%%%%%%%%%%%%%%%%%%%%%%%%%%%%%%%%%%%%%%%%%

%%%%%%%%%%%%%%%%%%%%%%%%%%%%%%%%%%%%%%%%%%%%%%%%%%%%%%%%%%%%%%%%%%  Table 15
\begin{table}
\caption{({\it Continued from Table~\ref{Bs_PP_DelS0_2}})~
 Singlet weak annihilation contributions.}
\begin{tabular}{c||c|c|c|c|c|c|c}
\hline
$\bar B_s \to P_1 P_2$ & factor & {\footnotesize $SE_{P_1 ~[P_2]}^{(\zeta)}$}
  & {\footnotesize $SA_{P_1 ~[P_2]}^{(\zeta)}$} & {\footnotesize $SPE_{P_1 ~[P_2]}^{(\zeta)}$}
  & {\footnotesize $SPA_{P_1 ~[P_2]}^{(\zeta)}$}
  & {\footnotesize $SPE_{{\rm EW}, ~P_1 ~[P_2]}^{(\zeta)}$}
  & {\footnotesize $SPA_{{\rm EW}, ~P_1 ~[P_2]}^{(\zeta)}$}
\\ \hline
$\bar B_s \to K^0 \eta^{(\prime)}$ & $\frac{1}{\sqrt{2}}$ & 0 & 0 & $2 (q) +\sqrt{2} (s)$ & 0
  & {\footnotesize $-\frac{2}{3} (q) -\frac{\sqrt{2}}{3} (s)$} & 0  \\
  &  & [0] & [0] & [0] & [0] & [0] & [0]
\\ \hline
\end{tabular}
\label{Bs_PP_DelS0_3}
\end{table}
%%%%%%%%%%%%%%%%%%%%%%%%%%%%%%%%%%%%%%%%%%%%%%%%%%%%%%%%%%%%%%%%%%

%%%%%%%%%%%%%%%%%%%%%%%%%%%%%%%%%%%%%%%%%%%%%%%%%%%%%%%%%%%%%%%%%%  Table 16
\begin{table}
\caption{Coefficients of SU(3)$_{\rm F}$ amplitudes in $\bar B_s \to P_1 P_2$ ( $|\Delta S| = 1$ ).}
\begin{tabular}{c||c|c|c|c|c|c|c}
\hline
$\bar B_s \to P_1 P_2$ & factor & {\footnotesize $T_{P_1 ~[P_2]}^{\prime (\zeta)}$}
  & {\footnotesize $C_{P_1 ~[P_2]}^{\prime (\zeta)}$}
  & {\footnotesize $S_{P_1 ~[P_2]}^{\prime (\zeta)}$}
  & {\footnotesize $P_{P_1 ~[P_2]}^{\prime (\zeta)}$}
  & {\footnotesize $P_{{\rm EW}, ~P_1 ~[P_2]}^{\prime (\zeta)}$}
  & {\footnotesize $P_{{\rm EW}, ~P_1 ~[P_2]}^{C \prime , ~(\zeta)}$}
\\ \hline
$\bar B_s \to \pi^+ \pi^-$ & 1 & 0 & 0 & 0 & 0 & 0 & 0 \\
  &  & [0] & [0] & [0] & [0] & [0] & [0] \\
$\bar B_s \to \pi^0 \pi^0$ & $\frac{1}{2}$ & 0 & 0 & 0 & 0 & 0 & 0 \\
  &  & [0] & [0] & [0] & [0] & [0] & [0] \\
$\bar B_s \to \bar K^0 K^0$ & 1 & 0 & 0 & 0 & 0 & 0 & 0 \\
  &  & [0] & [0] & [0] & [1] & [0] & $[-\frac{1}{3}]$ \\
$\bar B_s \to K^- K^+$ & 1 & 0 & 0 & 0 & 0 & 0 & 0 \\
  &  & [1] & [0] & [0] & [1] & [0] & $[\frac{2}{3}]$  \\
$\bar B_s \to \pi^0 \eta^{(\prime)}$ & $\frac{1}{2}$ & 0 & 0 & 0 & 0 & 0 & 0 \\
  &  & [0] & $[\sqrt{2} (s)]$ & [0] & [0] & $[\sqrt{2} (s)]$ & [0]  \\
$\bar B_s \to \eta^{(\prime)} \eta^{(\prime)}$ & $\frac{1}{2}$ & 0
  & {\footnotesize $\sqrt{2} (s,q)$} & {\footnotesize $2 \sqrt{2} (s,q) +2 (s,s)$}
  & {\footnotesize $2 (s,s)$} & {\footnotesize $\frac{\sqrt{2}}{3} (s,q)$}
  & {\footnotesize $- \frac{2}{3} (s,s)$}  \\
  &  &  & $+2 (s,c)$ & $+2 (s,c)$ & & $- \frac{2}{3} (s,s)$ &  \\
  &  & [0]
  & {\footnotesize $[\sqrt{2} (s,q)$} & {\footnotesize $[2 \sqrt{2} (s,q) +2 (s,s)$}
  & {\footnotesize $[2 (s,s)]$} & {\footnotesize $[\frac{\sqrt{2}}{3} (s,q)$}
  & {\footnotesize $[- \frac{2}{3} (s,s)]$}  \\
  &  &  & $+2 (s,c)]$ & $+2 (s,c)]$ & & $- \frac{2}{3} (s,s)]$ &
\\ \hline
\end{tabular}
\label{Bs_PP_DelS1_1}
\end{table}
%%%%%%%%%%%%%%%%%%%%%%%%%%%%%%%%%%%%%%%%%%%%%%%%%%%%%%%%%%%%%%%%%%

%%%%%%%%%%%%%%%%%%%%%%%%%%%%%%%%%%%%%%%%%%%%%%%%%%%%%%%%%%%%%%%%%%  Table 17
\begin{table}
\caption{({\it Continued from Table~\ref{Bs_PP_DelS1_1}})~ Weak annihilation contributions.}
\begin{tabular}{c||c|c|c|c|c|c|c}
\hline
$\bar B_s \to P_1 P_2$ & factor & {\footnotesize $E_{P_1 ~[P_2]}^{\prime (\zeta)}$}
  & {\footnotesize $A_{P_1 ~[P_2]}^{\prime (\zeta)}$}
  & {\footnotesize $PE_{P_1 ~[P_2]}^{\prime (\zeta)}$}
  & {\footnotesize $PA_{P_1 ~[P_2]}^{\prime (\zeta)}$}
  & {\footnotesize $PE_{{\rm EW}, ~P_1 ~[P_2]}^{\prime (\zeta)}$}
  & {\footnotesize $PA_{{\rm EW}, ~P_1 ~[P_2]}^{\prime (\zeta)}$}
\\ \hline
$\bar B_s \to \pi^+ \pi^-$ & 1 & 0 & 0 & 0 & 1 & 0 & $-\frac{1}{3}$ \\
  &  & [1] & [0] & [0] & [1] & [0] & $[\frac{2}{3}]$ \\
$\bar B_s \to \pi^0 \pi^0$ & $\frac{1}{2}$ & 1 & 0 & 0 & 2 & 0 & $\frac{1}{3}$ \\
  &  & [1] & [0] & [0] & [2] & [0] & $[\frac{1}{3}]$ \\
$\bar B_s \to \bar K^0 K^0$ & 1 & 0 & 0 & 0 & 1 & 0 & $-\frac{1}{3}$ \\
  &  & [0] & [0] & [1] & [1] & $[-\frac{1}{3}]$ & $[-\frac{1}{3}]$ \\
$\bar B_s \to K^- K^+$ & 1 & 1 & 0 & 0 & 1 & 0 & $\frac{2}{3}$ \\
  &  & [0] & [0] & [1] & [1] & $[-\frac{1}{3}]$ & $[-\frac{1}{3}]$ \\
$\bar B_s \to \pi^0 \eta^{(\prime)}$ & $\frac{1}{2}$ & $1 (q)$ & 0 & 0 & 0 & 0 & $1 (q)$ \\
  &  & $[1 (q)]$ & [0] & [0] & [0] & [0] & $[1 (q)]$  \\
$\bar B_s \to \eta^{(\prime)} \eta^{(\prime)}$ & $\frac{1}{2}$ & $1 (q,q)$
  & 0 & {\footnotesize $2 (s,s)$}
  & {\footnotesize $2 (q,q)$} & {\footnotesize $-\frac{2}{3} (s,s)$}
  & {\footnotesize $\frac{1}{3} (q,q)$}  \\
  &  &  &  &  & $+2 (s,s)$ &  & $- \frac{2}{3} (s,s)$  \\
  &  & $[1 (q,q)]$
  & 0 & {\footnotesize $[2 (s,s)]$}
  & {\footnotesize $[2 (q,q)$} & {\footnotesize $[-\frac{2}{3} (s,s)]$}
  & {\footnotesize $[\frac{1}{3} (q,q)$}  \\
  &  &  &  &  & $+2 (s,s)]$ &  & $- \frac{2}{3} (s,s)]$
\\ \hline
\end{tabular}
\label{Bs_PP_DelS1_2}
\end{table}
%%%%%%%%%%%%%%%%%%%%%%%%%%%%%%%%%%%%%%%%%%%%%%%%%%%%%%%%%%%%%%%%%%

%%%%%%%%%%%%%%%%%%%%%%%%%%%%%%%%%%%%%%%%%%%%%%%%%%%%%%%%%%%%%%%%%%  Table 18
{\squeezetable
\begin{table}
\caption{({\it Continued from Table~\ref{Bs_PP_DelS1_2}})~
 Singlet weak annihilation contributions.}
\begin{tabular}{c||c|c|c|c|c|c|c}
\hline
$\bar B_s \to P_1 P_2$ & factor & $SE_{P_1 ~[P_2]}^{\prime (\zeta)}$
  & $SA_{P_1 ~[P_2]}^{\prime (\zeta)}$ & $SPE_{P_1 ~[P_2]}^{\prime (\zeta)}$
  & $SPA_{P_1 ~[P_2]}^{\prime (\zeta)}$ & $SPE_{{\rm EW}, ~P_1 ~[P_2]}^{\prime (\zeta)}$
  & $SPA_{{\rm EW}, ~P_1 ~[P_2]}^{\prime (\zeta)}$
\\ \hline
$\bar B_s \to \pi^0 \eta^{(\prime)}$ & $\frac{1}{2}$ & $2 (q) +\sqrt{2} (s)$
  & 0 & 0 & 0 & 0 & $2 (q) +\sqrt{2} (s)$ \\
  &  & [0] & [0] & [0] & [0] & [0] & [0] \\
$\bar B_s \to \eta^{(\prime)} \eta^{(\prime)}$
  & $\frac{1}{2}$ & $2 (q,q)$ & 0 & $2 \sqrt{2} (s,q)$ & $4 (q,q) +2 \sqrt{2} (q,s)$
  & $-\sqrt{2} (s,q)$ & $\frac{2}{3} (q,q) +\frac{\sqrt{2}}{3} (q,s)$  \\
  &  & $+\sqrt{2} (q,s)$ &  & $+2 (s,s)$ & $+2 \sqrt{2} (s,q) +2 (s,s)$
  & $-\frac{2}{3} (s,s)$ & $-\sqrt{2} (s,q) -\frac{2}{3} (s,s)$  \\
  &  & $[2 (q,q)$ & [0] & $[2 \sqrt{2} (s,q)$ & $[4 (q,q) +2 \sqrt{2} (q,s)$
  & $[-\sqrt{2} (s,q)$ & $[\frac{2}{3} (q,q) +\frac{\sqrt{2}}{3} (q,s)$  \\
  &  & $+\sqrt{2} (q,s)]$ &  & $+2 (s,s)]$ & $+2 \sqrt{2} (s,q) +2 (s,s)]$
  & $-\frac{2}{3} (s,s)]$ & $-\sqrt{2} (s,q) -\frac{2}{3} (s,s)]$
\\ \hline
\end{tabular}
\label{Bs_PP_DelS1_3}
\end{table}
}
%%%%%%%%%%%%%%%%%%%%%%%%%%%%%%%%%%%%%%%%%%%%%%%%%%%%%%%%%%%%%%%%%%

%%%%%%%%%%%%%%%%%%%%%%%%%%%%%%%%%%%%%%%%%%%%%%%%%%%%%%%%%%%%%%%%%%  Table 19
\begin{table}
\caption{Coefficients of SU(3)$_{\rm F}$ amplitudes in $\bar B_s \to PV$ ( $\Delta S = 0$ ).
When ideal mixing for $\omega$ and $\phi$ is assumed, for $\bar B_s \to K^0 \omega ~(K^0 \phi)$,
set the coefficients of SU(3)$_{\rm F}$ amplitudes with the subscript $K$ and the superscript
$\zeta = s ~(q)$ to zero: {\it i.e.}, for $\bar B_s \to K^0 \omega$,
$S_K^{(s)} = P_{{\rm EW}, K}^{(s)} = 0$, and for $\bar B_s \to K^0 \phi$,
$C_K^{(q)} = S_K^{(q)} = P_K^{(q)} = \cdots = 0$.}
\begin{tabular}{c||c|c|c|c|c|c|c}
\hline
$\bar B_s \to PV$ & factor & {\footnotesize $T_{P ~[V]}^{(\zeta)}$}
  & {\footnotesize $C_{P ~[V]}^{(\zeta)}$}
  & {\footnotesize $S_{P ~[V]}^{(\zeta)}$}
  & {\footnotesize $P_{P ~[V]}^{(\zeta)}$}
  & {\footnotesize $P_{{\rm EW}, ~P ~[V]}^{(\zeta)}$}
  & {\footnotesize $P_{{\rm EW}, ~P ~[V]}^{C, ~(\zeta)}$}
\\ \hline
$\bar B_s \to \pi^- K^{*+}$ & 1 & 0 & 0 & 0 & 0 & 0 & 0 \\
  &  & [1] & [0] & [0] & [1] & [0] & $[\frac{2}{3}]$ \\
$\bar B_s \to \pi^0 K^{*0}$ & $\frac{1}{\sqrt{2}}$ & 0 & 0 & 0 & 0 & 0 & 0 \\
  &  & [0] & [1] & [0] & $[-1]$ & [1] & $[\frac{1}{3}]$ \\
$\bar B_s \to \eta^{(\prime)} K^{*0}$ & $\frac{1}{\sqrt{2}}$
  & 0 & 0 & 0 & $\sqrt{2} (s)$ & 0 & $-\frac{\sqrt{2}}{3} (s)$  \\
  &  & [0]
  & {\footnotesize $[1 (q) +\sqrt{2} (c)]$}
  & {\footnotesize $[2 (q) +\sqrt{2} (s) +\sqrt{2} (c)]$} & $[1 (q)]$
  & {\footnotesize $[\frac{1}{3} (q) - \frac{\sqrt{2}}{3} (s)]$} & $[- \frac{1}{3} (q)]$  \\
$\bar B_s \to K^{+} \rho^-$ & 1 & 1 & 0 & 0 & 1 & 0 & $\frac{2}{3}$ \\
  &  & [0] & [0] & [0] & [0] & [0] & [0] \\
$\bar B_s \to K^{0} \rho^0$ & $\frac{1}{\sqrt{2}}$ & 0 & 1 & 0 & $-1$ & 1 & $\frac{1}{3}$ \\
  &  & [0] & [0] & [0] & $[0]$ & [0] & [0] \\
$\bar B_s \to K^{0} ~\omega / \phi$ & $\frac{1}{\sqrt{2}}$ & 0
  & {\footnotesize $1 (q)$}
  & {\footnotesize $2 (q) +\sqrt{2} (s)$} & $1 (q)$
  & {\footnotesize $\frac{1}{3} (q) - \frac{\sqrt{2}}{3} (s)$} & $- \frac{1}{3} (q)$  \\
  &  & [0] & [0] & [0] & $[\sqrt{2} (s)]$ & [0] & $[-\frac{\sqrt{2}}{3} (s)]$
\\ \hline
\end{tabular}
\label{Bs_PV_DelS0_1}
\end{table}
%%%%%%%%%%%%%%%%%%%%%%%%%%%%%%%%%%%%%%%%%%%%%%%%%%%%%%%%%%%%%%%%%%

%%%%%%%%%%%%%%%%%%%%%%%%%%%%%%%%%%%%%%%%%%%%%%%%%%%%%%%%%%%%%%%%%%  Table 20
\begin{table}
\caption{({\it Continued from Table~\ref{Bs_PV_DelS0_1}})~ Weak annihilation contributions.
When ideal mixing for $\omega$ and $\phi$ is assumed, the same rules as used in
{\it Table~\ref{Bs_PV_DelS0_1}} are applied. }
\begin{tabular}{c||c|c|c|c|c|c|c}
\hline
$\bar B_s \to PV$ & factor & $E_{P ~[V]}^{(\zeta)}$ & $A_{P ~[V]}^{(\zeta)}$
  & $PE_{P ~[V]}^{(\zeta)}$ & $PA_{P ~[V]}^{(\zeta)}$
  & $PE_{{\rm EW}, ~P ~[V]}^{(\zeta)}$ & $PA_{{\rm EW}, ~P ~[V]}^{(\zeta)}$
\\ \hline
$\bar B_s \to \pi^- K^{*+}$ & 1 & 0 & 0 & 0 & 0 & 0 & 0 \\
  &  & [0] & [0] & [1] & [0] & $[-\frac{1}{3}]$ & [0] \\
$\bar B_s \to \pi^0 K^{*0}$ & $\frac{1}{\sqrt{2}}$ & 0 & 0 & 0 & 0 & 0 & 0 \\
  &  & [0] & [0] & $[-1]$ & [0] & $[\frac{1}{3}]$ & [0] \\
$\bar B_s \to \eta^{(\prime)} K^{*0}$ & $\frac{1}{\sqrt{2}}$
  & 0 & 0 & $\sqrt{2} (s)$ & 0 & $-\frac{\sqrt{2}}{3} (s)$ & 0 \\
  &  & [0] & [0] & $[1 (q)]$ & [0] & {\footnotesize $[-\frac{1}{3} (q)]$} & [0]  \\
$\bar B_s \to K^+ \rho^-$ & 1 & 0 & 0 & 1 & 0 & $-\frac{1}{3}$ & 0 \\
  &  & [0] & [0] & [0] & [0] & [0] & [0] \\
$\bar B_s \to K^0 \rho^0$ & $\frac{1}{\sqrt{2}}$ & 0 & 0 & $-1$ & 0 & $\frac{1}{3}$ & 0 \\
  &  & [0] & [0] & [0] & $[0]$ & [0] & [0] \\
$\bar B_s \to K^0 ~\omega / \phi$ & $\frac{1}{\sqrt{2}}$ & 0
  & 0 & $1 (q)$ & 0 & {\footnotesize $-\frac{1}{3} (q)$} & 0  \\
  &  & [0] & [0] & $[\sqrt{2} (s)]$ & [0] & $[-\frac{\sqrt{2}}{3} (s)]$ & [0]
\\ \hline
\end{tabular}
\label{Bs_PV_DelS0_2}
\end{table}
%%%%%%%%%%%%%%%%%%%%%%%%%%%%%%%%%%%%%%%%%%%%%%%%%%%%%%%%%%%%%%%%%%

%%%%%%%%%%%%%%%%%%%%%%%%%%%%%%%%%%%%%%%%%%%%%%%%%%%%%%%%%%%%%%%%%%  Table 21
\begin{table}
\caption{({\it Continued from Table~\ref{Bs_PV_DelS0_2}})~
Singlet weak annihilation contributions.
When ideal mixing for $\omega$ and $\phi$ is assumed, the same rules as used in
{\it Table~\ref{Bs_PV_DelS0_1}} are applied. }
\begin{tabular}{c||c|c|c|c|c|c|c}
\hline
$\bar B_s \to PV$ & factor & {\footnotesize $SE_{P ~[V]}^{(\zeta)}$}
  & {\footnotesize $SA_{P ~[V]}^{(\zeta)}$} & {\footnotesize $SPE_{P ~[V]}^{(\zeta)}$}
  & {\footnotesize $SPA_{P ~[V]}^{(\zeta)}$}
  & {\footnotesize $SPE_{{\rm EW}, ~P ~[V]}^{(\zeta)}$}
  & {\footnotesize $SPA_{{\rm EW}, ~P ~[V]}^{(\zeta)}$}
\\ \hline
$\bar B_s \to \eta^{(\prime)} K^{*0}$ & $\frac{1}{\sqrt{2}}$
  & 0 & 0 & 0 & 0 & 0 & 0 \\
  &  & [0] & [0] & $[2 (q) +\sqrt{2} (s)]$ & [0]
  & {\footnotesize $[-\frac{2}{3} (q) -\frac{\sqrt{2}}{3} (s)]$} & [0]  \\
$\bar B_s \to K^0 ~\omega / \phi$ & $\frac{1}{\sqrt{2}}$ & 0 & 0 & $2 (q) +\sqrt{2} (s)$ & 0
  & {\footnotesize $-\frac{2}{3} (q) -\frac{\sqrt{2}}{3} (s)$} & 0  \\
  &  & [0] & [0] & [0] & [0] & [0] & [0]
\\ \hline
\end{tabular}
\label{Bs_PV_DelS0_3}
\end{table}
%%%%%%%%%%%%%%%%%%%%%%%%%%%%%%%%%%%%%%%%%%%%%%%%%%%%%%%%%%%%%%%%%%

\clearpage
%%%%%%%%%%%%%%%%%%%%%%%%%%%%%%%%%%%%%%%%%%%%%%%%%%%%%%%%%%%%%%%%%%  Table 22
\begin{table}
\caption{Coefficients of SU(3)$_{\rm F}$ amplitudes in $\bar B_s \to PV$ ( $|\Delta S| = 1$ ).
When ideal mixing for $\omega$ and $\phi$ is assumed, i) for $\bar B_s \to \pi^0 \omega ~(\pi^0 \phi)$,
set the coefficients of SU(3)$_{\rm F}$ amplitudes with the subscript $\pi$ and the superscript
$\zeta = s ~(q)$ to zero: {\it i.e.}, for $\bar B_s \to \pi^0 \omega$,
$SE_{\pi}^{\prime (s)} = SPA_{{\rm EW}, \pi}^{\prime (s)} = 0$ [See {\it Tabel~\ref{Bs_PV_DelS1_3}}.],
and for $\bar B_s \to \pi^0 \phi$, $E_{\pi}^{\prime (q)} = PA_{\pi}^{\prime (q)} = \cdots = 0$
[See {\it Tabel~\ref{Bs_PV_DelS1_2}}.], and
ii) for $\bar B_s \to \eta^{(\prime)} \omega ~[\eta^{(\prime)} \phi]$, set the coefficients of
SU(3)$_{\rm F}$ amplitudes with the superscript $\zeta = (s,s) ~{\rm or}~ (q,s)$
$[(s,q) ~{\rm or}~ (q,q)]$ to zero: {\it i.e.}, for $\bar B_s \to \eta^{(\prime)} \omega$,
$C_{\omega}^{\prime (q,s)} = S_{\eta^{(\prime)}}^{\prime (s,s)} = S_{\omega}^{\prime (q,s)}
= S_{\omega}^{\prime (s,s)} = \cdots = 0$, and
for $\bar B_s \to \eta^{(\prime)} \phi$, $C_{\eta^{(\prime)}}^{\prime (s,q)}
= S_{\eta^{(\prime)}}^{\prime (s,q)} = P_{{\rm EW}, \eta^{(\prime)}}^{\prime (s,q)} = 0$. }
\begin{tabular}{c||c|c|c|c|c|c|c}
\hline
$\bar B_s \to PV$ & factor & {\footnotesize $T_{P ~[V]}^{\prime (\zeta)}$}
  & {\footnotesize $C_{P ~[V]}^{\prime (\zeta)}$}
  & {\footnotesize $S_{P ~[V]}^{\prime (\zeta)}$}
  & {\footnotesize $P_{P ~[V]}^{\prime (\zeta)}$}
  & {\footnotesize $P_{{\rm EW}, ~P ~[V]}^{\prime (\zeta)}$}
  & {\footnotesize $P_{{\rm EW}, ~P ~[V]}^{C \prime , ~(\zeta)}$}
\\ \hline
$\bar B_s \to \pi^+ \rho^-$ & 1 & 0 & 0 & 0 & 0 & 0 & 0 \\
  &  & [0] & [0] & [0] & [0] & [0] & [0] \\
$\bar B_s \to \pi^- \rho^+$ & 1 & 0 & 0 & 0 & 0 & 0 & 0 \\
  &  & [0] & [0] & [0] & [0] & [0] & [0] \\
$\bar B_s \to \pi^0 \rho^0$ & $\frac{1}{2}$ & 0 & 0 & 0 & 0 & 0 & 0 \\
  &  & [0] & [0] & [0] & [0] & [0] & [0] \\
$\bar B_s \to \bar K^0 K^{*0}$ & 1 & 0 & 0 & 0 & 0 & 0 & 0 \\
  &  & [0] & [0] & [0] & [1] & [0] & $[-\frac{1}{3}]$ \\
$\bar B_s \to K^{0} \bar K^{*0}$ & 1 & 0 & 0 & 0 & 1 & 0 & $-\frac{1}{3}$ \\
  &  & [0] & [0] & [0] & [0] & [0] & [0] \\
$\bar B_s \to K^- K^{*+}$ & 1 & 0 & 0 & 0 & 0 & 0 & 0 \\
  &  & [1] & [0] & [0] & [1] & [0] & $[\frac{2}{3}]$  \\
$\bar B_s \to K^+ K^{*-}$ & 1 & 1 & 0 & 0 & 1 & 0 & $\frac{2}{3}$  \\
  &  & [0] & [0] & [0] & [0] & [0] & [0] \\
$\bar B_s \to \pi^0 ~\omega / \phi$ & $\frac{1}{2}$ & 0 & 0 & 0 & 0 & 0 & 0 \\
  &  & [0] & $[\sqrt{2} (s)]$ & [0] & [0] & $[\sqrt{2} (s)]$ & [0]  \\
$\bar B_s \to \eta^{(\prime)} \rho^0$ & $\frac{1}{2}$
  & 0 & $\sqrt{2} (s)$ & 0 & 0 & $\sqrt{2} (s)$ & 0  \\
  &  & [0] & [0] & [0] & [0] & [0] & [0] \\
$\bar B_s \to \eta^{(\prime)} ~\omega / \phi$ & $\frac{1}{2}$ & 0
  & {\footnotesize $\sqrt{2} (s,q)$} & {\footnotesize $2 \sqrt{2} (s,q) +2 (s,s)$}
  & {\footnotesize $2 (s,s)$} & {\footnotesize $\frac{\sqrt{2}}{3} (s,q)$}
  & {\footnotesize $- \frac{2}{3} (s,s)$}  \\
  &  &  &  &  & & $- \frac{2}{3} (s,s)$ &  \\
  &  & [0]
  & {\footnotesize $[\sqrt{2} (q,s)$} & {\footnotesize $[2 \sqrt{2} (q,s) +2 (s,s)$}
  & {\footnotesize $[2 (s,s)]$} & {\footnotesize $[\frac{\sqrt{2}}{3} (q,s)$}
  & {\footnotesize $[- \frac{2}{3} (s,s)]$}  \\
  &  &  & $+2 (c,s)]$ & $+2 (c,s)]$ & & $- \frac{2}{3} (s,s)]$ &
\\ \hline
\end{tabular}
\label{Bs_PV_DelS1_1}
\end{table}
%%%%%%%%%%%%%%%%%%%%%%%%%%%%%%%%%%%%%%%%%%%%%%%%%%%%%%%%%%%%%%%%%%

%%%%%%%%%%%%%%%%%%%%%%%%%%%%%%%%%%%%%%%%%%%%%%%%%%%%%%%%%%%%%%%%%%  Table 23
\begin{table}
\caption{({\it Continued from Table~\ref{Bs_PV_DelS1_1}})~ Weak annihilation contributions.
When ideal mixing for $\omega$ and $\phi$ is assumed, the same rules as used in
{\it Table~\ref{Bs_PV_DelS1_1}} are applied. }
\begin{tabular}{c||c|c|c|c|c|c|c}
\hline
$\bar B_s \to PV$ & factor & {\footnotesize $E_{P ~[V]}^{\prime (\zeta)}$}
  & {\footnotesize $A_{P ~[V]}^{\prime (\zeta)}$}
  & {\footnotesize $PE_{P ~[V]}^{\prime (\zeta)}$}
  & {\footnotesize $PA_{P ~[V]}^{\prime (\zeta)}$}
  & {\footnotesize $PE_{{\rm EW}, ~P ~[V]}^{\prime (\zeta)}$}
  & {\footnotesize $PA_{{\rm EW}, ~P ~[V]}^{\prime (\zeta)}$}
\\ \hline
$\bar B_s \to \pi^+ \rho^-$ & 1 & 0 & 0 & 0 & 1 & 0 & $-\frac{1}{3}$ \\
  &  & [1] & [0] & [0] & [1] & [0] & $[\frac{2}{3}]$ \\
$\bar B_s \to \pi^- \rho^+$ & 1 & 1 & 0 & 0 & 1 & 0 & $\frac{2}{3}$ \\
  &  & [0] & [0] & [0] & [1] & [0] & $[-\frac{1}{3}]$ \\
$\bar B_s \to \pi^0 \rho^0$ & $\frac{1}{2}$ & 1 & 0 & 0 & 2 & 0 & $\frac{1}{3}$ \\
  &  & [1] & [0] & [0] & [2] & [0] & $[\frac{1}{3}]$ \\
$\bar B_s \to \bar K^0 K^{*0}$ & 1 & 0 & 0 & 0 & 1 & 0 & $-\frac{1}{3}$ \\
  &  & [0] & [0] & [1] & [1] & $[-\frac{1}{3}]$ & $[-\frac{1}{3}]$ \\
$\bar B_s \to K^0 \bar K^{*0}$ & 1 & 0 & 0 & 1 & 1 & $-\frac{1}{3}$ & $-\frac{1}{3}$ \\
  &  & [0] & [0] & [0] & [1] & [0] & $[-\frac{1}{3}]$ \\
$\bar B_s \to K^- K^{*+}$ & 1 & 1 & 0 & 0 & 1 & 0 & $\frac{2}{3}$ \\
  &  & [0] & [0] & [1] & [1] & $[-\frac{1}{3}]$ & $[-\frac{1}{3}]$ \\
$\bar B_s \to K^+ K^{*-}$ & 1 & 0 & 0 & 1 & 1 & $-\frac{1}{3}$ & $-\frac{1}{3}$ \\
  &  & [1] & [0] & [0] & [1] & [0] & $[\frac{2}{3}]$ \\
$\bar B_s \to \pi^0 ~\omega / \phi$ & $\frac{1}{2}$ & $1 (q)$ & 0 & 0 & 0 & 0 & $1 (q)$ \\
  &  & $[1 (q)]$ & [0] & [0] & [0] & [0] & $[1 (q)]$  \\
$\bar B_s \to \eta^{(\prime)} \rho^0$ & $\frac{1}{2}$ & $1 (q)$ & 0 & 0 & 0 & 0 & $1 (q)$ \\
  &  & $[1 (q)]$ & [0] & [0] & [0] & [0] & $[1 (q)]$  \\
$\bar B_s \to \eta^{(\prime)} ~\omega / \phi$ & $\frac{1}{2}$ & $1 (q,q)$
  & 0 & {\footnotesize $2 (s,s)$}
  & {\footnotesize $2 (q,q)$} & {\footnotesize $-\frac{2}{3} (s,s)$}
  & {\footnotesize $\frac{1}{3} (q,q)$}  \\
  &  &  &  &  & $+2 (s,s)$ &  & $- \frac{2}{3} (s,s)$  \\
  &  & $[1 (q,q)]$
  & 0 & {\footnotesize $[2 (s,s)]$}
  & {\footnotesize $[2 (q,q)$} & {\footnotesize $[-\frac{2}{3} (s,s)]$}
  & {\footnotesize $[\frac{1}{3} (q,q)$}  \\
  &  &  &  &  & $+2 (s,s)]$ &  & $- \frac{2}{3} (s,s)]$
\\ \hline
\end{tabular}
\label{Bs_PV_DelS1_2}
\end{table}
%%%%%%%%%%%%%%%%%%%%%%%%%%%%%%%%%%%%%%%%%%%%%%%%%%%%%%%%%%%%%%%%%%

%%%%%%%%%%%%%%%%%%%%%%%%%%%%%%%%%%%%%%%%%%%%%%%%%%%%%%%%%%%%%%%%%%  Table 24
{\squeezetable
\begin{table}
\caption{({\it Continued from Table~\ref{Bs_PV_DelS1_2}})
Singlet weak annihilation contributions.
When ideal mixing for $\omega$ and $\phi$ is assumed, the same rules as used in
{\it Table~\ref{Bs_PV_DelS1_1}} are applied. }
\begin{tabular}{c||c|c|c|c|c|c|c}
\hline
$\bar B_s \to PV$ & factor & $SE_{P ~[V]}^{\prime (\zeta)}$ & $SA_{P ~[V]}^{\prime (\zeta)}$
  & $SPE_{P ~[V]}^{\prime (\zeta)}$ & $SPA_{P ~[V]}^{\prime (\zeta)}$
  & $SPE_{{\rm EW}, ~P ~[V]}^{\prime (\zeta)}$ & $SPA_{{\rm EW}, ~P ~[V]}^{\prime (\zeta)}$
\\ \hline
$\bar B_s \to \pi^0 ~\omega / \phi$ & $\frac{1}{2}$
  & $2 (q) +\sqrt{2} (s)$ & 0 & 0 & 0 & 0 & $2 (q) +\sqrt{2} (s)$ \\
  &  & [0] & [0] & [0] & [0] & [0] & [0] \\
$\bar B_s \to \eta^{(\prime)} \rho^0$ & $\frac{1}{2}$
  & 0 & 0 & 0 & 0 & 0 & 0 \\
  &  & $[2 (q) +\sqrt{2} (s)]$ & [0] & [0] & [0] & [0] & $[2 (q) +\sqrt{2} (s)]$ \\
$\bar B_s \to \eta^{(\prime)} ~\omega / \phi$
  & $\frac{1}{2}$ & $2 (q,q)$ & 0 & $2 \sqrt{2} (s,q)$ & $4 (q,q) +2 \sqrt{2} (q,s)$
  & $-\frac{2 \sqrt{2}}{3} (s,q)$ & $\frac{2}{3} (q,q) +\frac{\sqrt{2}}{3} (q,s)$  \\
  &  & $+\sqrt{2} (q,s)$ &  & $+2 (s,s)$ & $+2 \sqrt{2} (s,q) +2 (s,s)$ & $-\frac{2}{3} (s,s)$
  & $-\frac{2 \sqrt{2}}{3} (s,q) -\frac{2}{3} (s,s)$  \\
  &  & $[2 (q,q)$ & [0] & $[2 \sqrt{2} (q,s)$ & $[4 (q,q) +2 \sqrt{2} (s,q)$
  & $[-\frac{2 \sqrt{2}}{3} (q,s)$ & $[\frac{2}{3} (q,q) +\frac{\sqrt{2}}{3} (s,q)$  \\
  &  & $+\sqrt{2} (s,q)]$ &  & $+2 (s,s)]$ & $+2 \sqrt{2} (s,q) +2 (s,s)]$ & $-\frac{2}{3} (s,s)]$
  & $-\frac{2 \sqrt{2}}{3} (q,s) -\frac{2}{3} (s,s)]$
\\ \hline
\end{tabular}
\label{Bs_PV_DelS1_3}
\end{table}
}
%%%%%%%%%%%%%%%%%%%%%%%%%%%%%%%%%%%%%%%%%%%%%%%%%%%%%%%%%%%%%%%%%%

%%%%%%%%%%%%%%%%%%%%%%%%%%%%%%%%%%%%%%%%%%%%%%%%%%%%%%%%%%%%%%%%%%  Table 25
\begin{table}
\caption{Numerical values of the SU(3)$_{\rm F}$ amplitudes of $\bar B_{u,d} \to P_1 P_2$ decays with
$\Delta S = 0$ and $|\Delta S| = 1$ calculated in QCD factorization.
The magnitude (in units of $10^{-9}$ GeV) and strong phase (in degrees) of each SU(3)$_{\rm F}$
amplitude are shown in order within the parenthesis : e.g., for the tree amplitude
$T_P \equiv |T_P| ~e^{i (\delta_P +\theta_P)}$ with $\delta_P$ and $\theta_P$ being the strong and
weak phases, respectively, its magnitude and strong phase are shown as $(|T_P|, ~ \delta_P)$. }
\vspace{0.4cm}
\begin{tabular}{l|l||l|l}
\hline
~$\Delta S = 0$  & ~ Numerical values ~ & ~$|\Delta S| = 1$ & ~ Numerical values
\\ \hline
~$T_P$ & \quad $(24.52,~ 0.9^{\circ})$ & ~$T'_P$ & \quad $(6.90,~ 0.9^{\circ})$  \\
~$C_P$ & \quad $(15.47,~ -54.8^{\circ})$ & ~$C'_P$ & \quad $(4.48,~ -56.6^{\circ})$   \\
~$S^{(q)}_P$ & \quad $(0.87,~ 159.1^{\circ})$
  & ~$S^{\prime (q)}_P$ & \quad $(4.11,~ 150.3^{\circ})$ \\
~$S^{(s)}_P$ & \quad $(0.89,~ 159.1^{\circ})$
  & ~$S^{\prime (s)}_P$ & \quad $(4.21,~ 150.3^{\circ})$ \\
~$S^{(c)}_P$ & \quad $(0.02,~ 159.1^{\circ})$
  & ~$S^{\prime (c)}_P$ & \quad $(0.09,~ 150.3^{\circ})$ \\
~$P_P$ & \quad $(5.59,~ -157.7^{\circ})$ & ~$P'_P$ & \quad $(34.25,~ -157.4^{\circ})$ \\
~$P_{{\rm EW}, ~P}$ & \quad $(0.82,~ -178.9^{\circ})$
  & ~$P'_{{\rm EW}, ~P}$ & \quad $(5.48,~ -178.9^{\circ})$ \\
~$P^C_{{\rm EW}, ~P}$ & \quad $(0.17,~ 163.9^{\circ})$
  & ~$P^{C \prime}_{{\rm EW}, ~P}$ & \quad $(1.05,~ 163.1^{\circ})$ \\
~$E_P$ & \quad $(1.96,~ 52.7^{\circ})$ & ~$E'_P$ & \quad $(0.54,~ 52.9^{\circ})$ \\
~$A_P$ & \quad $(0.61,~ -127.3^{\circ})$ & ~$A'_P$ & \quad $(0.17,~ -127.1^{\circ})$ \\
~$PE_P$ & \quad $(3.79,~ -146.2^{\circ})$ & ~$PE'_P$ & \quad $(21.19~,~ -146.2^{\circ})$ \\
~$PA_P$ & \quad $(0.61,~ -127.3^{\circ})$ & ~$PA'_P$ & \quad $(3.46,~ -127.1^{\circ})$ \\
~$PE_{{\rm EW}, ~P}$ & \quad $(0.02,~ -34.1^{\circ})$
  & ~$PE'_{{\rm EW}, ~P}$ & \quad $(0.13,~ -36.1^{\circ})$ \\
~$PA_{{\rm EW}, ~P}$ & \quad $(0.03,~ 52.7^{\circ})$
  & ~$PA'_{{\rm EW}, ~P}$ & \quad $(0.15,~ 52.9^{\circ})$
\\ \hline
\end{tabular}
\label{Abs_B_PP}
\end{table}
%%%%%%%%%%%%%%%%%%%%%%%%%%%%%%%%%%%%%%%%%%%%%%%%%%%%%%%%%%%%%%%%%%

%%%%%%%%%%%%%%%%%%%%%%%%%%%%%%%%%%%%%%%%%%%%%%%%%%%%%%%%%%%%%%%%%%  Table 26
{\squeezetable
\begin{table}
\caption{Same as Table~\ref{Abs_B_PP} except for $\bar B_{u,d} \to PV$ decays :
e.g., for the tree amplitudes $(T_P ~;~ T_V )$ where
$T_{P,V} \equiv |T_{P,V}| ~e^{i (\delta_{P,V} +\theta_{P,V})}$ with $\delta_{P,V}$ and $\theta_{P,V}$
being the strong and weak phases, respectively, their magnitudes (in units of $10^{-9}$ GeV) and
strong phases (in degrees) are shown as $( |T_P|, ~ \delta_P ~;~ |T_V|, ~ \delta_V )$. }
\vspace{0.4cm}
\begin{tabular}{l|l||l|l}
\hline
~ \quad $\Delta S = 0$  & ~~  Numerical values ~~ & ~ \quad $|\Delta S| = 1$ & ~~  Numerical values
\\ \hline
~( $T_P$~; ~$T_V$ )~ & ~( $40.82 ,~ 0.8^{\circ}~; ~30.16 ,~ 0.8^{\circ}$)~
  & ~( $T'_P$~; ~$T'_V$ )~ & ~( $9.63 ,~ 0.8^{\circ}~; ~8.54 ,~ 0.8^{\circ}$)  \\
~( $C_P$~; ~$C_V$ )~ & ~( $12.30 ,~ -16.0^{\circ}~; ~11.84 ,~ -51.7^{\circ}$)~
  & ~( $C'_P$~; ~$C'_V$ )~ & ~( $3.38 ,~ -19.0^{\circ}~; ~2.79 ,~ -55.4^{\circ}$)  \\
~( $S^{(q)}_P$~; ~$S^{(q)}_V$ )~ & ~( $0.50 ,~ -4.7^{\circ}~; ~0.79 ,~ 151.5^{\circ}$)~
  & ~( $S^{\prime (q)}_P$~; ~$S^{\prime (q)}_V$ )~
  & ~( $2.75 ,~ -5.6^{\circ}~; ~3.04 ,~ 134.2^{\circ}$)~  \\
~( $S^{(s)}_P$~; ~$S^{(s)}_V$ )~ & ~( $0.59 ,~ -4.6^{\circ}~; ~0.81 ,~ 151.5^{\circ}$)~
  & ~( $S^{\prime (s)}_P$~; ~$S^{\prime (s)}_V$ )~
  & ~( $3.22 ,~ -5.5^{\circ}~; ~3.11 ,~ 134.2^{\circ}$)~  \\
~( $S^{(c)}_P$~; ~$S^{(c)}_V$ )~ & ~( $-~; ~0.02 ,~ 151.5^{\circ}$)~
  & ~( $S^{\prime (c)}_P$~; ~$S^{\prime (c)}_V$ )~
  & ~( $-~; ~0.07 ,~ 134.2^{\circ}$)~  \\
~( $P_P$~; ~$P_V$ )~ & ~( $3.04 ,~ -144.7^{\circ}~; ~3.12 ,~ 8.4^{\circ}$)~
  & ~( $P'_P$~; ~$P'_V$ )~ & ~( $17.99 ,~ -144.4^{\circ}~; ~17.02 ,~ 7.9^{\circ}$)  \\
~( $P_{{\rm EW}, ~P}$~; ~$P_{{\rm EW}, ~V}$)~
  & ~( $1.41 ,~ -179.4^{\circ}~; ~1.01 ,~ -178.9^{\circ}$)~
  & ~( $P'_{{\rm EW}, ~P}$~; ~$P'_{{\rm EW}, ~V}$)~
  & ~( $9.39 ,~ -179.4^{\circ}~; ~6.00 ,~ -178.9^{\circ}$)  \\
~( $P^C_{{\rm EW}, ~P}$~; ~$P^C_{{\rm EW}, ~V}$)~
  & ~( $0.38 ,~ 165.0^{\circ}~; ~0.34 ,~ 164.7^{\circ}$)~
  & ~( $P^{C \prime}_{{\rm EW}, ~P}$~; ~$P^{C \prime}_{{\rm EW}, ~V}$)~
  & ~( $1.76 ,~ 165.1^{\circ}~; ~1.87 ,~ 160.6^{\circ}$)  \\
~( $E_P$~; ~$E_V$ )~ & ~( $2.46 ,~ 70.1^{\circ}~; ~2.29 ,~ 38.9^{\circ}$)~
  & ~( $E'_P$~; ~$E'_V$ )~ & ~( $0.60 ,~ 69.4^{\circ}~; ~0.64 ,~ 39.2^{\circ}$)  \\
~( $A_P$~; ~$A_V$ )~ & ~( $0.77 ,~ -109.9^{\circ}~; ~0.72 ,~ -141.1^{\circ}$)~
  & ~( $A'_P$~; ~$A'_V$ )~ & ~( $0.19 ,~ -110.6^{\circ}~; ~0.20 ,~ -140.8^{\circ}$)  \\
~( $PE_P$~; ~$PE_V$ )~ & ~( $3.82 ,~ -123.6^{\circ}~; ~3.83 ,~ 5.3^{\circ}$)~
  & ~( $PE'_P$~; ~$PE'_V$ )~ & ~( $19.83 ,~ -124.2^{\circ}~; ~21.20 ,~ 5.0^{\circ}$)  \\
~( $PA_P$~; ~$PA_V$ )~ & ~( $0.12 ,~ 70.1^{\circ}~; ~0.11 ,~ 38.9^{\circ}$)~
  & ~( $PA'_P$~; ~$PA'_V$ )~ & ~( $0.61 ,~ 69.4^{\circ}~; ~0.66 ,~ 39.2^{\circ}$)  \\
~( $PE_{{\rm EW}, ~P}$~; $PE_{{\rm EW}, ~V}$)~
  & ~( $0.02 ,~ -57.8^{\circ}~; ~0.14 ,~ -157.3^{\circ}$)~
  & ~( $PE'_{{\rm EW}, ~P}$~; $PE'_{{\rm EW}, ~V}$)~
  & ~( $0.10 ,~ -50.7^{\circ}~; ~0.77 ,~ -157.0^{\circ}$)  \\
~( $PA_{{\rm EW}, ~P}$~; $PA_{{\rm EW}, ~V}$)~
  & ~( $0.02 ,~ 70.1^{\circ}~; ~0.02 ,~ 38.9^{\circ}$)~
  & ~( $PA'_{{\rm EW}, ~P}$~; $PA'_{{\rm EW}, ~V}$)~
  & ~( $0.08 ,~ 69.4^{\circ}~; ~0.09 ,~ 39.2^{\circ}$)
\\ \hline
\end{tabular}
\label{Abs_B_PV}
\end{table}
}
%%%%%%%%%%%%%%%%%%%%%%%%%%%%%%%%%%%%%%%%%%%%%%%%%%%%%%%%%%%%%%%%%%

%%%%%%%%%%%%%%%%%%%%%%%%%%%%%%%%%%%%%%%%%%%%%%%%%%%%%%%%%%%%%%%%%%  Table 27
\begin{table}
\caption{Same as Table~\ref{Abs_B_PP} except for $\bar B_s \to P_1 P_2$ decays. }
\vspace{0.4cm}
\begin{tabular}{l|l||l|l}
\hline
~$\Delta S = 0$  & ~ Numerical values ~ & ~$|\Delta S| = 1$ & ~ Numerical values
\\ \hline
~$T_P$ & \quad $(23.54,~ 0.9^{\circ})$ & ~$T'_P$ & \quad $(6.61,~ 0.8^{\circ})$  \\
~$C_P$ & \quad $(19.57,~ -51.6^{\circ})$ & ~$C'_P$ & \quad $(5.72,~ -50.5^{\circ})$ \\
~$S^{(q)}_P$ & \quad $(1.31,~ 166.6^{\circ})$
  & ~$S^{\prime (s,q)}_P$ & \quad $(3.53,~ 162.5^{\circ})$  \\
~$S^{(s)}_P$ & \quad $(1.35,~ 166.6^{\circ})$
  & ~$S^{\prime (s,s)}_P$ & \quad $(3.62,~ 162.5^{\circ})$  \\
~$S^{(c)}_P$ & \quad $(0.03,~ 166.6^{\circ})$
  & ~$S^{\prime (s,c)}_P$ & \quad $(0.08,~ 162.5^{\circ})$  \\
~$P_P$ & \quad $(5.64,~ -158.1^{\circ})$ & ~$P'_P$ & \quad $(34.69,~ -157.9^{\circ})$  \\
~$P_{{\rm EW}, ~P}$ & \quad $(0.78,~ -178.9^{\circ})$
  & ~$P'_{{\rm EW}, ~P}$ & \quad $(4.46,~ -178.8^{\circ})$ \\
~$P^C_{{\rm EW}, ~P}$ \quad & \quad $(0.25,~ 170.0^{\circ})$
  & ~$P^{C \prime}_{{\rm EW}, ~P}$ & \quad $(1.58,~ 171.0^{\circ})$  \\
~$E_P$ & \quad $(2.35,~ 51.2^{\circ})$ & ~$E'_P$ & \quad $(0.65,~ 51.4^{\circ})$  \\
~$A_P$ & \quad $(0.74,~ -128.8^{\circ})$ & ~$A'_P$ & \quad $(0.20,~ -128.6^{\circ})$  \\
~$PE_P$ & \quad $(4.31,~ -149.1^{\circ})$ & ~$PE'_P$ & \quad $(24.12,~ -149.1^{\circ})$  \\
~$PA_P$ & \quad $(0.74,~ -128.8^{\circ})$ & ~$PA'_P$ & \quad $(4.18,~ -128.6^{\circ})$  \\
~$PE_{{\rm EW}, ~P}$ & \quad $(0.03,~ -46.2^{\circ})$
  & ~$PE'_{{\rm EW}, ~P}$~ & \quad $(0.15,~ -48.4^{\circ})$  \\
~$PA_{{\rm EW}, ~P}$ & \quad $(0.03,~ 51.2^{\circ})$
  & ~$PA'_{{\rm EW}, ~P}$~ & \quad $(0.18,~ 51.4^{\circ})$
\\ \hline
\end{tabular}
\label{Abs_Bs_PP}
\end{table}
%%%%%%%%%%%%%%%%%%%%%%%%%%%%%%%%%%%%%%%%%%%%%%%%%%%%%%%%%%%%%%%%%%

%%%%%%%%%%%%%%%%%%%%%%%%%%%%%%%%%%%%%%%%%%%%%%%%%%%%%%%%%%%%%%%%%%  Table 28
{\squeezetable
\begin{table}
\caption{Same as Table~\ref{Abs_B_PV} except for $\bar B_s \to PV$ decays. }
\vspace{0.4cm}
\begin{tabular}{l|l||l|l}
\hline
~ \quad $\Delta S = 0$  & ~~ Numerical values ~~ & ~ \quad $|\Delta S| = 1$ & ~~  Numerical values
\\ \hline
~( $T_P$~; ~$T_V$ )~ & ~( $39.33,~ 0.9^{\circ} ~; ~30.59,~ 0.8^{\circ}$)~
  & ~( $T'_P$~; ~$T'_V$ )~ & ~( $9.28,~ 0.9^{\circ} ~; ~8.60,~ 0.9^{\circ}$)  \\
~( $C_P$~; ~$C_V$ )~ & ~( $15.49,~ -12.5^{\circ} ~; ~12.90,~ -50.6^{\circ}$)~
  & ~( $C'_P$~; ~$C'_V$ )~ & ~( $3.59,~ -13.0^{\circ} ~; ~3.82,~ -50.7^{\circ}$)  \\
~( $S^{(q)}_P$~; ~$S^{(q)}_V$ )~ & ~( $0.64,~ -3.6^{\circ} ~; ~0.91,~ 155.3^{\circ}$)~
  & ~( $S^{\prime (s,q)}_P$~; ~$S^{\prime (q,s)}_V$ )~
  & ~( $1.90,~ -4.3^{\circ} ~; ~4.56,~ 155.1^{\circ}$)~  \\
~( $S^{(s)}_P$~; ~$S^{(s)}_V$ )~ & ~( $0.76,~ -3.6^{\circ} ~; ~0.94,~ 155.3^{\circ}$)~
  & ~( $S^{\prime (s,s)}_P$~; ~$S^{\prime (s,s)}_V$ )~
  & ~( $2.24,~ -4.2^{\circ} ~; ~4.68,~ 155.1^{\circ}$)~  \\
~( $S^{(c)}_P$~;~$S^{(c)}_V$ )~ & ~( $-$~; ~$0.02,~ 155.3^{\circ}$)~
  & ~( $S^{\prime (s,c)}_P$~; ~$S^{\prime (c,s)}_V$ )~
  & ~( $-$~; ~$0.10,~ 155.1^{\circ}$)~  \\
~( $P_P$~; ~$P_V$ )~ & ~( $2.82,~ -142.7^{\circ} ~; ~3.51,~ 7.7^{\circ}$ )~
  & ~( $P'_P$~; ~$P'_V$ )~ & ~( $16.71,~ -142.1^{\circ} ~; ~19.83,~ 6.7^{\circ}$)  \\
~( $P_{{\rm EW}, ~P}$~; ~$P_{{\rm EW}, ~V}$ )~
  & ~( $1.38,~ -179.4^{\circ} ~; ~1.03,~ -178.9^{\circ}$)~
  & ~( $P'_{{\rm EW}, ~P}$~; ~$P'_{{\rm EW}, ~V}$ )~
  & ~( $6.61,~ -179.4^{\circ} ~; ~5.89,~ -178.9^{\circ}$)  \\
~( $P^C_{{\rm EW}, ~P}$~; ~$P^C_{{\rm EW}, ~V}$ )~
  & ~( $0.48,~ 168.6^{\circ} ~; ~0.36,~ 164.9^{\circ}$)~
  & ~( $P^{C \prime}_{{\rm EW}, ~P}$~; ~$P^{C \prime}_{{\rm EW}, ~V}$ )~
  & ~( $2.30,~ 168.2^{\circ} ~; ~2.10,~ 164.6^{\circ}$)  \\
~( $E_P$~; ~$E_V$ )~ & ~( $3.24,~ 70.5^{\circ} ~; ~2.44,~ 45.0^{\circ}$)~
  & ~( $E'_P$~; ~$E'_V$ )~ & ~( $0.79,~ 69.8^{\circ} ~; ~0.68,~ 45.4^{\circ}$)  \\
~( $A_P$~; ~$A_V$ )~ & ~( $1.01,~ -109.5^{\circ} ~; ~0.76,~ -135.0^{\circ}$)~
  & ~( $A'_P$~; ~$A'_V$ )~ & ~( $0.25,~ -110.2^{\circ} ~; ~0.21,~ -134.6^{\circ}$)  \\
~( $PE_P$~; ~$PE_V$ )~ & ~( $4.87,~ -123.4^{\circ} ~; ~4.01,~ 16.3^{\circ}$)~
  & ~( $PE'_P$~; ~$PE'_V$ )~ & ~( $25.33,~ -124.0^{\circ} ~; ~22.29,~ 16.1^{\circ}$)  \\
~( $PA_P$~; ~$PA_V$ )~ & ~( $0.16,~ 70.5^{\circ} ~; ~0.12,~ 45.0^{\circ}$)~
  & ~( $PA'_P$~; ~$PA'_V$ )~ & ~( $0.81,~ 69.8^{\circ} ~; ~0.70,~ 45.4^{\circ}$)  \\
~( $PE_{{\rm EW}, ~P}$~; ~$PE_{{\rm EW}, ~V}$ )~
  & ~( $0.03,~ -63.1^{\circ} ~; ~0.15,~ -148.7^{\circ}$)~
  & ~( $PE'_{{\rm EW}, ~P}$~; ~$PE'_{{\rm EW}, ~V}$ )~
  & ~( $0.13,~ -56.1^{\circ} ~; ~0.82,~ -148.4^{\circ}$) \\
~( $PA_{{\rm EW}, ~P}$~; ~$PA_{{\rm EW}, ~V}$ )~
  & ~( $0.02,~ 70.5^{\circ} ~; ~0.02,~ 45.0^{\circ}$)~
  & ~( $PA'_{{\rm EW}, ~P}$~; ~$PA'_{{\rm EW}, ~V}$ )~
  & ~( $0.11,~ 69.8^{\circ} ~; ~0.09,~ 45.4^{\circ}$)
\\ \hline
\end{tabular}
\label{Abs_Bs_PV}
\end{table}
}
%%%%%%%%%%%%%%%%%%%%%%%%%%%%%%%%%%%%%%%%%%%%%%%%%%%%%%%%%%%%%%%%%%

%%%%%%%%%%%%%%%%%%%%%%%%%%%%%%%%%%%%%%%%%%%%%%%%%%%%%%%%%%%%%%%%%%%%%%%%%%%%%%%%%%%%%%%
%%%%%%%%%%%%%%%%%%%%%%%%%%%%%%%%%%%%%%%%%%%%%%%%%%%%%%%%%%%%%%%%%%%%%%%%%%%%%%%%%%%%%%%
\clearpage
%==============================================================================
%                               References
%==============================================================================

\end{document}